# Anomalous deflection of an electron after passing through magnetic-field zero point

Changggen Zou[1], graduated from Department of Radio Engineering, Southeast University.
Wanshou Jiang, State Key Laboratory of Information Engineering in Surveying, Mapping and Remote Sensing, Wuhan University.

(July 6, 2026 )

**Abstract:** This paper reports an experiment about anomalous deflection of cathode ray in odd-symmetric magnetic field. The experiments shows that after passing through the magnetic-field zero point formed by two opposing magnetic fields, an electron deflects in the direction opposite to the Lorentz force within several tens of micrometres from the zero point. It can be explained by the inertial effect of the electron rotating on its axis in magnetic field, and Lorentz force is similar to the Magnus effect in fluid. In this paper, a mechanical model that replaces potential energy with rotational kinetic energy is used to calculate the force exerted on an electron and a proton under different conditions, and the Maxwell's equations of electromagnetic field are derived.

**Key words:** Lorentz force; electron rotation; inertial effect; Magnus effect; Maxwell's equations

## 1. Introduction.

Lorentz force is the basis of classical electromagnetism, but there is no ideal mechanical model for Lorentz force. Cathode-ray tube is easy to generate high-speed electron beam, and its speed can reach one tenth or more of the speed of light. It is an ideal device for measuring Lorentz force. Based on the experiment of measuring electron charge-to-mass ratio with cathode-ray tube, this experiment uses modern digital photography technology to measure the characteristics of moving electron deflected by Lorentz force in magnetic field more accurately, and differential measurement results is used to improve reliability of the measurement.

The measurement results show that after passing through the magnetic-field zero point of an odd-symmetric magnetic field, the cathode ray deflects in the direction opposite to the Lorentz force within a very short distance. This result, which is contradictory to electromagnetism, can be explained by the inertial effect of the electron rotating on its axis in magnetic field, and Lorentz force is similar to the Magnus effect in fluid. In this paper, a mechanical model that replaces potential energy with rotational kinetic energy is used to calculate the force exerted on an electron and a proton under different conditions, and the Maxwell's equations of electromagnetic field are derived. The experimental measurement, related analysis and calculation are helpful to reveal the mechanical essence of Lorentz force.

## 2. Theoretical deflection of cathode ray.

Firstly, the method of measuring the charge-to-mass ratio of electrons by magnetic focusing is briefly described; it is improved from the charge-to-mass ratio experiment. As shown in Fig-1, an electron with electric charge of $q$ and mass of $m$, its three-dimensional

1 Corresponding author. Email addresses: jws@whu.edu.cn

components of initial velocity are $V_X$ $V_Y$ $V_Z$ respectively and $V_Z = 0$, when the electron enters a constant magnetic field parallel to the X-axis with magnetic induction intensity $B$, it does a uniform linear motion in the X-axis direction and a uniform circular motion in the YZ plane, trajectory of the electron is a helical line.

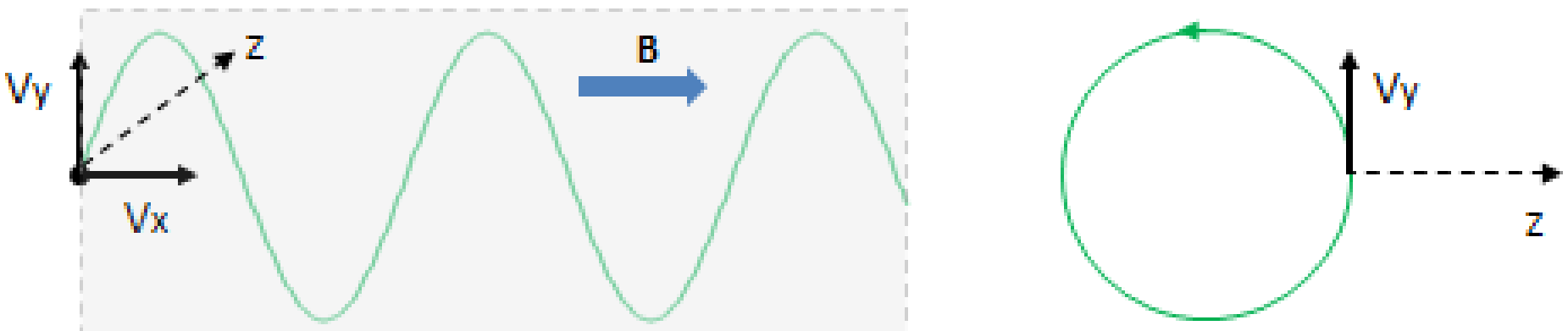

**Fig-1** Helical trajectory of an electron

Assume that $r$ is radius of the uniform circular motion of the electron, the centripetal force is equal to the Lorentz force, so

$$\frac{mV_Y^2}{r} = qV_YB$$

The time required for an electron to circle in YZ plane is $T$ and the angular velocity of uniform circular motion is $\omega_e$, so

$$T = \frac{2\pi r}{V_Y} \qquad \omega_e = \frac{2\pi}{T} = \frac{q}{m} * B$$

$\omega_e$ is proportional to the charge-to-mass ratio $q/m$ of the electron and has nothing to do with the value of $V_Y$. The direction of the uniform circular motion of the electron in the YZ plane changes while the direction of $V_Y$ changes; the radius $r$ of the uniform circular motion of the electron in the YZ plane changes while the magnitude of $V_Y$ changes. All electrons with different $V_Y$ have the same angular velocity $\omega_e$ moving around their respective centres, all electrons starting from a same point will arrive at another same point after a period of T, and this phenomenon is called magnetic focusing. Projection of the trajectories of electrons on the YZ plane is shown in Fig-2.

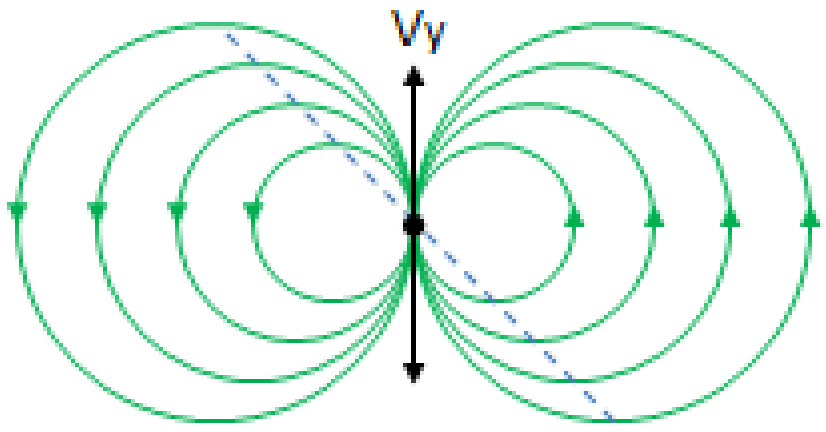

**Fig-2** Projection of the trajectories of electrons

In the experiment of measuring the charge-to-mass ratio of electron with cathode-ray tube, the velocity component $V_X$ of electron is generated by a constant accelerating voltage $U_X$, and the velocity component $V_Y$ is generated by a periodically alternating voltage $U_Y$, as shown in Fig-3. The alternating voltage $U_Y$ causes the change of magnitude and direction of $V_Y$, but while electrons reach screen of the cathode-ray tube, they revolve the same angle on the YZ plane and therefore the image on screen is always a straight line. Adjust the magnitude of the magnetic induction intensity B, the line image will rotate and be gradually shortened into a point. The above is the method of measuring charge-to-mass ratio of electron by magnetic focusing.

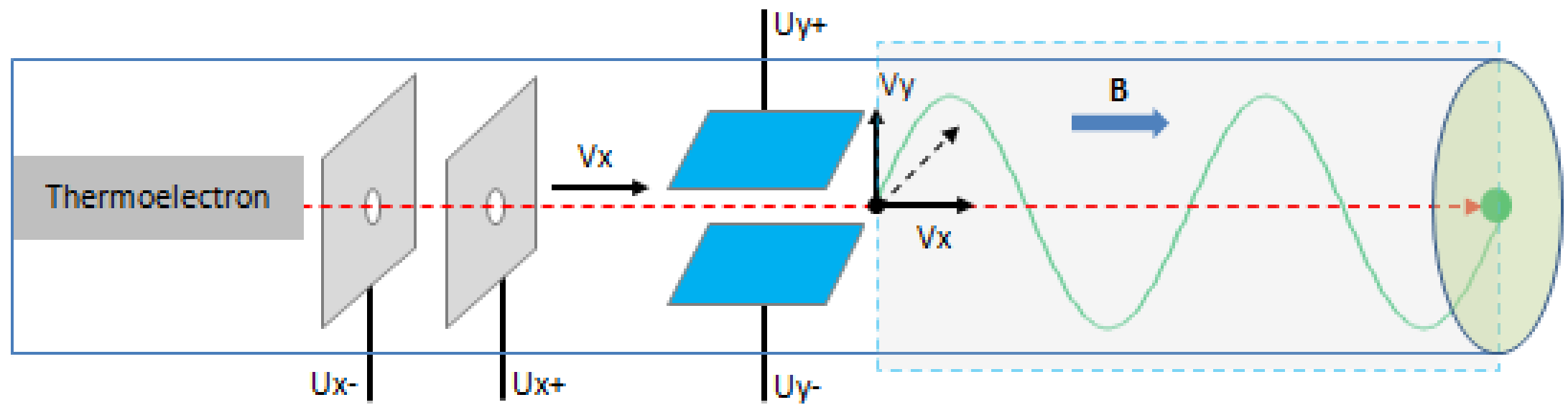

**Fig-3** Accelerating and alternating voltage of the cathode-ray tube

Assume that the cathode-ray tube in Fig-3 can be lengthened in the X-axis direction, the line image on screen will rotate with a tiny angle $d\theta$ when the distance of electrons movement increases by $dx$.

$$d\theta = \frac{\omega_e}{2} * \frac{dx}{V_X} = \frac{qB}{2m} * \frac{dx}{V_X} = \frac{q}{2mV_X} * Bdx$$

Suppose that the direction of magnetic induction intensity B is still parallel to the X-axis but its magnitude is a function $B_X(x)$ that varies with $x$, and then $d\theta$ becomes

$$d\theta \approx \frac{q}{2mV_X} * B(x)dx$$

Fig-4 shows the main device used to measure deflection angle of the image. The diameter D of the screen of cathode-ray tube is much smaller than the distance L between electron gun and the screen. Near the screen of the cathode-ray tube, two identical and reverse-connected coils are used as excitation coils; they generate axial magnetic fields of equal magnitude in opposite directions. These two coils are called a coil-pair, it can be moved left and right, and its axis overlaps with that of the cathode-ray tube.

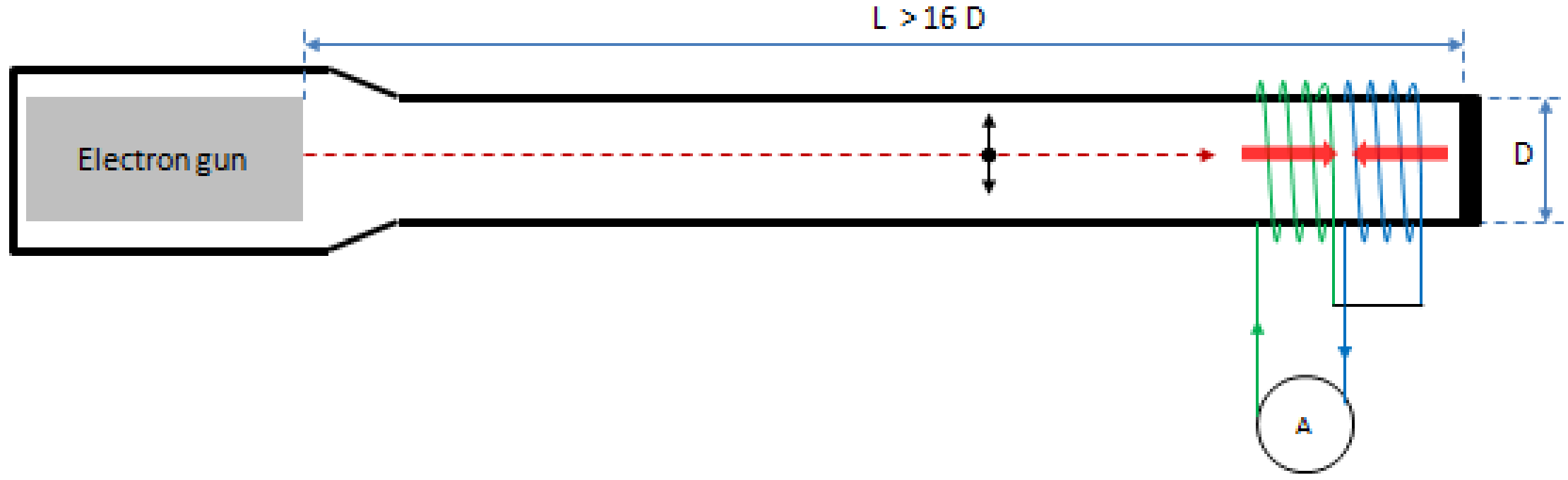

**Fig-4** The cathode-ray tube and the coil-pair

Taking axis of the cathode-ray tube as the X-axis and direction of electron motion as the positive direction of the X-axis, the midpoint of the coil-pair is the contact surface position of the two coils, taking the midpoint of the coil-pair as the origin of X-axis. When the current of constant current source remains unchanged, axial magnetic induction intensity $B_X(x)$ of the coil-pair is shown in Fig-5, $B_X(x)$ is an odd function.

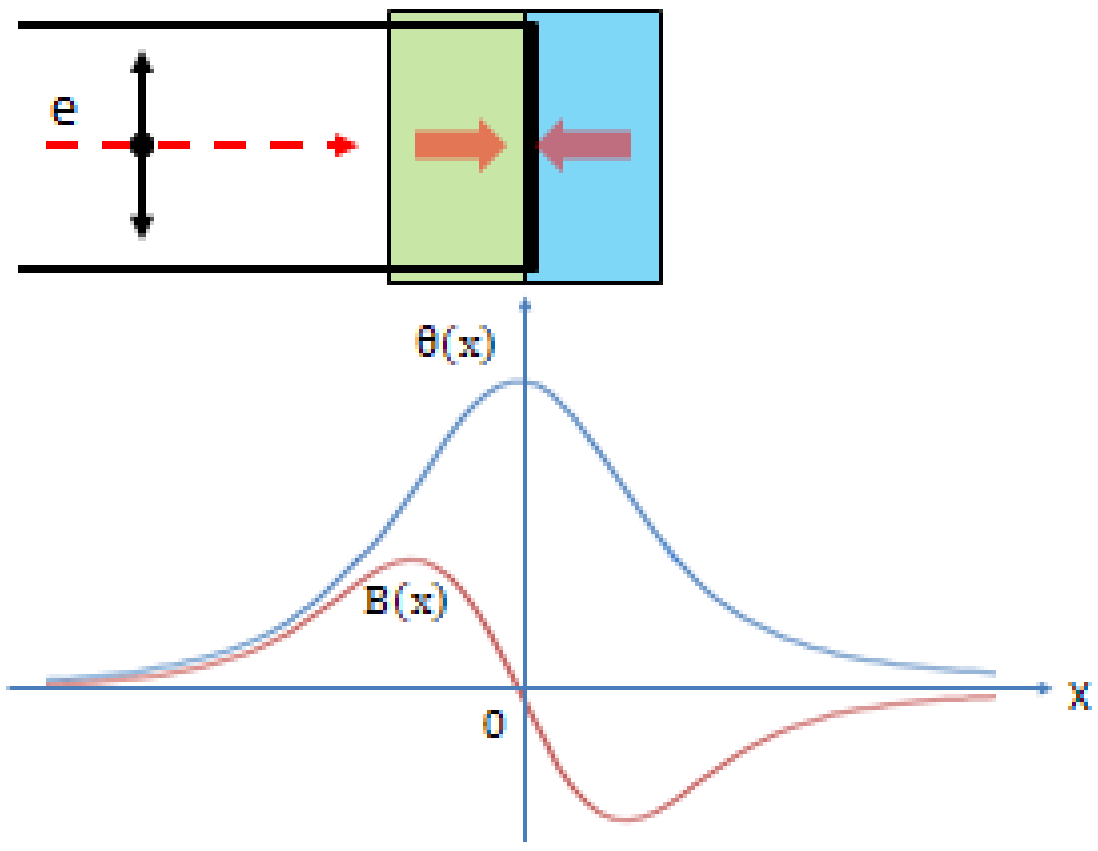

**Fig-5** Theoretical curves of magnetic induction intensity and deflection angle $\theta$

The line image of the cathode ray is used as the reference line while the current of constant current source is zero. When position of the inner surface of the screen of the cathode-ray tube is $x$, deflection angle $\theta(x)$ of the line image is a function as follows

$$\theta(x) \approx \int_{-\infty}^{x} d\theta = \frac{q}{2mV_X} \int_{-\infty}^{x} B(\tau) d\tau$$

$B_X(x)$ is an odd function, therefore $\theta(x)$ is an even function as shown in Fig-5, $x = 0$ is its extreme point. Assume that the cathode-ray tube can be stretched in X-axis direction, while position of the inner surface of the screen moves from $x < 0$ to $x > 0$, it can be found that deflection angle of the line image increases gradually and reaches the maximum at $x = 0$, and then the line image begins to reverse and deflection angle decreases gradually, as shown in Fig-6.

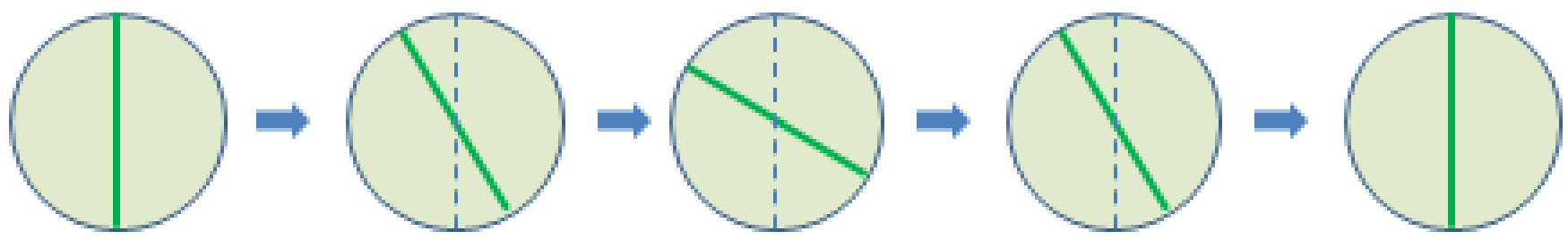

**Fig-6** Theoretical deflection angle variation

Keeping magnitude of the current of constant current source unchanged and changing the direction of the current, rotation characteristics of the line image should be exactly the same except direction of rotation. These results can be predicted by existing knowledge of electromagnetics.

The motion of electrons in a non-uniform constant magnetic field is complex, and the value of deflection angle $\theta(x)$ cannot be calculated directly by integration. The magnitude $V_Y$ of velocity component of electron in the Y-axis direction is marked as the speed variable $v$. When deflection voltage in Y-axis direction is a periodic alternating signal, the image on cathode-ray tube screen is composed of afterglow generated by electrons with different speed $v$ impacting the fluorescent screen after being accelerated by voltage $U_X$ and deflected by magnetic field. Since the axial magnetic field magnitude inside the coil-pair varies with the distance from the central axis, the image is not a straight line but an odd-symmetric S-shaped curve.

Fig-6 shows that the deflection angle $\theta(x)$ of the straight-line image reaches an extremum at $x = 0$. Similarly, the tangent angle $\psi(x)$ of the S-shaped line pattern reaches an extremum near $x = 0$. Choose an appropriate excitation current such that the line pattern

shortens with increasing $x$. Let the maximum of $\psi(x)$ be located at $x = x_1$, we have $x_1 \leq 0$, meaning that the maximum point of $\psi(x)$ lies to the left of the coil-pair centre. For its mathematical proof, please refer to appendix E "Calculation of the point where the axial magnetic field drops to zero".

## 3. Actual deflection of cathode ray.

If we improve measuring accuracy, we will find that the measurement results are different from the predicted value discussed in previous. The extreme point of $\psi(x)$ will change to $x > 0$. While magnitude of the current of constant current source is unchanged but direction is changed, the distance from the extreme point of $\psi(x)$ to the position of $x = 0$ also changes, and the extreme point of $\psi(x)$ deviates more from $x = 0$ when the direction of magnetic field which electrons enter first is opposite to the direction of electrons motion, as shown in Fig-7, $b > a$ .

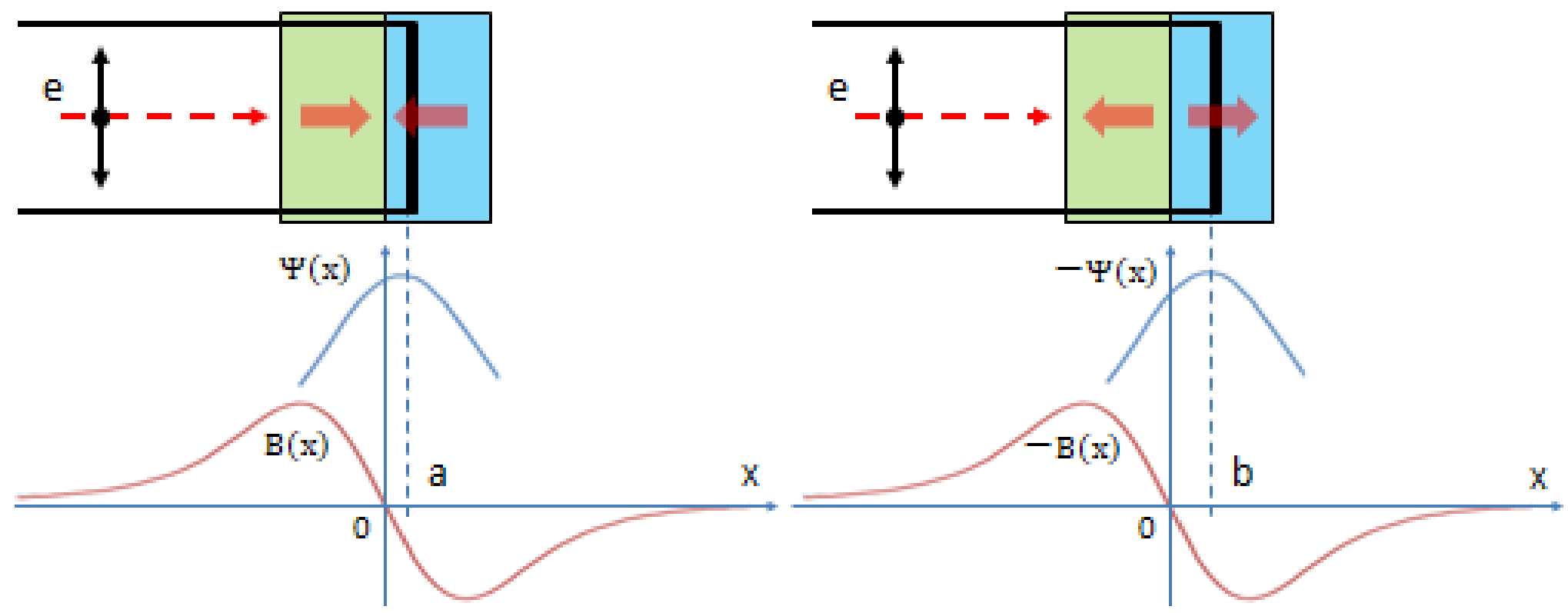

**Fig-7** Actual curves of magnetic induction intensity and inclination angle $\psi$ of tangent

Moving the cathode-ray tube so that position of the inner surface of the screen moves from $x < 0$ to $x > 0$, because the direction of Lorentz force related to $V_Y$ on electrons changes at $x = 0$, the line pattern should have begun to reverse at $x = 0$. However, the measurement results shown in Fig-7 indicate that the line pattern continues to rotate for a distance in original direction, indicating that the electron is exerted a force opposite to Lorentz force within this distance, and this force is called as "reversed Lorentz force".

## 4. Method for measuring cathode ray deflection.

Although the measurement result is difficult to explain with the existing electromagnetic knowledge, it is the result predicted in appendix A. In order to verify this prediction, we spent a lot of time in customizing several special cathode-ray tubes. The shape of the cathode-ray tube that can be used to measure is slender (about 470 mm in length, but the screen is only 20 mm in diameter, equivalent to a coin), and the exact value of thickness of the screen must be known (for example 1.20 mm), its photo is shown in Fig-8. The reason why diameter $D$ of the screen of the cathode-ray tube is much smaller than its length $L$ ($L > 16D$) is that the magnetic induction intensity $B(x)$ of coil-pair must have enough decreasing distance to ensure that $B(x)$ is an odd function.

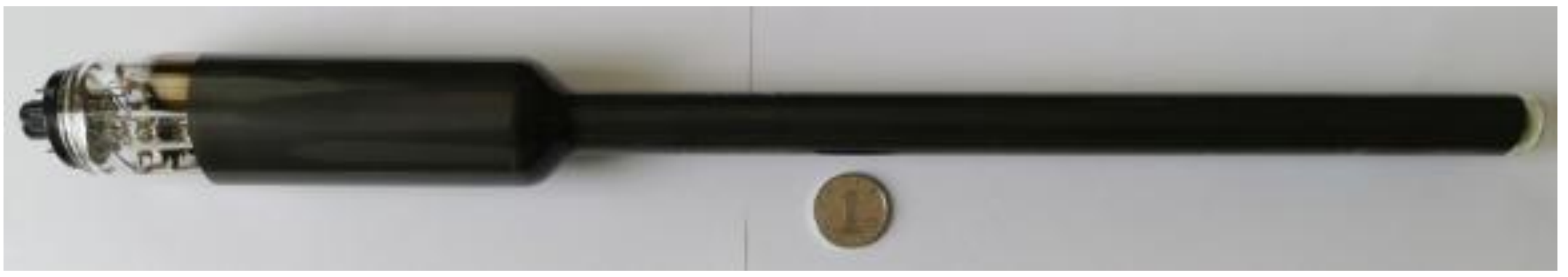

**Fig-8** Physical photo of the cathode-ray tube

The deflection of line pattern caused by the reversed Lorentz force is very weak, so that it is difficult to be measured. We used a high-pixel industrial camera to take photos of the screen of the cathode-ray tube, and the inclination angle of tangent of line pattern can be calculated.

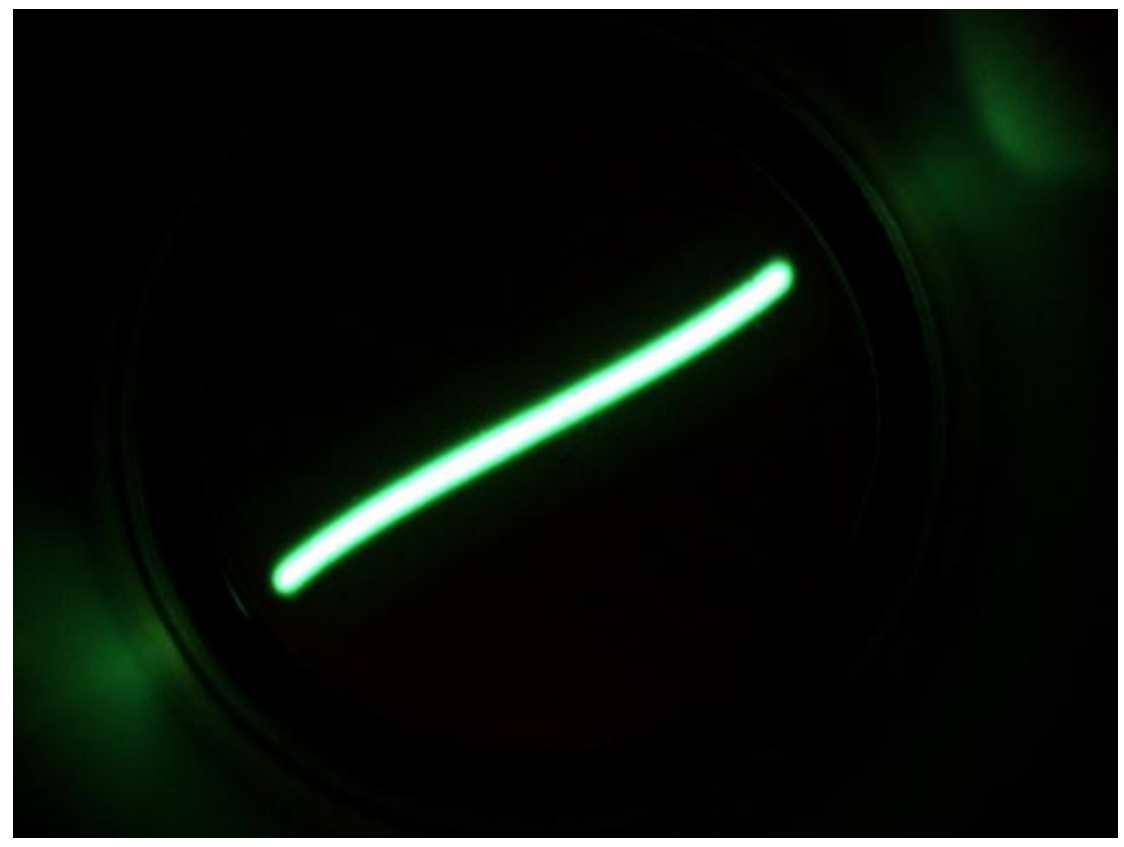

**Fig-9** A photo taken during experimental measurement

Fig-9 is a photo taken during the measurement, in order to facilitate calculation, the Y-axis is adjusted to a horizontal line, take it as an example to illustrate the method of calculating inclination angle of tangent. The line pattern is an odd symmetrical curve. The inclination angle of tangent of the curve at the midpoint is the inclination angle $\psi$ of tangent corresponding to $v = 0$. Due to mechanical error and interference, the midpoint of the odd-symmetric curve in Fig-9 cannot be measured directly, but the midpoint of the odd-symmetric curve in Fig-9 is also the inflection point of the curve, the exact value of the inflection point can be calculated. For example, the midpoint of the curve can be roughly calculated first, and then the middle part curve (as shown in Fig-10) is taken for cubic fitting to obtain the curve equation.

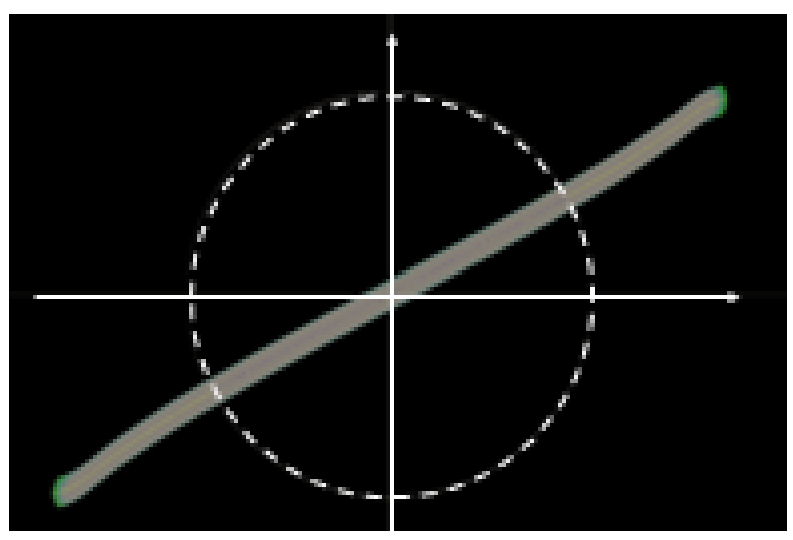

**Fig-10** Middle part of a photo

The inclination angle of the tangent line of the cubic fitting curve of Fig-9 at each point is shown as Fig-11, the inclination angle of tangent corresponding to the minimum point is 27.949539 degrees. For each photo, the inclination angle $\psi$ of tangent that meets measurement accuracy can be calculated.

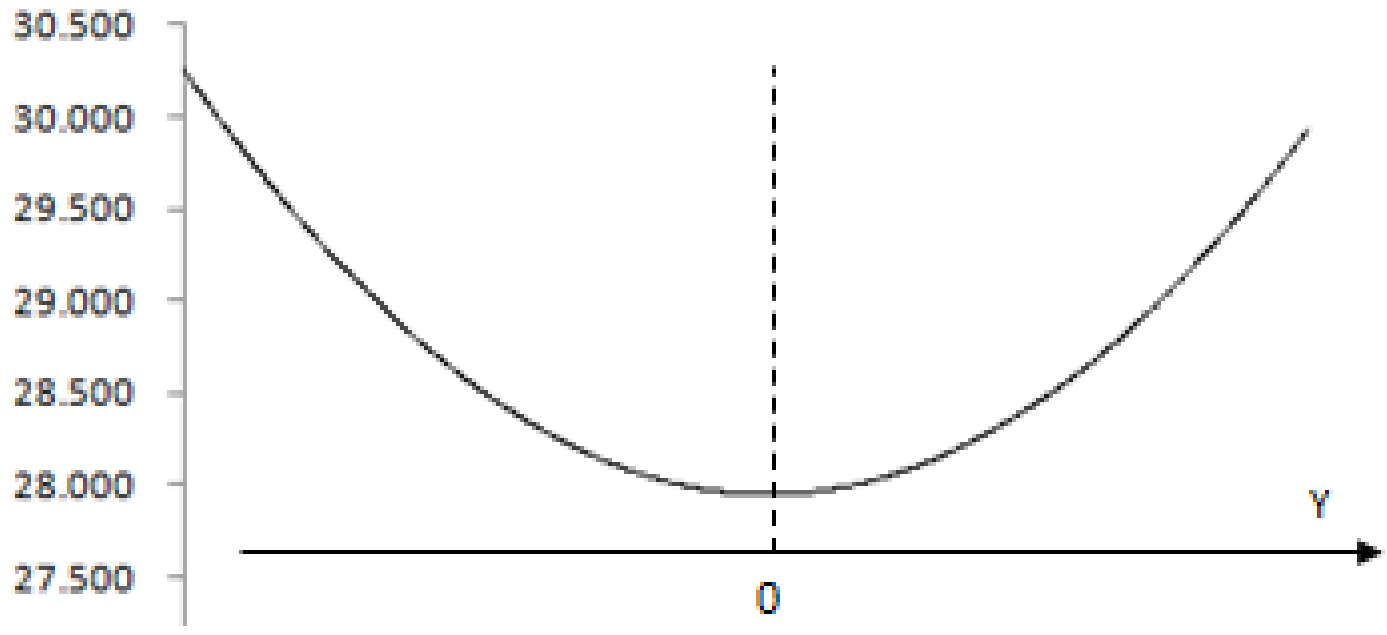

**Fig-11** Inclination angle of the tangent line at each point

The main experimental equipment includes: (1) A slender cathode-ray tube with a length of about 470mm and a screen with diameter of 20mm. (2) A coil-pair consists of two air core coils with inner diameter of 20 mm, length of 10.04 mm, line diameter of 0.20 mm and 448 turns. (3) An industrial camera with resolution of 16 megapixels. (4) A laser distance sensor with resolving power of 0.01mm and accuracy of 0.03mm. (5) High-precision sawtooth signal generator. (6) Current source and voltage source.

The main parameters includes: (1) Accelerating voltage is 2196 V. (2) Current of constant current source is 720 mA.

In order to improve measuring accuracy, following 32 measurement combinations were considered, and a total of 672 photos were taken.

(1) Four directions of electron motion: in the same direction as geomagnetic field (including geomagnetic declination and geomagnetic dip), in the opposite direction to geomagnetic field, in the same direction as earth rotation and in the opposite direction to earth rotation.

(2) Four combinations of connection and direction of coil A and coil B, as shown in Fig-12.

(3) Two ways to connect with constant current source.

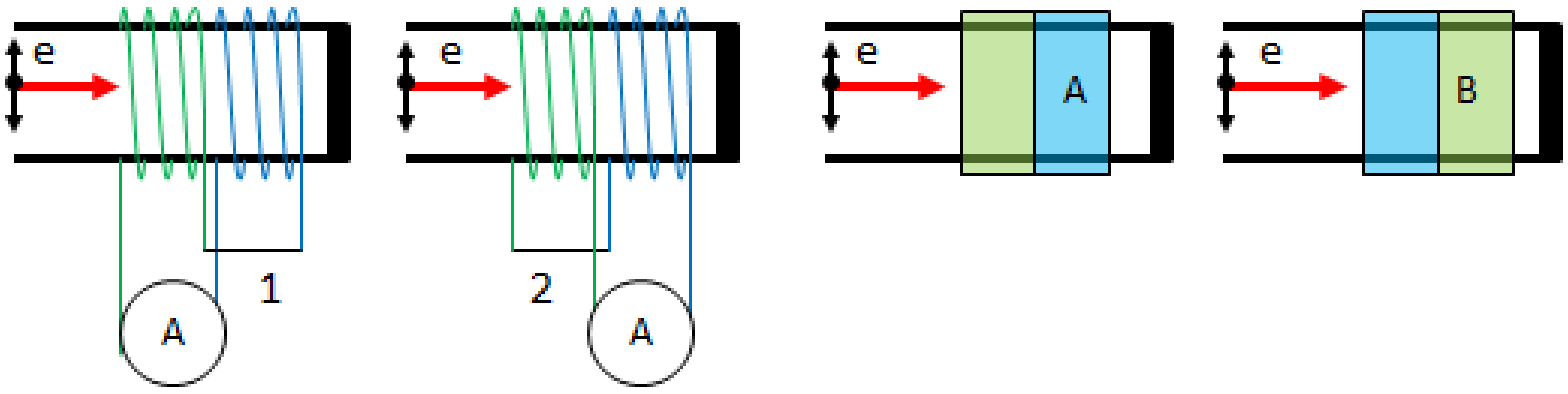

**Fig-12** Four combinations of connection and direction of coils

There are a total of $4*4*2=32$ combinations. In order to facilitate measurement, keep position of the cathode-ray tube fixed and move the coil-pair for measurement. Taking the axis of the cathode-ray tube as the X-axis and the direction of electron motion as the positive direction of the X-axis, the position of inner surface of the screen of the cathode-ray tube is $x=0$ (the reference selected in experimental measurement is different from the reference selected in previous theoretical analysis, will cause the meaning of the sign of $x$ to change). Moving the coil-pair makes midpoint of the coil-pair move step by step from $x=-2.00mm$ to $x=+2.00mm$ with a step length of 0.20mm. One photo is taken every step, 21 photos are taken each combination, $21*32=672$ photos are taken in total.

In order to improve reliability of the measurement, differential inclination angle $\psi$ of tangent is used for analysis. For each combination, inclination angle $\psi$ of tangent at $x=+2.00mm$ is used as the reference angle $\psi$ of tangent, absolute inclination angle $\psi$ of tangent minus the reference angle $\psi$ of tangent is taken as relative inclination angle $\psi_r$ of tangent for every measurement point.

## 5. Measurement result.

The calculation results of these 672 photos show that the geomagnetic direction or the earth rotation direction or the connection method of the coil-pair or AB direction of the coil-pair have little influence on quantitative analysis, but the current direction of constant current source can influence qualitative analysis. The current direction of constant current source can be divided into two categories, one causes that direction of magnetic field which electrons enter first is the same as direction of electrons motion (as shown in Fig-7 left side), and the other causes that direction of magnetic field which electrons enter first is opposite to direction of electrons motion (as shown in Fig-7 right side). The calculation results of 672 photos are divided into two types according to the direction category of the current of constant current source, and mathematical average is used for each type of data, which can filter out most of noise and reduce error of final measurement results. After cubic fitting the mathematical average, the final measurement results can be obtained, as shown in Tab-1 and Fig-13.

**Tab-1** Measurement result table

| Measure Point | Direction of magnetic field which electrons enter first is opposite to direction of electrons motion | | Direction of magnetic field which electrons enter first is the same as direction of electrons motion | |
|---|---|---|---|---|
| | Average of relative inclination angle of tangent | Fitting value of relative inclination angle of tangent | Average of relative inclination angle of tangent | Fitting value of relative inclination angle of tangent |
| -10 | 0.17763 | 0.17868 | 0.07829 | 0.07542 |
| -9 | 0.32319 | 0.31644 | 0.22246 | 0.22203 |
| -8 | 0.43593 | 0.43807 | 0.35306 | 0.352 |
| -7 | 0.54449 | 0.54368 | 0.46535 | 0.46547 |
| -6 | 0.63006 | 0.6334 | 0.55451 | 0.56256 |
| -5 | 0.69912 | 0.70733 | 0.63561 | 0.64343 |
| -4 | 0.75981 | 0.7656 | 0.70981 | 0.70819 |
| -3 | 0.8129 | 0.80831 | 0.75724 | 0.75698 |
| -2 | 0.83886 | 0.83559 | 0.79627 | 0.78995 |
| -1 | 0.85079 | 0.84755 | 0.81532 | 0.80721 |
| 0 | 0.85085 | 0.84431 | 0.81403 | 0.80892 |
| 1 | 0.83359 | 0.82598 | 0.79372 | 0.79519 |
| 2 | 0.78961 | 0.79267 | 0.76457 | 0.76617 |
| 3 | 0.74157 | 0.74451 | 0.71917 | 0.72199 |
| 4 | 0.67564 | 0.68161 | 0.66127 | 0.66279 |
| 5 | 0.6029 | 0.60409 | 0.58171 | 0.58869 |
| 6 | 0.50933 | 0.51206 | 0.50132 | 0.49984 |
| 7 | 0.40471 | 0.40563 | 0.39694 | 0.39637 |
| 8 | 0.2867 | 0.28493 | 0.27978 | 0.27841 |

| 9 | 0.15401 | 0.15007 | 0.14682 | 0.14609 |
|---|---|---|---|---|
| 10 | 0 | 0.00116 | 0 | -0.00044 |

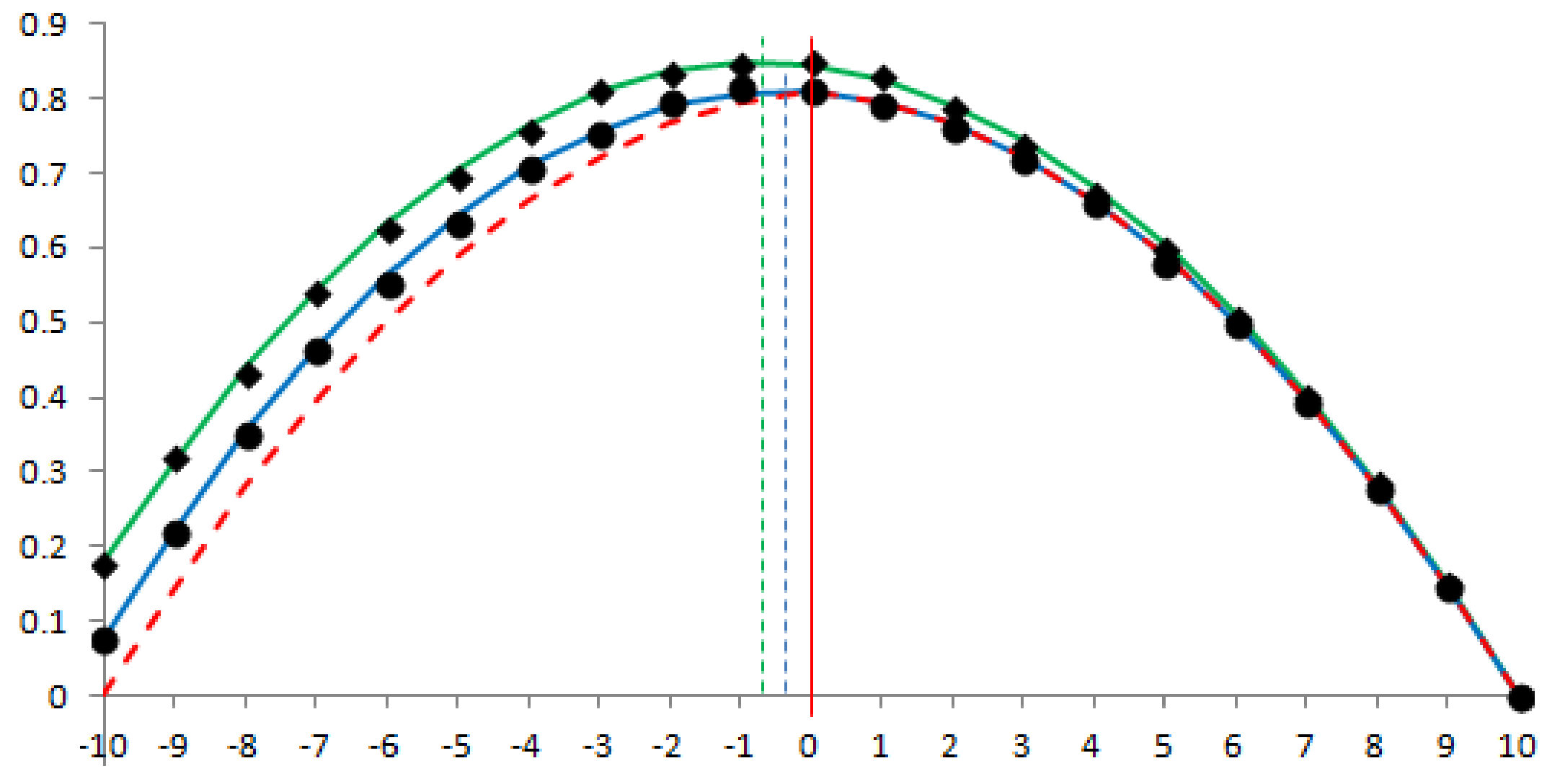

**Fig-13** Measurement result curves

As shown in Fig-13, the left curve is the relative inclination angle $\psi_r(x)$ of tangent when direction of the magnetic field which electrons enter first is opposite to direction of electrons motion, its cubic fitting equation is $0.0000193123667x^3 - 0.00754385862x^2 - 0.0108071624x + 0.844307228$, and its extreme point is $x = -0.714$, that is to say, when midpoint of the coil-pair moves $0.714 * 0.20 = 0.143mm$ from position of inner surface of the screen of the cathode-ray tube to direction of electron gun, relative inclination angle $\psi_r(x)$ of tangent of the line pattern reaches its maximum value. The middle curve is the relative inclination angle $\psi_r(x)$ of tangent when direction of the magnetic field which electrons enter first is the same as direction of electrons motion, its cubic fitting equation is $0.0000224023665x^3 - 0.00771426295x^2 - 0.00603336354x + 0.80891566$, and its extreme point is $x = -0.390$, that is to say, when midpoint of the coil-pair moves $0.390 * 0.20 = 0.078mm$ from position of inner surface of the screen of the cathode-ray tube to direction of electron gun, inclination angle $\psi_r(x)$ of tangent of the line pattern reaches its maximum value. The dashed line is the reference curve symmetric about $x = 0$.

## 6. Conclusions.

As shown in Fig-7, the extreme point of the inclination angle $\psi(x)$ of tangent changes to $x > 0$ and $b > a$. According to the appendix A, the reason why extreme point of $\psi(x)$ changes to $x > 0$ is as follows, electrons rotate on their own axes in magnetic field, at the point of $x = 0$, the inertial effect of rotating electron (proportional to the moment of inertia of electron) causes that line pattern continues to rotate for a distance in original direction. When the line pattern shortens with increasing x, the deviation of extreme point of $\psi(x)$ from $x = 0$ cannot be caused by the inertial effect of the electron's tangential motion (which is proportional to the electron mass). The shift of extreme point of $\psi(x)$ to $x > 0$ cannot be caused by the deflection of axially moving electrons in radial magnetic field, because near $x = 0$, the deflection direction of axially moving electrons in radial magnetic field is opposite to that of radially moving electrons in axial magnetic field of left coil, result of the deflection of axially moving electrons in the radial magnetic field is $x < 0$. Assuming that the extreme point of $\psi(x)$ shifted to $x > 0$ is caused by the mechanical error of cathode-ray tube or coils, then it should be $b = a$ instead of $b > a$. The result $b > a$ was not anticipated before the experimental measurement, and it indicates that the electron has an internal structure.

672 photos taken during experiment measurement and related calculation results, as well as the code of Python program for calculation, can be downloaded at https://pan.baidu.com/s/1Eete__RbaZz3pYM3DzYQ_A, the download code is **24pf** and the URL is case-sensitive. It is suggested that laboratories with good conditions should make more accurate measurements and error analysis.

## Appendix A: Mechanical model of electromagnetic field

**Abstract:** In this paper, a mechanical model that replaces potential energy with rotational kinetic energy is proposed, it is based on rotating physical particle and can be used to explain why like charges repel but opposite charges attract. According to this mechanical model, electromagnetic field is only physical property generated by motions of the particle. Based on this mechanical model and Newtonian mechanics, the force exerted on an electron and a proton under different conditions is calculated, and therefore, electrostatic force and Lorentz force are calculated, mathematical expression of permittivity of vacuum and Maxwell's equations of electromagnetic field are derived, the mechanical essence of induced electric field and displacement current is described, it is explained that there is no causal relationship between changing electric field and changing magnetic field but concomitant relationship. Based on the mechanics model, it can be predicted that "after electrons pass through the magnetic-field zero point formed by two opposing magnetic fields, deflection opposite to Lorentz force will occur due to inertial effect". The prediction is verified by experimental measurement of cathode ray.

**Key words**: U-particle; electron; proton; electric field; magnetic field; electrostatic force; Lorentz force; Magnus effect; permittivity of vacuum; permeability of vacuum; induced electric field; displacement current; Maxwell’s equations

### 1 Mechanical model of U-particle

The purpose of this mechanical model is to replace potential energy with rotational kinetic energy, use Newtonian mechanics to analyze and calculate electromagnetic fields from the perspective of momentum, making electromagnetics more intuitive and easier to understand. Correspondingly, Maxwell used analytical mechanics to analyze and calculate the electromagnetic field from the perspective of energy, requiring more abstract knowledge such as calculus of variations and potential energy.

U-1: Assuming that electromagnetic field is physical properties generated by different motions of unknown physical particle, take a point A in three-dimensional space as reference point, and use linear time and linear space to calculate velocity. For any point in three-dimensional space, if momentum density of the particle is zero and translational kinetic energy density of the particle remains unchanged, then point A is called a stationary point, otherwise point A is called a moving point.

Explanation: The analysis of this mechanical model is limited to the case of a stationary frame of reference and low-speed movement of electrons/protons. The accurate calculation of this mechanical model is limited to the motion of electrons/protons in a stationary reference frame, where they are completely stationary or have zero velocity but non-zero acceleration.

U-2: The physical particle assumed in U-1 is called universal particle, referred to as U-particle. Suppose that U-particle has following characteristics: (1) inertial mass of a single U-particle is a constant $M_U$, $M_U$ is much less than the inertial mass of an electron; geometric size of a single U-particle is a constant which is much smaller than that of an electron; (2) U-particle is uniformly distributed in three-dimensional space, and motion state of U-particle will change after collision between U-particles or collision between U-particle and electron/proton; (3) a U-particle with no translational motion is isotropic and has rotational kinetic energy inside; (4) the sum of translational kinetic energy and rotational kinetic energy of a U-particle is a constant $E_U$.

Explanation: Absolute vacuum does not exist, the space is full of U-particles, and traditional vacuum just has no gas molecules. In addition to electron and proton, neutron can be seen as a combination of electron and proton, and motion state of U-particle will also change after collision between U-particle and neutron. Because electromagnetic field is physical properties generated by motion of U-particle, U-particle has only kinetic energy, and any other form of energy, including potential energy and heat energy, is different manifestation of kinetic energy of U-particle.

U-3: Because a U-particle with no translational motion is isotropic and has rotational kinetic energy inside, it can be assumed that there are two structural models of U-particle, Ue and Up. U-particle is an equivalent sphere O with following structure, the sphere is composed of many high-speed rotating tiny balls, and extension line of the tiny ball's rotation axis passes through the centre of sphere O. From the point of view of the centre of sphere O, all tiny balls rotate anticlockwise and angular momentum points to centre O, this kind of U-particle is called Ue. From the point of view of the centre of sphere O, all tiny balls rotate clockwise and angular momentum is backward to centre O, this kind of U-particle is called Up, as shown in Fig-A1.

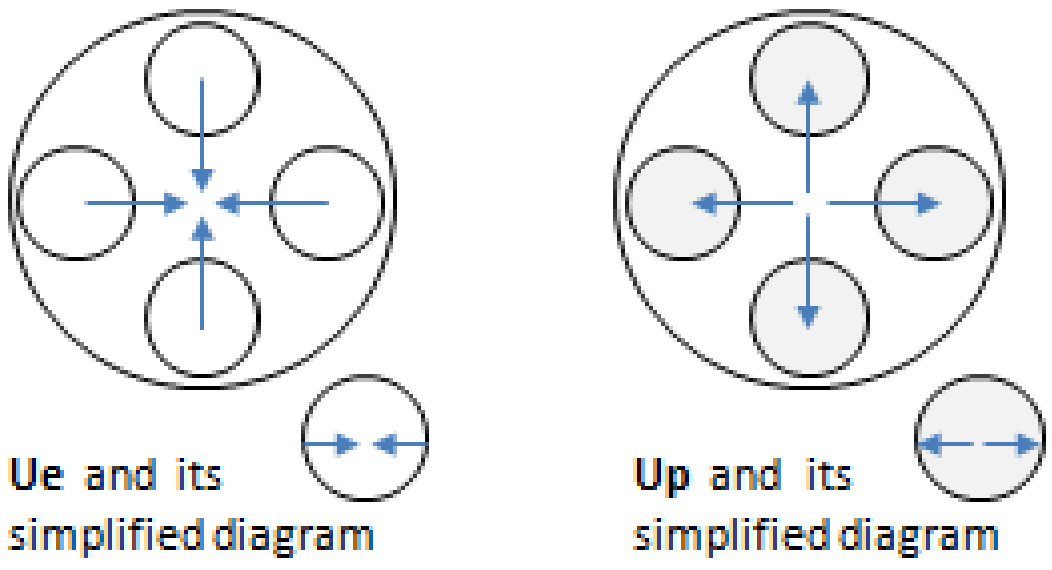

**Fig-A1** Structural models of U-particle

Explanation: The structural model of U-particle is a little like dandelion flower ball.

U-4: An electron or a proton with no translational motion is isotropic, they have large amount of rotational kinetic energy stored inside. Suppose that the electron is an equivalent sphere O with the following structure: its radius is a constant R, and its outermost layer consists of many small balls rotating at high speed. Each small ball is similar in structure to a U-particle and has the same mass as the entire U-particle, but is not isotropic. Each small ball rotates about an axis whose extension passes through the centre of sphere O. From the point of view of centre of the electron, all small balls rotate anticlockwise and angular momentum points to centre of the electron. A proton is a sphere with following structure: the sphere is composed of an inner layer and an outer layer. In process of U-particle crossing the outer layer, translational and rotational kinetic energy of the U-particle are not subject to additional effects. Radius of the inner layer is a constant R, and structure of the inner layer is similar to that of an electron, but rotation direction of small ball is opposite, that's to say, from the point of view of centre of the proton, all small balls rotate clockwise and angular momentum is backward to centre of the proton, as shown in Fig-A2. The inner layer of a proton can also be called an anti-electron.

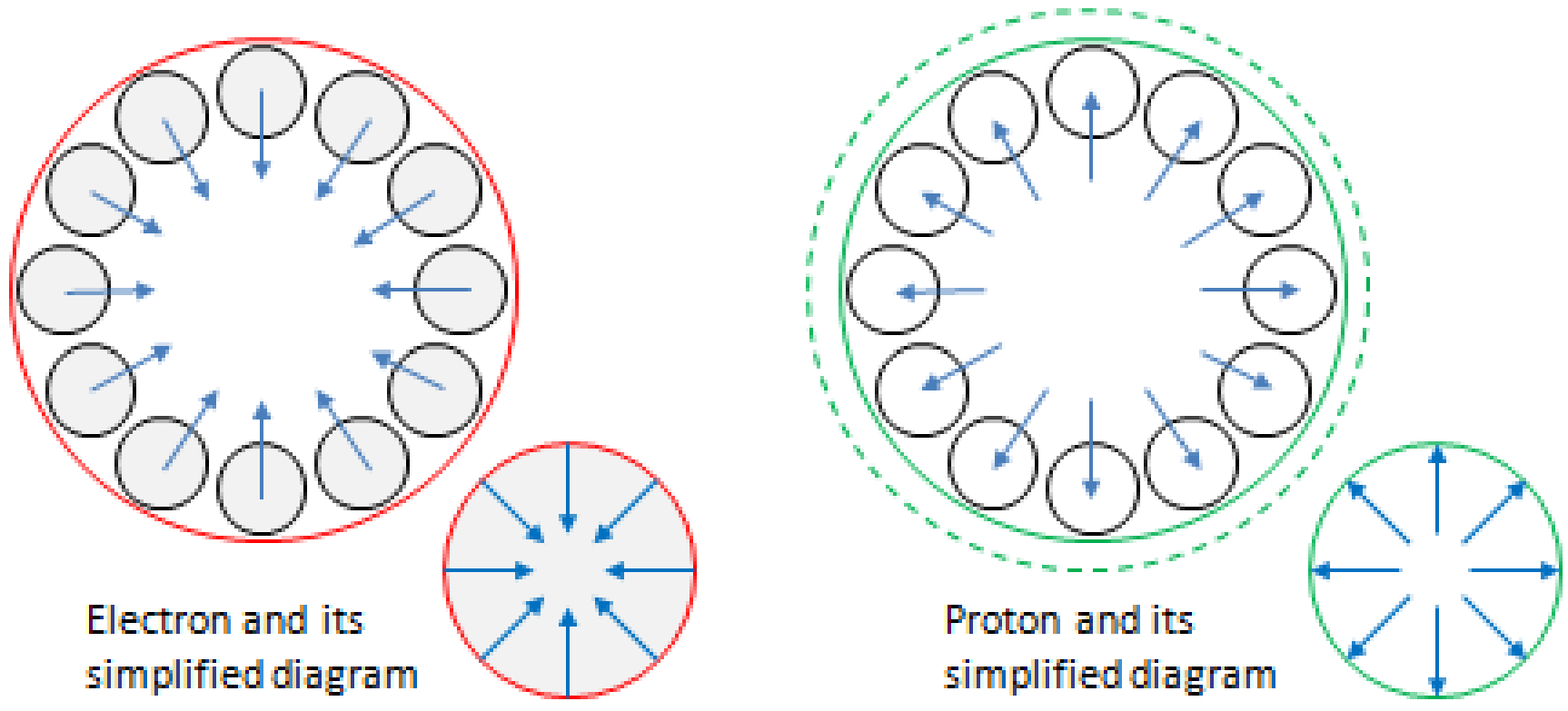

**Fig-A2** Structural model of electron/proton

U-5: There are two U-particles, one of which has rotational kinetic energy of $U_1$ and the other has rotational kinetic energy of $U_2$, characteristics of collision between these two U-particles are as follows: after collision, each U-particle has a rotational kinetic energy of $(U_1+U_2)/2$, and the particle types remain unchanged.

Explanation: After collision, Up remains Up, and Ue remains Ue.

U-6: The characteristics of collision between U-particle and static electron/proton are as follows: (1) A Up with rotational kinetic energy $U_1$ collides with an electron. After the collision, the small ball on the electron surface becomes a second-hand Up and sinks into the electron, while the incident particle remains a Up. The rotational kinetic energy of both Up becomes $(E_U-U_1)/2$. A Ue with rotational kinetic energy $U_1$ collides with an electron. After the collision, the small ball on the electron surface becomes a second-hand Up and sinks into the electron, while the incident particle becomes a Up. The rotational kinetic energy of both Up becomes $(E_U+U_1)/2$. (2) A Ue with rotational kinetic energy $U_2$ collides with a proton. After the collision, the small ball on the proton surface becomes a second-hand Ue and sinks into the proton, while the incident particle remains a Ue. The rotational kinetic energy of both Ue becomes $(E_U-U_2)/2$. A Up with rotational kinetic energy $U_2$ collides with a proton. After the collision, the small ball on the proton surface becomes a second-hand Ue and sinks into the proton, while the incident particle becomes a Ue. The rotational kinetic energy of both Ue becomes $(E_U+U_2)/2$. Electron/proton is the converter that converts translational kinetic energy of U-particle into rotational kinetic energy of U-particle without the influence of other electron/proton.

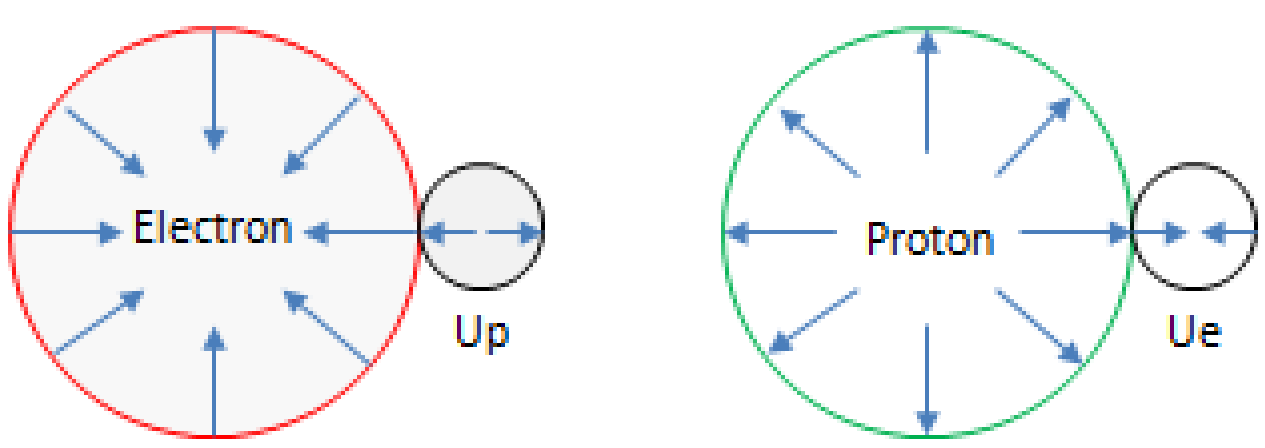

**Fig-A3** Collision between U-particle and electron/proton

Explanation: A Ue becomes Up upon collision with an electron, while a Up remains Up; a Up becomes Ue upon collision with a proton, while a Ue remains Ue.

U-7: A Ue becomes a Up after colliding with an electron; a Up remains a Up after colliding with an electron. These Up particles then collide with other U-particles and converge to a stable diffusive equilibrium state. For an isolated stationary electron, the rotational kinetic energy of a Up at a certain point is inversely proportional to the distance between the point and centre of the electron. A Up becomes a Ue after colliding with a proton; a Ue remains a Ue after colliding with a proton. These Ue particles then collide with other U-particles and converge to a stable diffusive equilibrium state. For an isolated stationary proton, the rotational kinetic energy of a Ue at a certain point is inversely proportional to the distance between the point and centre of the proton. Assume $r$ is the distance between the Up and centre of the electron, then the decreasing function of rotational kinetic energy $E_R(r)$ of Up is as follows

$$E_R(r) = K_U E_U * \frac{R}{r} \qquad (1)$$

$K_U$ is a coefficient related to the diffusion motion of U-particle and $K_U \ll 1$. The gradient of rotational kinetic energy $E_R(r)$ of Up is

$$\boldsymbol{\nabla} E_R(r) = -K_U E_U * \frac{R}{r^2} * \boldsymbol{e}_r = -\frac{K_U M_U C_U^2}{2} * \frac{R}{r^2} * \boldsymbol{e}_r \qquad (2)$$

$C_U$ is the maximum speed of U-particle translational motion. The flux of rotational kinetic energy $E_R(r)$ is proportional to $\boldsymbol{\nabla} E_R(r)$. The decreasing function of rotational kinetic energy of Ue around a proton is similar.

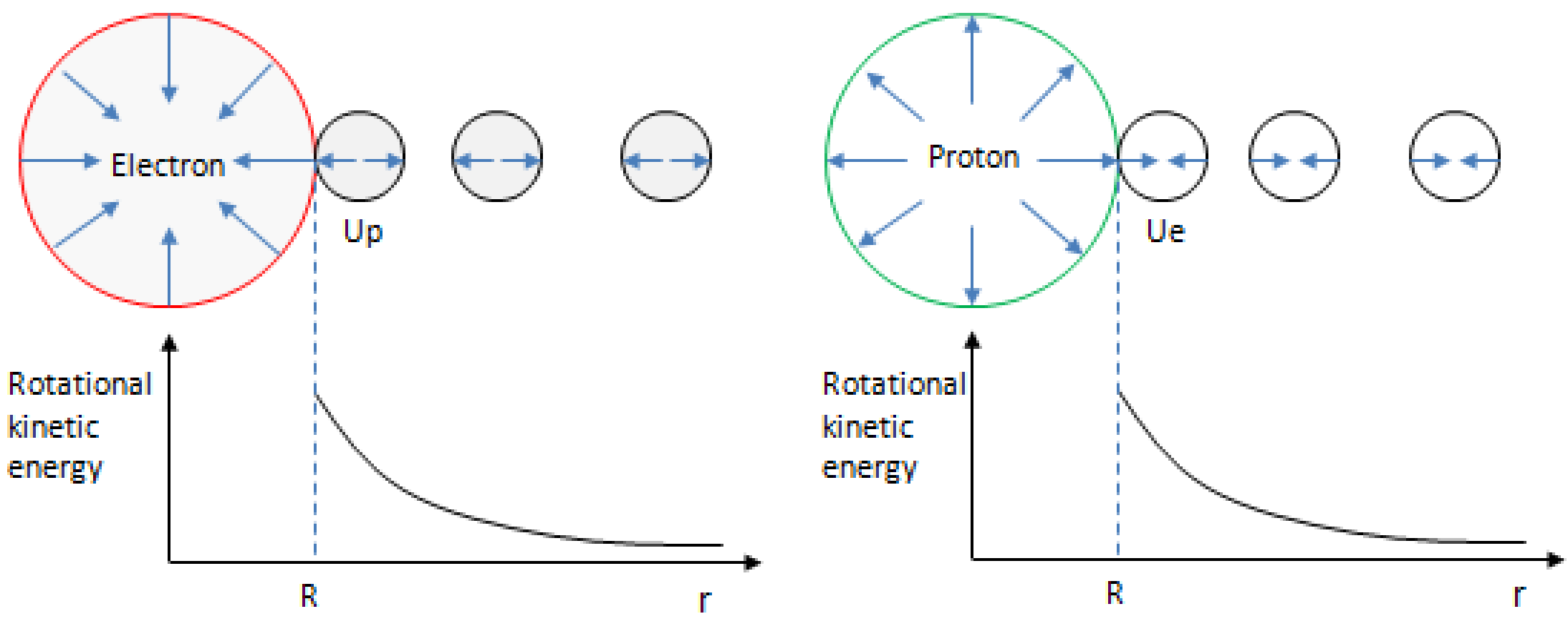

**Fig-A4** Decreasing curves of rotational kinetic energy of U-particle

Explanation: Fig-A4 shows decreasing curves of rotational kinetic energy of Up around an electron and Ue around a proton, for mathematical proof of the curve, please refer to B-1 of appendix B "Mathematical calculation of random collision of U-particle", it is consistent with Fick's law of diffusion. According to decreasing function of rotational kinetic energy of U-particle $E_R(r) = K_U E_U * R/r$, when $r = \infty$, $E_R = 0$, U-particle has only translational kinetic energy. $E_U = M_U * C_U^2/2$, the following sections calculate $C_U$ is the speed of light. Since the radius R of an electron is very small, assuming that R is $10^{-16}$m, according to equation (1), when distance between U-particle and centre of the electron is 1 nm, rotational kinetic energy of the U-particle is far less than $E_U * 10^{-7}$, and speed of the U-particle is close to $C_U$. In spherical coordinates system, the gradient of $E_R(r)$ is

$$\boldsymbol{\nabla} E_R(r) = \frac{\partial E_R(r)}{\partial r} * \boldsymbol{e}_r = -K_U E_U * \frac{R}{r^2} * \boldsymbol{e}_r = -\frac{K_U M_U C_U^2}{2} * \frac{R}{r^2} * \boldsymbol{e}_r$$

$$\boldsymbol{\nabla}^2 E_R(r) = \frac{1}{r^2} * \frac{\partial}{\partial r}\left[r^2 * \left(-\frac{K_U M_U C_U^2}{2} * \frac{R}{r^2}\right)\right] = 0$$

U-8: Assume that the macroscopic velocity at the collision point of the electron or proton is $V$, the characteristics of collision between moving electron/proton and U-particle are as follows:

(1) After a Up with rotational kinetic energy $U_1$ collides with an electron, the rotational kinetic energy of the second-hand Up is $(E_U - M_U V^2/2 - U_1)/2$; after a Ue with rotational kinetic energy $U_1$ collides with an electron, the rotational kinetic energy of the second-hand Up is $(E_U - M_U V^2/2 + U_1)/2$. The second-hand Up sinks into the electron. A Ue becomes Up upon collision with an electron, while a Up remains Up. Since the collision process satisfies momentum conservation, these Up particles gain additional macroscopic momentum.

(2) After a Ue with rotational kinetic energy $U_2$ collides with a proton, the rotational kinetic energy of the second-hand Ue is $(E_U - M_U V^2/2 - U_2)/2$; after a Up with rotational kinetic energy $U_2$ collides with a proton, the rotational kinetic energy of the second-hand Ue is $(E_U - M_U V^2/2 + U_2)/2$. The second-hand Ue sinks into the proton. A Up becomes Ue upon collision with a proton, while a Ue remains Ue. Since the collision process satisfies momentum conservation, these Ue particles gain additional macroscopic momentum. The momentum of macro motion of U-particle is spread by random collision. The relationship between velocity $\boldsymbol{v}_0$ of an electron/proton moving at a low speed, macro velocity $\boldsymbol{v}_U$ of U-particle and distance $r$ between the U-particle and centre of the electron/proton is

$$\boldsymbol{v}_U(r) = K_U \boldsymbol{v}_0 * \frac{R}{r} \qquad (3)$$

Explanation: Macro velocity decreasing function of Up around an electron moving at a low speed is similar to decreasing function of rotational kinetic energy of Up around a static electron. The mathematical proof of this decreasing function is referred to B-2 of appendix B "Mathematical calculation of random collision of U-particle".

## 2 Electrostatic force, gravitation, induced force, Lorentz force

There will be more mathematical calculations in the following sections, in order to reduce the trouble of marking vector specially, for the same variable, if it is not bold, it represents a scalar, and if it is bold, it represents a vector. For example, $v$ represents magnitude of velocity, and $\boldsymbol{v}$ represents velocity vector. For the spatial structure of three-dimensional rectangular coordinate system $(x, y, z)$, cylindrical coordinate system $(r, \varphi, z)$, spherical coordinate system $(r, \theta, \varphi)$, and divergence and curl operation in cylindrical coordinate system, please refer to appendix C "Three dimensional coordinate system and simplified operation of Hamilton operator". The constants and variables commonly used in this paper are shown in the table below.

| Constant | Description |
|---|---|
| $M_U$ | Inertial mass of a single U-particle |
| $E_U$ | Total kinetic energy of a single U-particle |
| $R$ | Radius of electron |
| $C_U$ | Maximum translational speed of U-particle |
| $\rho$ | Inertial mass density of U-particle in space |
| $\rho_N$ | Quantity density of U-particle in space |
| $Q_e$ | Absolute value of the electric charge carried by a single electron |
| $K_U$ | Coefficient far less than 1, it is related to random diffusion of U-particle |
| $\varepsilon_0$ | permittivity of vacuum |
| $\mu_0$ | permeability of vacuum |
| **Variable** | **Description** |
| $\rho_{ER}$ | Rotational kinetic energy density of U-particle |
| $\rho_{PU}$ | Macro momentum density of U-particle |
| $E_R$ | Rotational kinetic energy of U-particle |
| $E_T$ | Translational kinetic energy of U-particle |
| $v_i$ | The speed at which electrons move in a wire |
| $v$ | The velocity of movement of an electron outside a wire |
| $v_U$ | The velocity of macro motion of U-particle, generally refers to Up |
| $v_{Ue}$ | The velocity of macro motion of U-particle generated by the movement of a single electron |
| $V_U$ | The microscopic translational velocity of U-particle |
| $E_h$ | Hidden energy, which is is equal to rotational kinetic energy density of U-particle times the volume of a sphere with radius R |
| $P_{em}$ | Electrical momentum, which is equal to macro momentum density of U-particle times the volume of a sphere with radius R, generally refers to Up |
| $\lambda$ | Number of electrons in directional motion in per unit length wire |
| $r$ | Distance between measuring point and electron/proton centre or a wire |
| $\omega$ | Angular velocity of an electron |
| $F_E$ | Electric field force |
| $F_B$ | Lorentz force |
| $F_i$ | Induced force |
| $F_g$ | Gravitational force |
| $q$ | Electric charge |
| $E$ | Electric field intensity |
| $B$ | Magnetic induction intensity |
| $\varphi_s$ | Electrostatic potential |
| $A$ | Magnetic vector potential |

U-9: As shown in Fig-A5, two static electrons A and B, after a U-particle collides with electron A, the particle becomes $\mathrm{Up_a}$. After collisions between these $\mathrm{Up_a}$ and other

U-particles, rotational kinetic energy of $\mathrm{Up_a}$ at a certain point is inversely proportional to the distance between the point and centre of electron A. After being struck by $\mathrm{Up_a}$, the small ball on the surface of electron B becomes a second-hand Us and sinks into electron B. Electron B then generates a new small ball to fill the collision point. The net force on electron B caused by the second-hand Us sinking into it is the electrostatic force exerted by electron A on electron B. Electrostatic force between an electron and a proton or between two protons is similar.

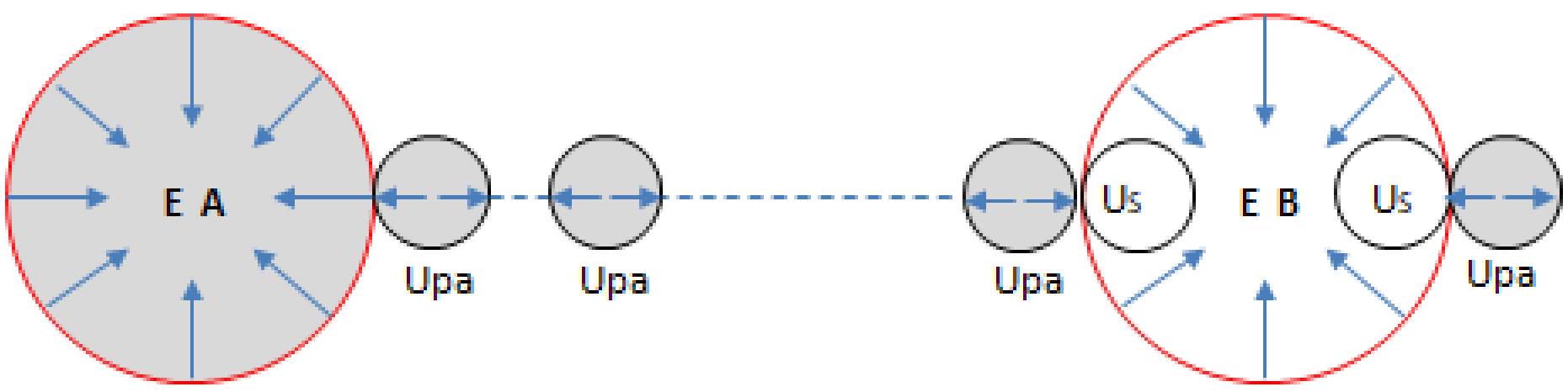

**Fig-A5** Electrostatic force is generated

U-10: The reason why two electrons repel each other is that the rotational kinetic energy of the Up particles between them is larger, leading to greater translational kinetic energy of the second-hand Us sunk into the electrons. The reason why two protons repel each other is that the rotational kinetic energy of the Ue particles between them is larger, leading to greater translational kinetic energy of the second-hand Us sunk into the protons. The reason why an electron and a proton attract each other is that the rotational kinetic energy of the U-particles between them is larger, resulting in smaller translational kinetic energy of the second-hand Us sunk into the proton or the electron.

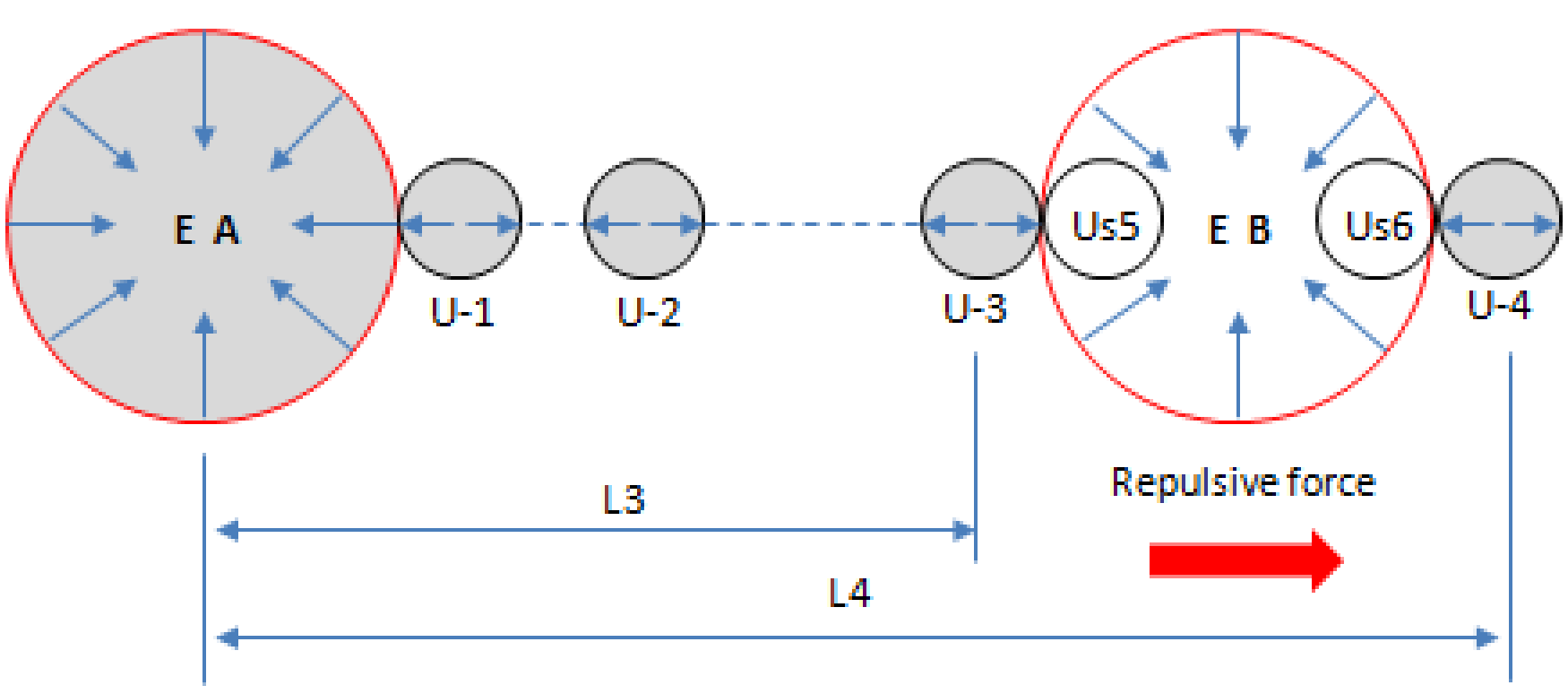

**Fig-A6** Electrostatic repulsion is generated

Explanation: As shown in Fig-A6, U-1, U-2, U-3, and U-4 are Up in equilibrium state around electron A when electron B does not exist, their rotational kinetic energy are $U_1$,$U_2$, $U_3$,$U_4$ respectively. Us5 is the second-hand U-particle produced after U-3 collides with electron B, its rotational kinetic energy is $(E_U - U_3)/2$ and its translational kinetic energy is $(E_U + U_3)/2$. Us6 is the second-hand U-particle produced after U-4 collides with electron B, its rotational kinetic energy is $(E_U - U_4)/2$ and its translational kinetic energy is $(E_U + U_4)/2$. For the reason of $U_3 > U_4$, translational kinetic energy of the Us on the left-side of electron B is more than that of right-side and therefore generates greater pressure, so electron B moves to the right, it shows that two electrons repel each other. The force between two protons is similar.

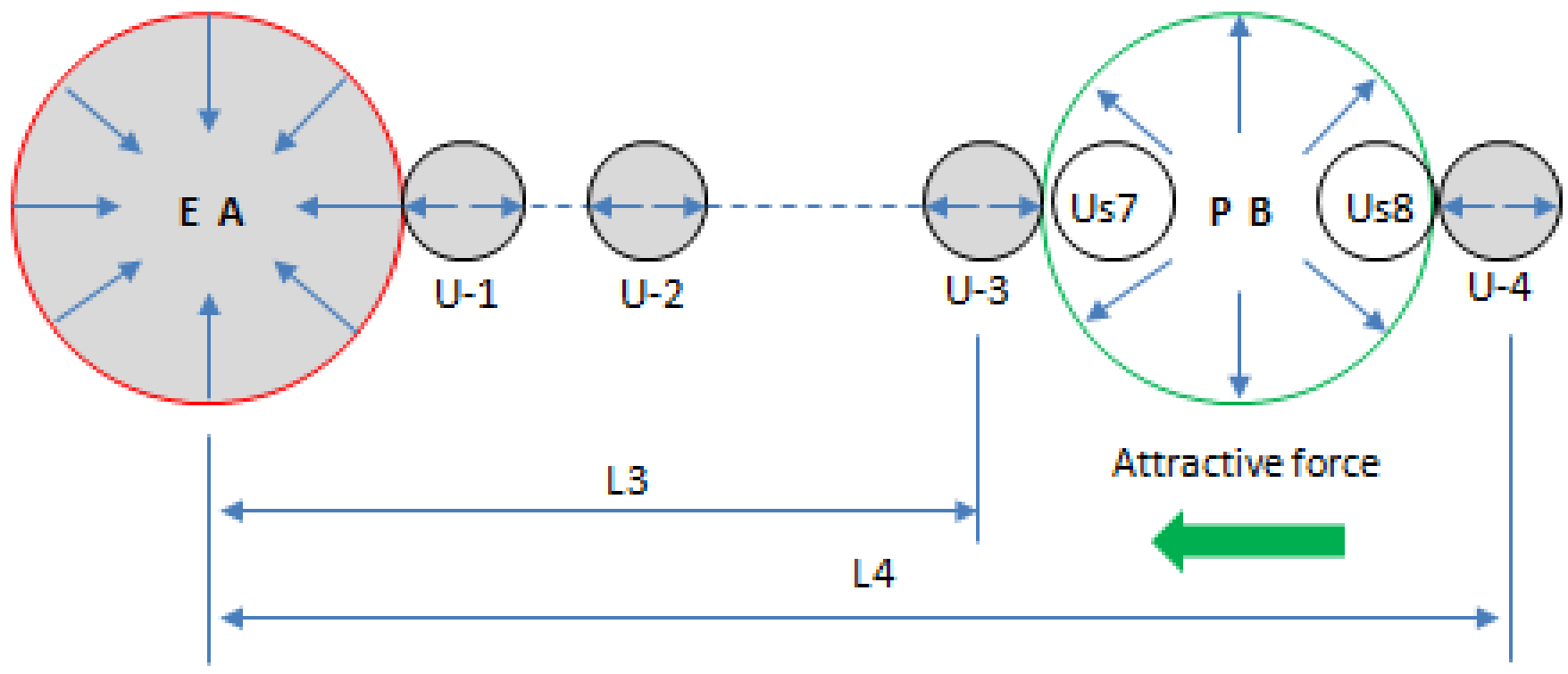

**Fig-A7** Electrostatic attraction is generated

As shown in Fig-A7, U-1, U-2, U-3, and U-4 are Up in equilibrium state around electron A when proton B does not exist, their rotational kinetic energy are $U_1$,$U_2$, $U_3$,$U_4$ respectively. Us7 is the second-hand U-particle produced after U-3 collides with proton B, its rotational kinetic energy is $(E_U + U_3)/2$ and its translational kinetic energy is $(E_U - U_3)/2$. Us8 is the second-hand U-particle produced after U-4 collides with proton B, its rotational kinetic energy is $(E_U + U_4)/2$ and its translational kinetic energy is $(E_U - U_4)/2$. For the reason of $U_3 > U_4$, translational kinetic energy of the Us on the left-side of proton B is less than that of right-side and therefore generates smaller pressure, so proton B moves to the left, it shows that an electron and a proton attracts each other.

U-11: The distance between centres of two static electrons/protons is L, electrostatic force between them is

$$F_e = \frac{2\pi R^4 K_U \rho C_U^2}{9} * \frac{1}{L^2} = \frac{4\pi R^3 \rho_N}{9} * \left|(\nabla E_R(L))_r\right| \qquad (4)$$

$\rho$ and $\rho_N$ are the inertial mass density and quantity density of U-particle in space respectively. Electrostatic force is inversely proportional to square of the distance between two static electrons/protons, or proportional to the gradient of rotational kinetic energy of U-particle. The direction of electrostatic force is the line between the electrons/protons, like charges repel but opposite charges attract. Electrostatic force is differential force, and unidirectional component force is much greater than resultant force. When the distance between two electrons or two protons is less than $2R$, the electrostatic repulsion between them drops to zero.

Explanation: As shown in Fig-A5, relationship between rotational kinetic energy of $\mathrm{Up_a}$ around electron A and distance $r$ between $\mathrm{Up_a}$ and electron A is equation (1). If the distance between a certain point on surface of electron B and centre of electron A is $r$, after the small ball at this point on electron B is struck by $\mathrm{Up_a}$, it becomes a second-hand Us, and rotational kinetic energy of the Us is

$$E_{RS}(r) = \frac{E_U - E_R(r)}{2} = \frac{E_U}{2} - \frac{K_U R E_U}{2r}$$

Translational kinetic energy of the Us is

$$E_{TS}(r) = E_U - E_{RS}(r) = \frac{E_U}{2} + \frac{K_U R E_U}{2r}$$

Show as Fig-A8.

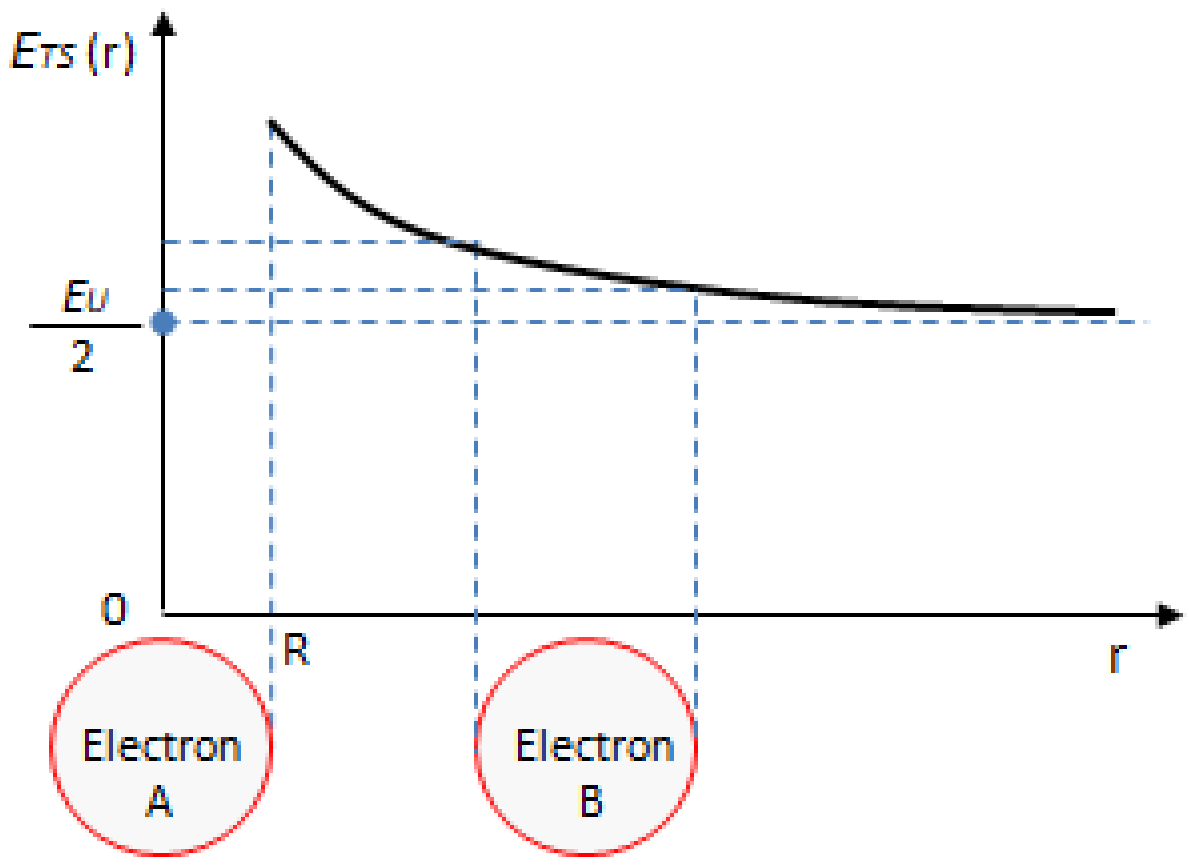

**Fig-A8** Translational kinetic energy decreasing curve of second-hand U-particle

If the distance between a certain point on the inner surface of proton B and centre of electron A is $r$, after the small ball at this point on proton B is struck by $\mathrm{Up_a}$, it becomes a second-hand Us, and rotational kinetic energy of the Us is

$$E_{RS}(r) = \frac{E_U + E_R(r)}{2} = \frac{E_U}{2} + \frac{K_U R E_U}{2r}$$

Translational kinetic energy of the Us is

$$E_{TS}(r) = E_U - E_{RS}(r) = \frac{E_U}{2} - \frac{K_U R E_U}{2r}$$

Show as Fig-A9.

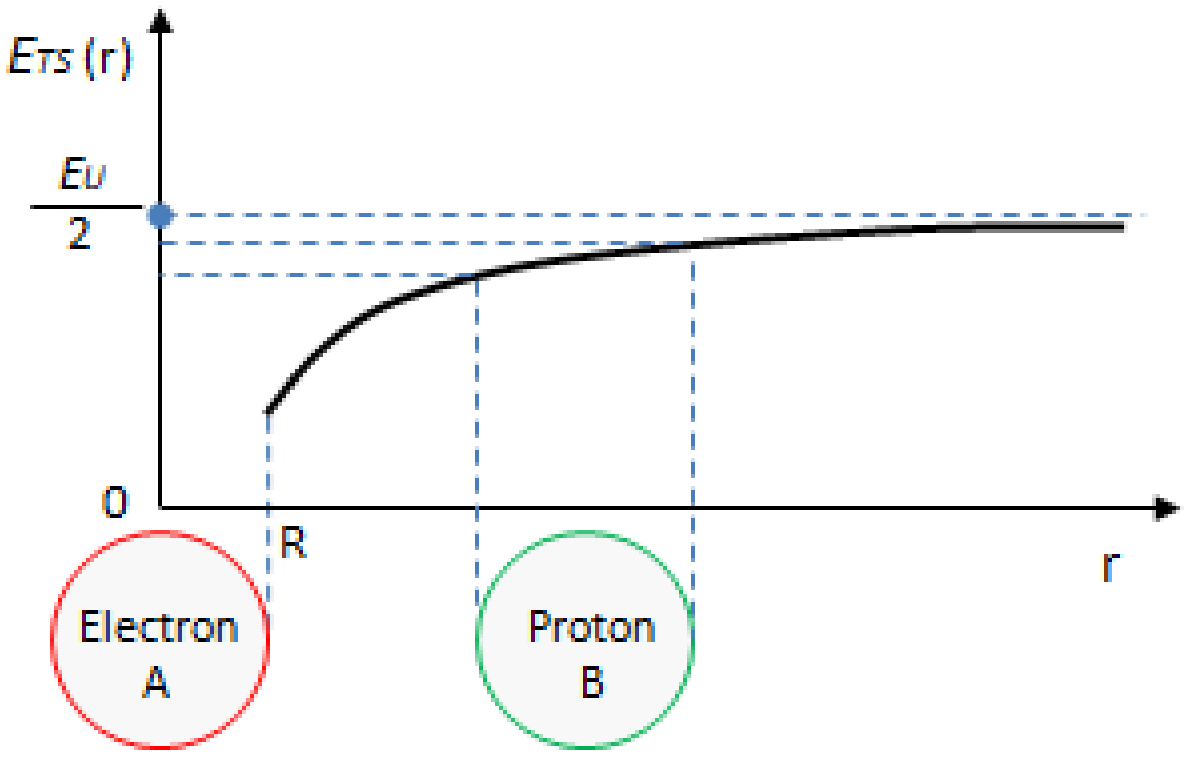

**Fig-A9** Translational kinetic energy enhancement curve of second-hand U-particle

Electrostatic repulsion between two electrons is calculated below. Electrostatic repulsion between two protons and electrostatic attraction between an electron and a proton can be calculated in the same way, magnitude of the force is the same.

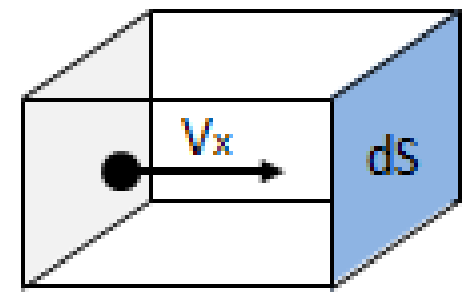

**Fig-A10** A cuboid for calculating pressure of second-hand U-particle

As shown in Fig-A10, the right plane of the cuboid is the surface of electron B (with the high-speed rotating small balls of the electron's outermost layer on its left), and its area is $dS$. The velocity component of the second-hand Us in the X direction is $V_x$. Assuming that the density of second-hand Us equals that of U-particles in space, the number of Us entering the electron surface during the time interval $dt$ is $N = V_x dtdS\rho_N$. After these Us particles sink into the electron, the electron fills the collision point with new small balls, all of which have zero momentum, so the momentum increment of electron B in the X-axis direction is $N * M_U V_x = V_x^2 dtdS\rho_N M_U$, it can be regarded as the result of the force $dF_e$ exerted on surface dS of electron B in $dt$ interval, so $dF_e\, dt = V_x^2 dtdS\rho_N M_U$. The pressure on the surface of electron B caused by Us sinking into it is $p_S = dF_e/dS = \rho_N M_U V_x^2$.Since Us is isotropic in three-dimensional space, if the speed of the Us is $V_S$, then $V_x^2 = V_S^2/3$, translational kinetic energy of the Us is $E_{TS} = M_U V_S^2/2$, so

$$p_S = \frac{\rho_N M_U V_S^2}{3} = \frac{\rho V_S^2}{3} = \frac{2\rho_N E_{TS}}{3} \qquad (5)$$

The Electrostatic repulsion on electron B generated by electron A is calculated as follows.

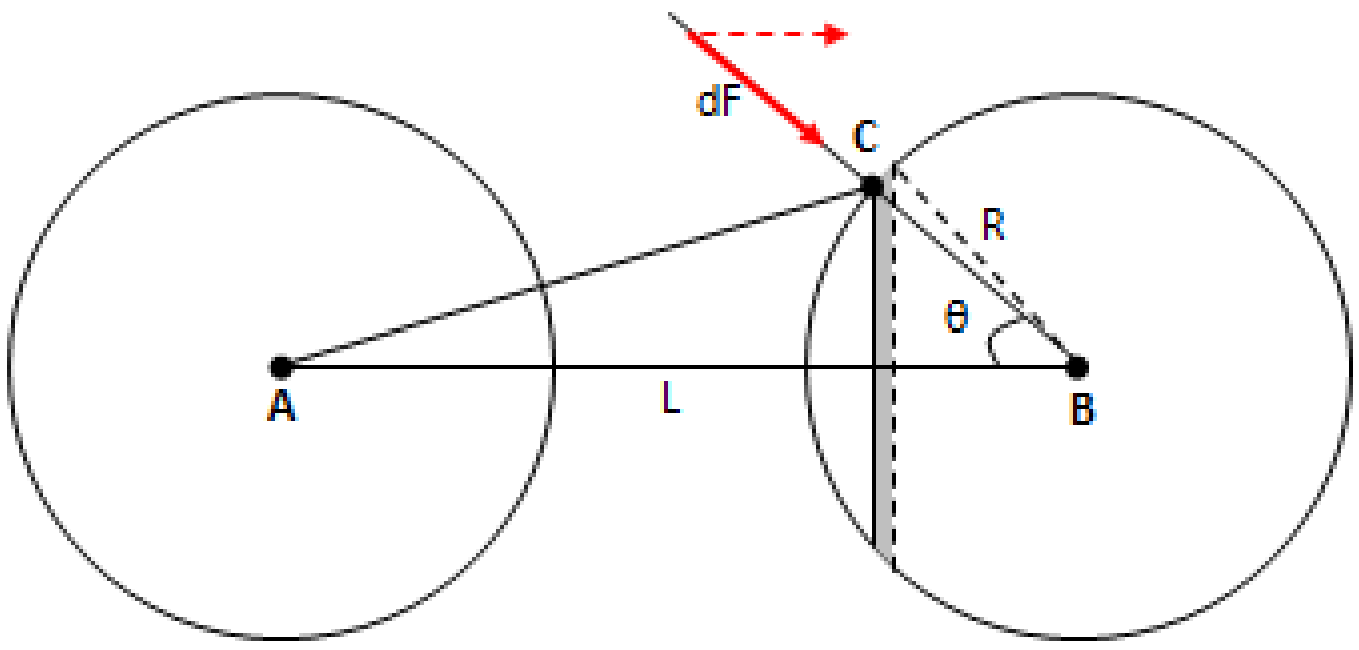

**Fig-A11** Repulsion between two electrons

As shown in Fig-A11, the distance between centre of electron A and centre of electron B is $AB = L$. The area of shadowed surface of electron B is $dS = 2\pi R^2 \sin\theta\, d\theta$, using the value of $E_{TS}$ in Fig-A8, according to equation (5), pressure on point C of electron B is

$$p_S = \frac{2\rho_N E_{TS}}{3} = \frac{2\rho_N}{3}\left(\frac{E_U}{2} + \frac{K_U R E_U}{2AC}\right) = \frac{\rho_N E_U}{3}\left(1 + \frac{K_U R}{AC}\right) = \frac{\rho_N M_U C_U^2}{6}\left(1 + \frac{K_U R}{AC}\right)$$
$$= \frac{\rho C_U^2}{6}\left(1 + \frac{K_U R}{AC}\right)$$

When $p_S$ applies on shadowed area $dS$, component force perpendicular to AB direction counteracts, and the horizontal force on electron B to the right is

$$F_e = \int_0^\pi p_S * dS * \cos\theta = \int_0^\pi \frac{\rho C_U^2}{6}\left(1 + \frac{K_U R}{AC}\right) * 2\pi R^2 \sin\theta\, d\theta * \cos\theta$$
$$= \frac{\pi R^2 \rho C_U^2}{3} \int_0^\pi \left(1 + \frac{K_U R}{\sqrt{(L - R\cos\theta)^2 + (R\sin\theta)^2}}\right) \sin\theta \cos\theta\, d\theta$$
$$= \frac{\pi R^2 \rho C_U^2}{3} \int_0^\pi \left(1 + \frac{K_U R}{\sqrt{L^2 + R^2 - 2LR\cos\theta}}\right) \sin\theta \cos\theta\, d\theta$$

Refer to D-1 of appendix D "Integral calculation related to electrostatic force", there is

$$F_e = \frac{2\pi R^4 K_U \rho C_U^2}{9} * \frac{1}{L^2}$$

According to equation (2)

$$\left(\nabla E_R(L)\right)_r = -\frac{K_U M_U C_U^2}{2} * \frac{R}{L^2}$$

$$F_e = \frac{2\pi R^4 K_U \rho C_U^2}{9} * \frac{1}{L^2} = \frac{4\pi R^3 \rho_N}{9} * \frac{K_U M_U C_U^2 R}{2L^2} = \frac{4\pi R^3 \rho_N}{9} * \left|\left(\boldsymbol{\nabla} E_R(L)\right)_r\right|$$

When L is much greater than R, the component force to the right on the left hemisphere of electron B is

$$hF_e = \int_0^{\pi/2} p_S * dS * \cos\theta = \frac{\pi R^2 \rho C_U^2}{3} \int_0^{\pi/2} \left(1 + \frac{K_U R}{\sqrt{L^2 + R^2 - 2LR\cos\theta}}\right) \sin\theta \cos\theta \, d\theta$$

$$\approx \frac{\pi R^2 \rho C_U^2}{6}$$

For detailed calculation, please refer to D-2 of appendix D "Integral calculation related to electrostatic force". Suppose that radius of electron is $R = 1.0 * 10^{-16}$m, when the distance between two electrons is 1 m, the ratio of component force on half sphere of an electron to resultant force on the whole electron is

$$\frac{hF_e}{F_e} \approx \frac{\pi R^2 \rho C_U^2}{6} * \frac{9L^2}{2\pi R^4 K_U \rho C_U^2} = \frac{3L^2}{4R^2 K_U} \gg 7.5 * 10^{31}$$

It shows that electrostatic force is differential pressure, and unidirectional component force is much greater than resultant force. When calculating repulsion force between two electrons, assuming that rotational kinetic energy of the Ue around electron is exactly the same, and then its influence on differential force can be ignored.

When the distance L between two electrons is less than 2R, the two electrons overlap in space, as shown in Fig-A12.

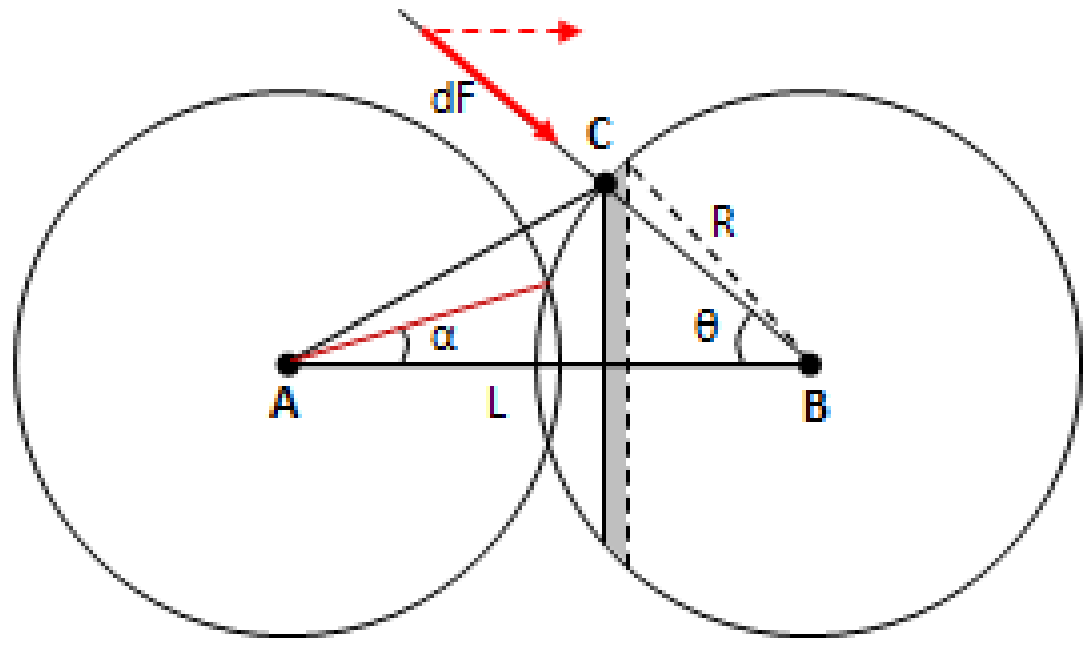

**Fig-A12** Distance between two electrons is less than diameter of an electron

When $p_S$ applies on dS, component force in upward and downward directions counteracts, and the horizontal force on electron B to the right is

$$F_e = \int_\alpha^{\pi} p_S * dS * \cos\theta = \frac{\pi R^2 \rho C_U^2}{3} \int_\alpha^{\pi} \left(1 + \frac{K_U R}{\sqrt{L^2 + R^2 - 2LR\cos\theta}}\right) \sin\theta \cos\theta \, d\theta$$

Assume $x = \cos\alpha = L/2R$, then

$$F_e = \frac{\pi R^2 \rho C_U^2}{18}(3x^2 + 4K_U x - 3 - 3K_U) = \frac{\pi R^2 K_U \rho C_U^2}{18} * \left(\frac{3x^2 - 3}{K_U} + 4x - 3\right)$$

For detailed calculation, please refer to D-3 of appendix D "Integral calculation related to electrostatic force".

Since $K_U \ll 1$, assume that $K_U = 10^{-3}$, solve the equation of $x^2 + 4K_U x - 3 - 3K_U = 0$, so $x \approx 0.99983$, that is to say when $L = 2R * x \approx 1.99966R$, $F_e = 0$. If $K_U = 10^{-30}$, when $L \approx 2R$, $F_e = 0$. Electrostatic repulsion between two protons is equal to that between two electrons. Therefore, when the distance between two protons is less than 2 times of the electron radius, electrostatic repulsion between two protons drops to zero, which maybe helps to explain the strong interaction in nucleus.

When $L > 2R$, assume $x = L/2R$ also, then the repulsion between two electrons is

$$F_e = \frac{2\pi R^4 K_U \rho C_U^2}{9} * \frac{1}{L^2} = \frac{\pi R^2 K_U \rho C_U^2}{18} * \frac{1}{x^2}$$

Assume

$$F_e = \frac{\pi R^2 K_U \rho C_U^2}{18} * f(x)$$

Then the curve of $f(x)$ is shown in Fig-A13. From $x = 1$ to the left, the smaller $K_U$, the faster $f(x)$ decreases, and $f(x)$ almost vertically decreases.

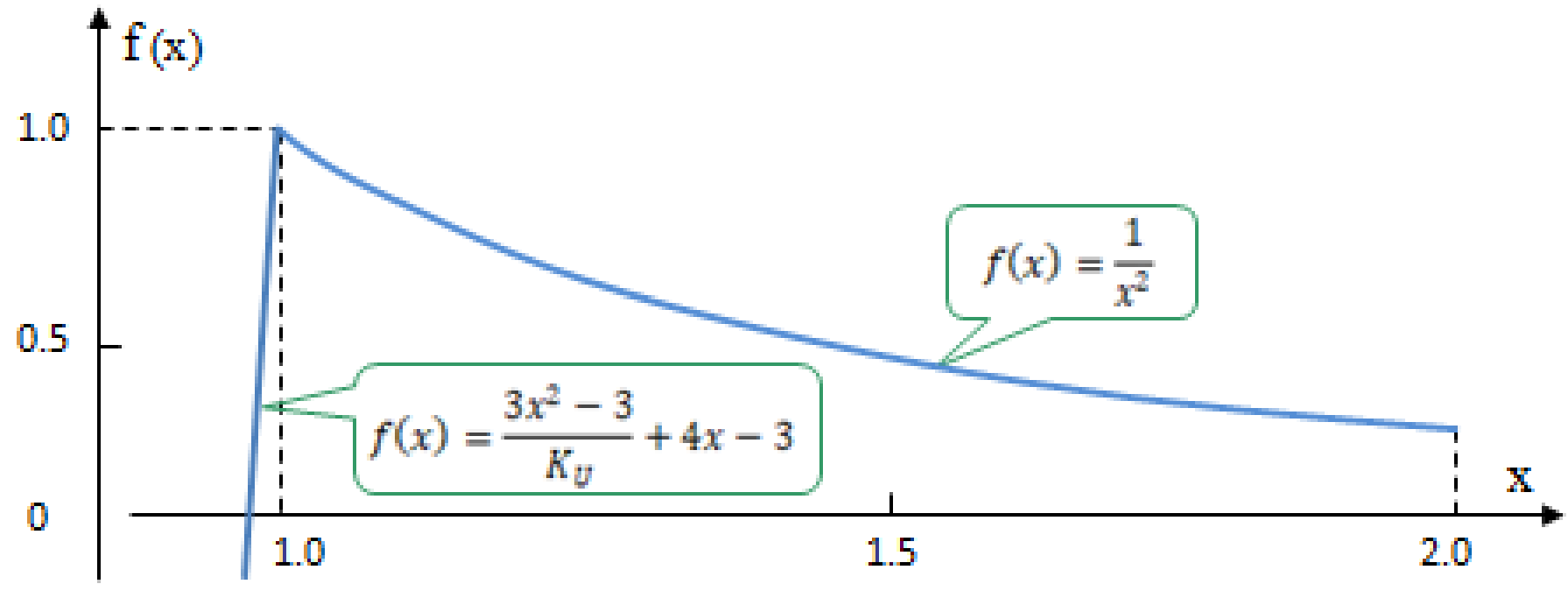

**Fig-A13** The change from electrostatic attraction to repulsion force

U-12: Assume that in addition to U-particle in vacuum, there is another type of particle with a mass $M_B$ much less than U-particle. This particle is called tiny-ball, abbreviated as TB. TB is not isotropic and can be regarded as a small ball constituting U-particle. An electron/proton swallows any TB it collides with, and then releases a TB with only rotational kinetic energy and zero translational kinetic energy at the collision point. The rotation direction of TB in space varies randomly with collisions between TBs. The sum of rotational kinetic energy $e_R$ and translational kinetic energy $e_T$ of a TB is a constant $e_B$, where $e_B = M_B C_U^2/2$. Assume $r$ is the distance between the TB and centre of the electron, then the decreasing function of rotational kinetic energy $e_R(r)$ of TB is $e_R(r) = K_B e_B * R/r$, where $K_B$ is a coefficient related to the diffusion motion of TB. Gravitation is the result of the direct collision between TB and electron/proton. Since the translational kinetic energy of TB released by an electron/proton increases with the distance between the TB and the electron/proton, the gravitation is always attractive force. Gravitation on a proton generated by another proton is inversely proportional to square of the distance between them, or proportional to the gradient of rotational kinetic energy of TB. Direction of gravitation is the line between the two protons. Gravitation is also differential force. If the distance between the two protons is $L$, then gravitation on a proton generated by another proton is

$$F_g \propto \frac{1}{L^2} \qquad (6)$$

Gravitation on a proton generated by an electron is similar. In addition to colliding with each other and electrons/protons, TB also collides with U-particle. TB is much smaller than the U-particle. In an electrically neutral environment, after TBs collide with a U-particle, the U-particle experiences a force from TBs that is proportional to the gradient of TB's rotational kinetic energy.

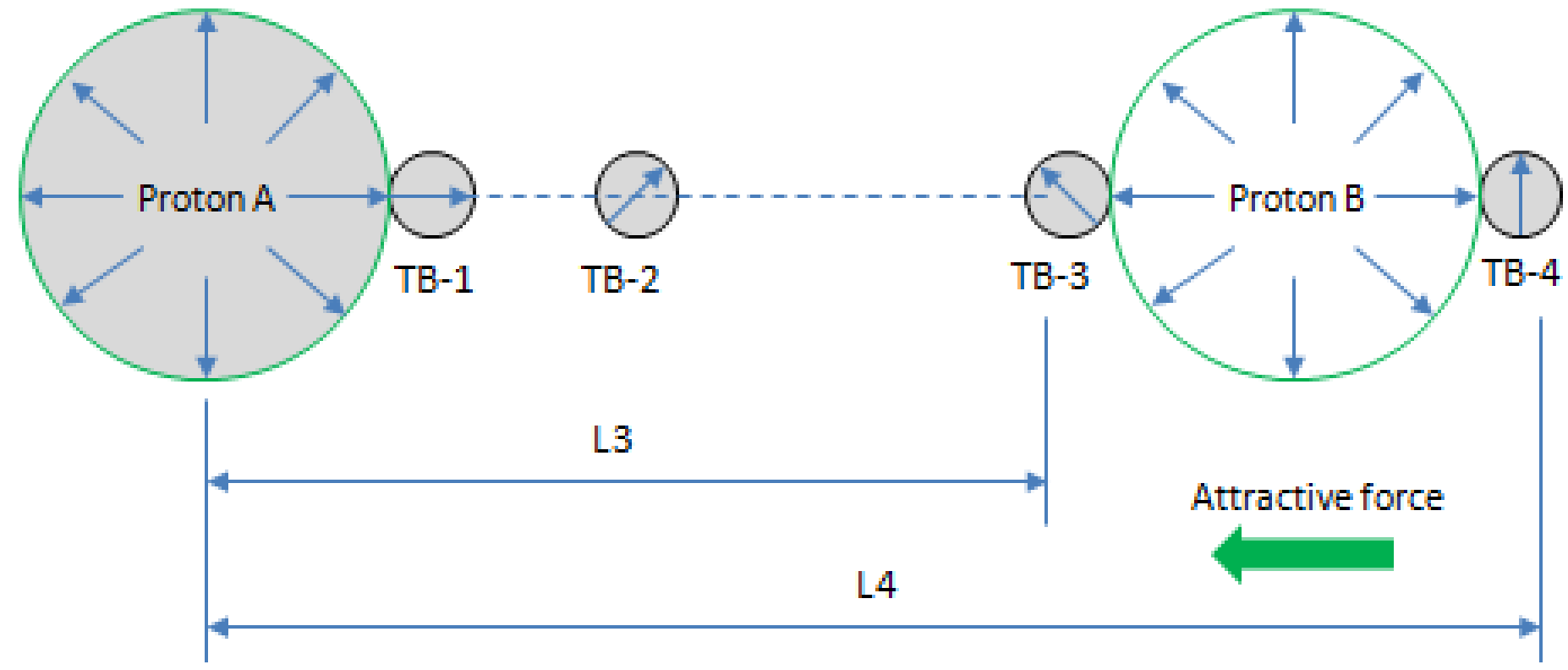

**Fig-A14** Gravitation is generated

Explanation: As shown in Fig-A14, TB-1, TB-2, TB-3, and TB-4 are TB in equilibrium state around proton A when proton B does not exist, their rotational kinetic energy are $B_1$,$B_2$, $B_3$,$B_4$ respectively, so their translational kinetic energy are $(e_B - B_1)$, $(e_B - B_2)$, $(e_B - B_3)$, $(e_B - B_4)$ respectively, gravitation is the result of their direct collision with proton B, for the reason of $B_3 > B_4$, translational kinetic energy of the TB colliding with proton B on the left-side is smaller than that of right-side and therefore generates smaller pressure, so proton B moves to the left, it shows that gravitation is always attractive force.

Similar to calculating electrostatic repulsion between electrons A and B, it can be calculated that gravitation between protons A and B satisfies the inverse square relationship of distance. In applications requiring high precision, equation (1) needs to be corrected to include the effect of TB.

U-13: A static electron/proton with zero translational velocity and angular velocity is pushed by U-particles with changing macro motion velocity, and this force is called the ideal induced force $\boldsymbol{F}_i$, its value is proportional to the rate of change of macro velocity $\boldsymbol{v}_U(t)$ of U-particles with respect to time. Direction of the force on a proton is the same as that of acceleration of Up and opposite to that of Ue. Direction of the force on an electron is the same as that of acceleration of Ue and opposite to that of Up. The force exerted on a proton is

$$\boldsymbol{F}_i = \frac{2\pi R^3 \rho}{9} * \frac{d\boldsymbol{v}_U(t)}{dt} \quad (7)$$

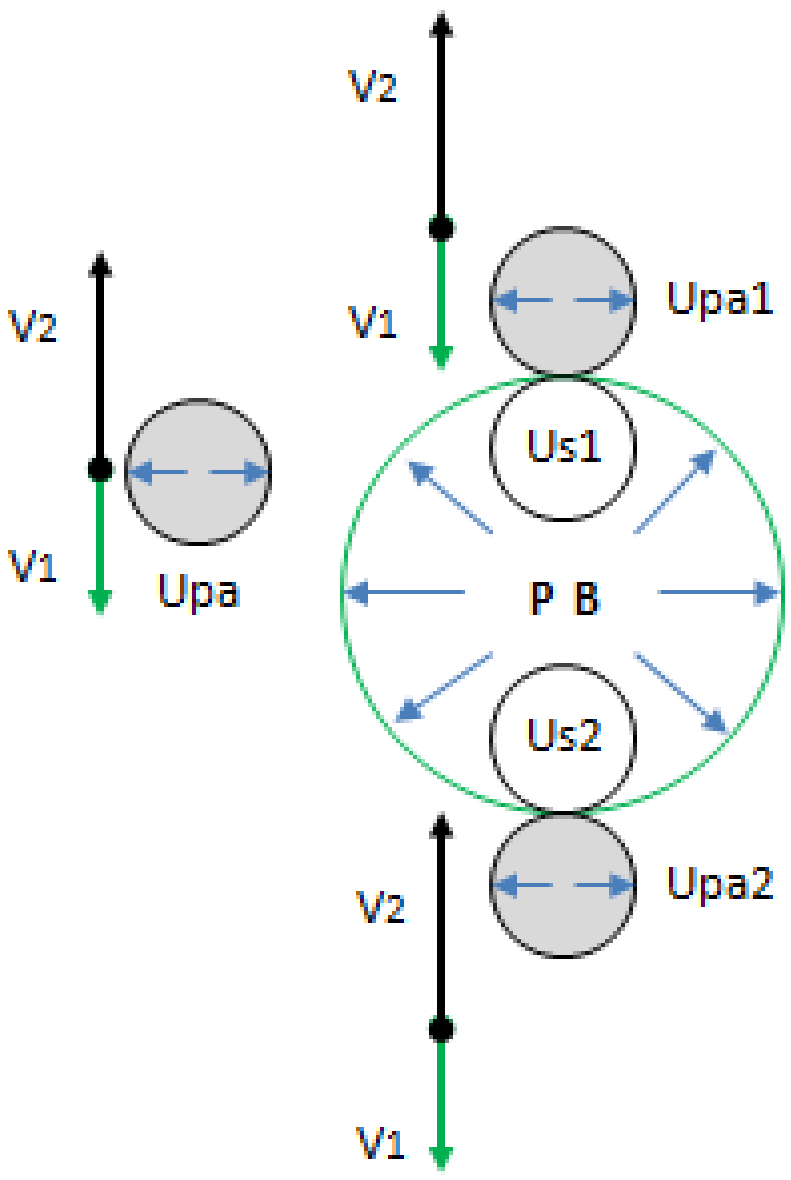

**Fig-A15** The collision between macro moving Up and a static proton

As shown in Fig-A15, the static proton B is in an environment that macro velocity of Up is $\boldsymbol{v}_U = \boldsymbol{V}_2 - \boldsymbol{V}_1$, direction of $\boldsymbol{V}_2$ is positive direction of Z-axis. Before colliding with the proton, Upa1 has translational velocity $\boldsymbol{V}_1$ and rotational kinetic energy $E_{R1}$. After the collision, the second-hand Us1 has rotational kinetic energy $E_{RS1}$ and translational kinetic energy $E_{TS1}$. Before colliding with the proton, Upa2 has translational velocity $\boldsymbol{V}_2$ and rotational kinetic energy $E_{R2}$. After the collision, the second-hand Us2 has rotational kinetic energy $E_{RS2}$ and translational kinetic energy $E_{TS2}$.

$$E_{R1} = E_U - \frac{M_U V_1^2}{2}$$

$$E_{RS1} = \frac{E_U + E_{R1}}{2} = E_U - \frac{M_U V_1^2}{4}$$

$$E_{TS1} = E_U - E_{RS1} = \frac{M_U V_1^2}{4}$$

In the same way, it can be calculated

$$E_{TS2} = \frac{M_U V_2^2}{4} = \frac{M_U (V_1 + v_U)^2}{4}$$

When $v_U > 0$, $E_{TS2} > E_{TS1}$, the pressure generated by second-hand U-particles on the bottom surface of the proton is greater than the pressure generated by second-hand U-particles on the top surface, therefore, proton B is pushed upward. No matter what kind of collision occurs between Upa and proton B, when the upward velocity of proton B increases from zero to $v_U$, the momentum change of proton B is the same.

When the macro velocity of Upa is $\boldsymbol{v}_U(t)$, the macro momentum of Upa contained in a virtual sphere Ω with radius R is

$$\boldsymbol{P}_V(t) = \iiint\limits_{\Omega} \boldsymbol{v}_U(t) * \rho d\Omega = \frac{4\pi R^3 \rho}{3} * \boldsymbol{v}_U(t)$$

When an object is stationary, the momentum entering its surface is equal to its momentum increment. According to B-3 of appendix B "Mathematical calculation of random collision of U-particle", the magnitude of momentum increment of proton B per unit time is

$$\frac{dP(t)}{dt}=\frac{dP_V(t)}{dt}*\frac{1}{6}=\frac{4\pi R^3\rho}{3}*\frac{dv_U(t)}{dt}*\frac{1}{6}=\frac{2\pi R^3\rho}{9}*\frac{dv_U(t)}{dt}$$

The upward pushing force on proton B is

$$F_i(t)=\frac{dP(t)}{dt}=\frac{2\pi R^3\rho}{9}*\frac{dv_U(t)}{dt}$$

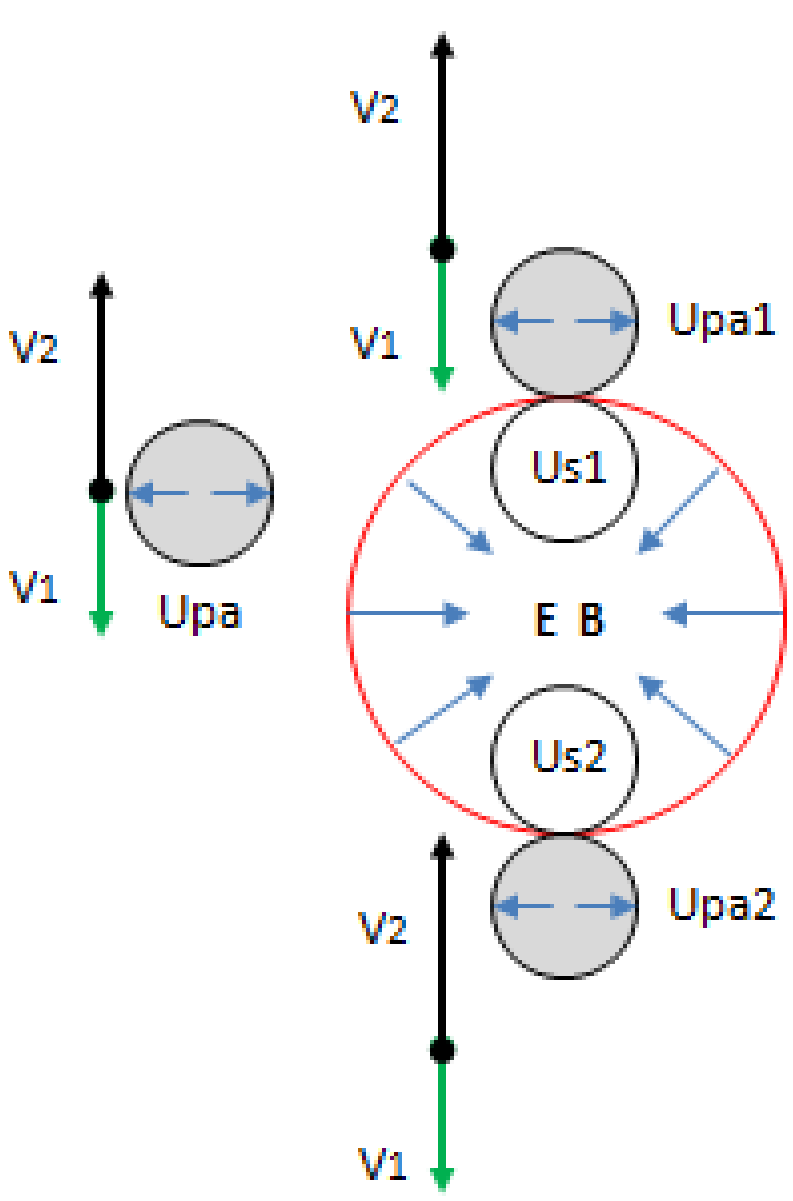

**Fig-A16** The collision between macro moving Up and a static electron

As shown in Fig-A16, the static electron B is in an environment that macro velocity of Up is $\boldsymbol{v}_U=\boldsymbol{V}_2-\boldsymbol{V}_1$, direction of $\boldsymbol{V}_2$ is positive direction of Z-axis. Before colliding with the electron, Upa1 has translational velocity $\boldsymbol{V}_1$ and rotational kinetic energy $E_{R1}$. After the collision, the second-hand Us1 has rotational kinetic energy $E_{RS1}$ and translational kinetic energy $E_{TS1}$. Before colliding with the electron, Upa2 has translational velocity $\boldsymbol{V}_2$ and rotational kinetic energy $E_{R2}$. After the collision, the second-hand Us2 has rotational kinetic energy $E_{RS2}$ and translational kinetic energy $E_{TS2}$.

$$E_{R1}=E_U-\frac{M_UV_1^2}{2}$$

$$E_{RS1}=\frac{E_U-E_{R1}}{2}=\frac{M_UV_1^2}{4}$$

$$E_{TS1}=E_U-E_{RS1}=E_U-\frac{M_UV_1^2}{4}$$

In the same way, it can be calculated

$$E_{TS2}=E_U-\frac{M_UV_2^2}{4}=E_U-\frac{M_U(V_1+v_U)^2}{4}$$

When $v_U>0$, $E_{TS2}<E_{TS1}$, the pressure generated by second-hand U-particles on the bottom surface of the electron is less than the pressure generated by second-hand U-particles on the top surface, therefore, electron B is pushed downward. No matter what kind of collision occurs between Upa and electron B, when the downward velocity of electron B increases from zero to $v_U$, the momentum change of electron B is the same. The momentum of electron B in the Z-axis direction decreases, but the sum of the momentum in the Z-axis direction of Upa1 and Upa2 increases after collision, so it does not violate the momentum conservation.

Similar to momentum change of proton B, the magnitude of momentum increment of electron B per unit time is

$$\frac{dP(t)}{dt}=\frac{dP_V(t)}{dt}*\frac{1}{6}=\frac{4\pi R^3\rho}{3}*\frac{dv_U(t)}{dt}*\frac{1}{6}=\frac{2\pi R^3\rho}{9}*\frac{dv_U(t)}{dt}$$

The downward pushing force on electron B is

$$F_i(t)=\frac{dP(t)}{dt}=\frac{2\pi R^3\rho}{9}*\frac{dv_U(t)}{dt}$$

For two parallel wires, when the current of one wire increases, the other wire will induce a current in the opposite direction, which seemingly violates the law of conservation of momentum. According to the above analysis, due to the existence of U-particles, the momentum is actually conserved. The phenomenon of electromagnetic induction seemingly violates the law of conservation of momentum, which makes it difficult for scientists in the nineteenth century to analyze electromagnetic field with Newtonian mechanics. Maxwell finally used analytical mechanics to analyze and calculate electromagnetic field from the perspective of energy. In the book of A Treatise on Electricity and Magnetism, he expressed his expectation for the mechanical model of electricity.

U-14: The motion state of U-particles around a electron/proton is the same, angular velocity of the electron/proton rotating on the axis of itself is a constant $\boldsymbol{\omega}$, if the electron/proton also has a translation velocity $\boldsymbol{v}$, then the electron/proton is pushed by U-particle and the force is

$$\boldsymbol{F}_B=-\frac{4\pi R^3\rho}{9}*(\boldsymbol{v}\times\boldsymbol{\omega}) \qquad (8)$$

This force is called ideal Lorentz force.

Explanation: As shown in Fig-A17, an electron rotates anticlockwise with a constant angular velocity $\boldsymbol{\omega}$ in the positive direction of Z-axis in XYZ coordinate system, and moves in the direction of X-axis with velocity $\boldsymbol{v}$. The electron is cut parallel to XY plane to form a circle with radius $r$, and $\varphi$ is the angle between 0 and $\pi$.

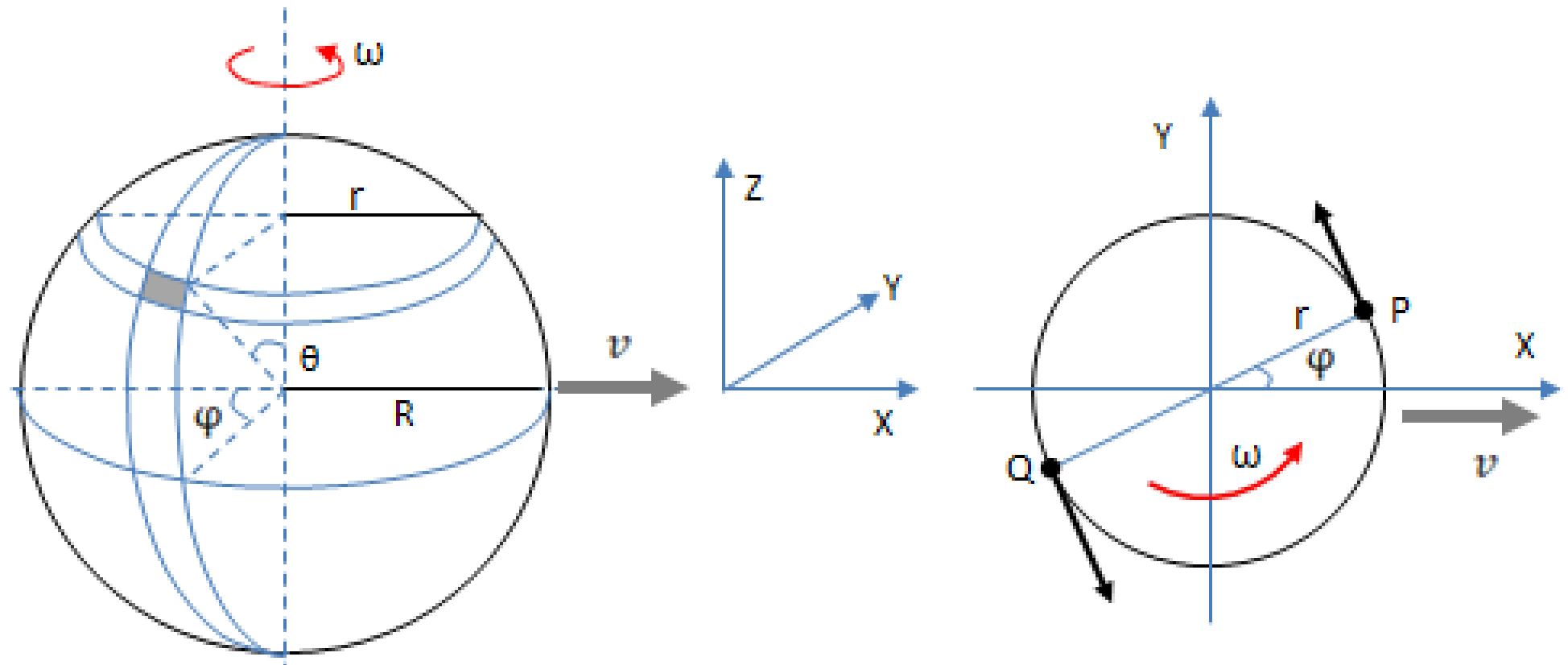

**Fig-A17** An electron rotating on its axis

In Fig-A17, on the cross-section of the electron, suppose that the velocity at point P is $\boldsymbol{V}_1$ and the velocity at point Q is $\boldsymbol{V}_2$, then we have

$$V_{1x} = v - \omega r \sin\varphi \qquad V_{1y} = \omega r \cos\varphi$$
$$V_{2x} = v + \omega r \sin\varphi \qquad V_{2y} = -\omega r \cos\varphi$$
$$V_2^2 - V_1^2 = \left(V_{2x}^2 + V_{2y}^2\right) - \left(V_{1x}^2 + V_{1y}^2\right) =$$
$$= [(v + \omega r \sin\varphi)^2 + (-\omega r \cos\varphi)^2] - [(v - \omega r \sin\varphi)^2 + (\omega r \cos\varphi)^2]$$
$$= 4v\omega r \sin\varphi = 4v\omega R \sin\theta \sin\varphi$$

The motion state of U-particle in surrounding environment is consistent, so the rotational kinetic energy of U-particles are the same，the electron with angular velocity $\omega$ enters the environment at translational velocity $\boldsymbol{v}$.

Case 1: rotational kinetic energy of Up in surrounding environment is $E_{Rp}$

Suppose that after a Up collides with the electron at point P, the resulting second-hand U-particle is $U_{S1}$, with rotational kinetic energy $E_{RS1}$ and translational kinetic energy $E_{TS1}$; and after a Up collides with the electron at point Q, the resulting second-hand U-particle is $U_{S2}$, with rotational kinetic energy $E_{RS2}$ and translational kinetic energy $E_{TS2}$. The pressure generated by $U_{S1}$ is $p_{S1}$ and the pressure generated by $U_{S2}$ is $p_{S2}$. According to U-8

$$E_{RS1} = \frac{(E_U - M_U V_1^2/2) - E_{Rp}}{2} = \frac{E_U}{2} - \frac{E_{Rp}}{2} - \frac{M_U V_1^2}{4}$$

$$E_{TS1} = E_U - E_{RS1} = \frac{E_U + E_{Rp}}{2} + \frac{M_U V_1^2}{4} = \frac{M_U V_{S1}^2}{2}$$

$$V_{S1}^2 = \frac{E_U + E_{Rp}}{M_U} + \frac{V_1^2}{2}$$

According to equation (5)

$$p_{S1} = \frac{\rho V_{S1}^2}{3} = \frac{\rho\left(E_U + E_{Rp}\right)}{3M_U} + \frac{\rho V_1^2}{6}$$

It can be calculated similarly

$$p_{S2} = \frac{\rho V_{S2}^2}{3} = \frac{\rho\left(E_U + E_{Rp}\right)}{3M_U} + \frac{\rho V_2^2}{6}$$

So

$$p_{S2} - p_{S1} = \frac{\rho(V_2^2 - V_1^2)}{6}$$

Case 2: rotational kinetic energy of Ue in the surrounding environment is $E_{Re}$

Suppose that after a Ue collides with the electron at point P, the resulting second-hand U-particle is $U_{S1}$, with rotational kinetic energy $E_{RS1}$ and translational kinetic energy $E_{TS1}$; and after a Ue collides with the electron at point Q, the resulting second-hand U-particle is $U_{S2}$, with rotational kinetic energy $E_{RS2}$ and translational kinetic energy $E_{TS2}$. The pressure generated by $U_{S1}$ is $p_{S1}$ and the pressure generated by $U_{S2}$ is $p_{S2}$. According to U-8

$$E_{RS1} = \frac{(E_U - M_U V_1^2/2) + E_{Re}}{2} = \frac{E_U}{2} + \frac{E_{Re}}{2} - \frac{M_U V_1^2}{4}$$

$$E_{TS1} = E_U - E_{RS1} = \frac{E_U - E_{Re}}{2} + \frac{M_U V_1^2}{4} = \frac{M_U V_{S1}^2}{2}$$

$$V_{S1}^2 = \frac{E_U - E_{Re}}{M_U} + \frac{V_1^2}{2}$$

According to equation (5)

$$p_{S1} = \frac{\rho V_{S1}^2}{3} = \frac{\rho(E_U - E_{Re})}{3M_U} + \frac{\rho V_1^2}{6}$$

It can be calculated similarly

$$p_{S2} = \frac{\rho V_{S2}^2}{3} = \frac{\rho(E_U - E_{Re})}{3M_U} + \frac{\rho V_2^2}{6}$$

So

$$p_{S2} - p_{S1} = \frac{\rho(V_2^2 - V_1^2)}{6}$$

The equation of $p_{S2} - p_{S1}$ is the same regardless of whether the surrounding environment is Up or Ue, so

$$p_{S2} - p_{S1} = \frac{\rho(V_2^2 - V_1^2)}{6} = \frac{\rho}{6} * 4v\omega R \sin\theta \sin\varphi = \frac{2R\rho v\omega \sin\theta \sin\varphi}{3}$$

In Fig-A17, area of shadowed surface is $dS = R^2 \sin\theta\, d\theta d\varphi$, when $p_S$ applies on $dS$, component force in positive direction of Y-axis is $dF_B = p_S * dS * \sin\theta \sin\varphi$. For the whole electron, component force in upward and downward directions counteracts. Suppose that front hemisphere of the electron is a curved surface Σ, and then the resultant force on the electron in positive direction of Y-axis is

$$F_B = \iint_\Sigma dF_B = \iint_\Sigma p_S * dS * \sin\theta \sin\varphi = \iint_\Sigma (p_{S2} - p_{S1}) * dS * \sin\theta \sin\varphi$$

$$= \iint_\Sigma \frac{2R\rho v\omega \sin\theta \sin\varphi}{3} * R^2 \sin\theta\, d\theta d\varphi * \sin\theta \sin\varphi$$

$$= \frac{2R^3\rho v\omega}{3} \int_0^\pi \sin^3\theta d\theta \int_0^\pi \sin^2\varphi\, d\varphi = \frac{2R^3\rho v\omega}{3} * \frac{4}{3} * \frac{\pi}{2}$$

$$= \frac{4\pi R^3 \rho}{9} * v * \omega$$

The resultant force on the electron in direction of X-axis is

$$\iint_\Sigma p_S * dS * \sin\theta \cos\varphi = \frac{2R^3\rho v\omega}{3} \int_0^\pi \sin^3\theta d\theta \int_0^\pi \sin\varphi \cos\varphi\, d\varphi = 0$$

The force $\boldsymbol{F}_B$ is Lorentz force in classical electromagnetism, it is similar to the result of Magnus effect in fluid, and the banana ball shot by football players is the result of similar force. When $\boldsymbol{v}$ is not perpendicular to $\boldsymbol{\omega}$,

$$\boldsymbol{F}_B = -\frac{4\pi R^3 \rho}{9} * (\boldsymbol{v} \times \boldsymbol{\omega})$$

A negative sign indicates that the direction of $\boldsymbol{F}_B$ is opposite to that of the vector product of $\boldsymbol{v} \times \boldsymbol{\omega}$. It can be seen from the derivation that equation (8) is the same for both electron and proton.

## 3 Superposition principle

From previous calculations of electrostatic force, induced force, and Lorentz force, it can be seen that all these forces are the rate of change of momentum of U-particle with respect to time. Since momentum satisfies linear addition, these forces also satisfy linear addition.

From the calculation process of electrostatic force, it can be seen that when Up's $\boldsymbol{\nabla} E_R$ or Ue's $\boldsymbol{\nabla} E_R$ remains unchanged, the force direction on an electron is opposite to that on a proton. Direction of the force generated by Up's $\boldsymbol{\nabla} E_R$ and Ue's $\boldsymbol{\nabla} E_R$ on the same electric charge is opposite. For the convenience of calculations, it can be assumed that the signs of proton and electron are opposite, and the signs of Up's $E_R$ and Ue's $E_R$ are opposite. In this paper, take proton as positive and electron as negative, take Up's $E_R$ as positive and Ue's $E_R$ as negative. This negative sign only indicates that direction of the force generated by Ue's $\boldsymbol{\nabla} E_R$ and Up's $\boldsymbol{\nabla} E_R$ on the same electric charge is opposite, not Ue's $E_R$ and Up's $E_R$ can cancel absolutely each other out.

From the calculation process of induced force, it can be seen that when Up's $d\boldsymbol{v}_U/dt$ or Ue's $d\boldsymbol{v}_U/dt$ remains unchanged, the force direction on an electron is opposite to that on a proton. Direction of the force generated by Up's $d\boldsymbol{v}_U/dt$ and Ue's $d\boldsymbol{v}_U/dt$ on the same electric charge is opposite. For the convenience of calculations, this paper takes Up's $\boldsymbol{v}_U$ as positive and Ue's $\boldsymbol{v}_U$ as negative. This negative sign indicates that direction of the force generated by Ue's $d\boldsymbol{v}_U/dt$ and Up's $d\boldsymbol{v}_U/dt$ on the same electric charge is opposite.

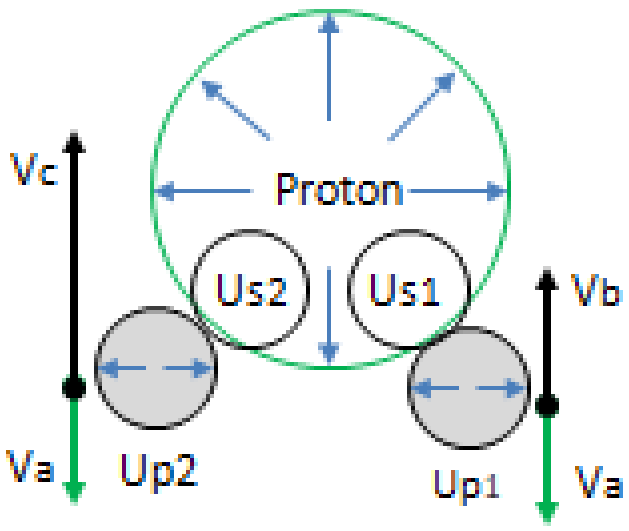

**Fig-A18** Rotation direction of a proton

As shown in Fig-A18, Direction of macro motion of Up is upward. Assume that the macro speed $(V_c - V_a)$ of Up on the lower left side of the proton is greater than the macro speed $(V_b - V_a)$ of Up on the lower right side of the proton. After Up1 collides with the proton, a second-hand Us1 is produced with translational kinetic energy $E_{TS1}$; after Up2 collides with the proton, a second-hand Us2 is produced with translational kinetic energy $E_{TS2}$. Referring to U-13 analysis, it can be seen that $E_{TS2} > E_{TS1}$, so the proton will rotate clockwise. If the proton in Fig-A18 is replaced by an electron, the electron rotates anticlockwise. Similarly, macro motion of Ue causes proton and electron to rotate in opposite directions. Although the Lorentz force on electrons and protons can be calculated by equation (8), the macro velocity of Up and the macro velocity of Ue cause the same electric charge to rotate in the opposite direction, resulting in the opposite direction of the Lorentz force. If rotation direction of an object is represented by right-hand rule, then, the rotational direction of proton is the same as the direction of curl $\boldsymbol{\nabla} \times \boldsymbol{v}_U$ of Up's macro velocity.

Therefore, if we take proton as positive and electron as negative, take rotational kinetic energy $E_R$ of Up as positive and rotational kinetic energy $E_R$ of Ue as negative, take macro velocity $\boldsymbol{v}_U$ of Up as positive and macro velocity $\boldsymbol{v}_U$ of Ue as negative, then electrostatic force, induced force, and Lorentz force satisfy linear calculations. Similarly, the gradient, divergence, and curl associated with them also satisfy linear operations.

In U-7, the decreasing function of rotational kinetic energy of U-particle is derived by applying mathematical theorem of random collision to an isolated electron and equilibrium system, but in the real physical environment, such as two or more electrons, it can neither satisfy the isolated electron nor keep the system in equilibrium state, therefore, the use of $E_R(r) = K_U E_U * R/r$ is limited.

When an isolated electron is in stable diffusion equilibrium state, the flux $\emptyset_e$ of rotational kinetic energy $E_{Re}$ of U-particle can be calculated in spherical coordinates according to equation (2) as follows, Σ is the closed surface encloses the electron.

$$\emptyset_e \propto \oiint_{\Sigma} \boldsymbol{\nabla} E_{Re}(r) \cdot d\boldsymbol{S} = \oiint_{\Sigma} -\frac{K_U E_U R}{r^2} * r^2 \sin\theta \, d\theta d\varphi = -4\pi R K_U E_U$$

It can be seen from the calculation results that the flux $\emptyset_e$ of rotational kinetic energy is a constant. According to U-6, an electron is an energy converter which can convert translational kinetic energy into rotational kinetic energy without the influence of other electrons/protons, then, in the case of multiple electrons, the equation is still suitable for each individual electron. In the case of N electrons, if the gradient of rotational kinetic energy of U-particle is $\boldsymbol{\nabla} E_R(r)$, then the flux of rotational kinetic energy of U-particle can be calculated as follows, Σ is a closed surface that encloses these N electrons.

$$\emptyset \propto \oiint_{\Sigma} \boldsymbol{\nabla} E_R(r) \cdot d\boldsymbol{S} = \oiint_{\Sigma} \left(\boldsymbol{\nabla} E_{Re1}(r) + \boldsymbol{\nabla} E_{Re2}(r) + \cdots + \boldsymbol{\nabla} E_{ReN}(r)\right) \cdot d\boldsymbol{S} \propto N * \emptyset_e = Constant$$

U-15: Electrostatic force, induced force, and Lorentz force are all the rate of change of momentum of U-particle with respect to time. Take proton as positive and electron as negative, take rotational kinetic energy $E_R$ of Up as positive and rotational kinetic energy $E_R$ of Ue as negative, take macro velocity $\boldsymbol{v}_U$ of Up as positive and macro velocity $\boldsymbol{v}_U$ of Ue as negative, then electrostatic force, induced force, and Lorentz force satisfy linear addition. Similarly, the gradient, divergence, and curl associated with them also satisfy linear addition. The principle of superposition in physics is to describe physical law by accurate mathematical addition, which has a small error with the actual physical results. The superposition principle can be used to describe low-speed motion, and it is assumed that the following equation is always correct for stationary electrons/protons.

$$\oiint_{\Sigma} \boldsymbol{\nabla} E_R(r) \cdot d\boldsymbol{S} = Constant \qquad (9)$$

Σ is a closed surface that encloses these electrons/protons. Similarly, gravitation also satisfies linear addition.

Suppose the number of protons is positive, the number of electrons is negative, and the number of charges is the number of protons plus the number of electrons. The distance between two objects with charge number N and M is L, then, electrostatic force between them is

$$F_E = N * M * F_e = \frac{2\pi R^4 K_U \rho C_U^2}{9} * \frac{NM}{L^2}$$

It's hard to understand superposition principle of Coulomb force in classical electromagnetism, if there are N+1 electrons on a line, electron A is on the left and others N electrons on the right, no matter how many the N is, repulsive force from electron A to every electron is unchangeable, is electron A green giant? The superposition principle of electrostatic force is easy to understand by mechanical model based on U-particle, the force exerted on the N electron is generated by U-particles suround electron A. It's similar to gas pressure.

Equation (3) is based on the assumption that electrons move at low speeds, so the superposition principle is only suitable for low speed motion.

## 4 Electrostatic field and constant magnetic field

In classical electromagnetism, the unit of current Ampere is defined as follows: two parallel infinite long straight wires in vacuum, distance between them is 1 meter, current on the two wires is equal in magnitude and in the same direction，if attractive force exerted on per meter wire is equal to $2 * 10^{-7}\ N$, then the current in each wire is 1 Ampere, equal to 1 Coulomb per second. The definition of Ampere is the bridge of quantitative calculation between Newtonian mechanics and electromagnetics. In classical electromagnetism, the electric charge of an electron $Q_e \approx 1.602 * 10^{-19} C$ and permittivity of vacuum $\varepsilon_0 = 8.854 * 10^{-12}\ C^2/(N.m^2)$ are measured by experiments. There is no concept of electric charge in mechanical model of U-particle. In order to verify correctness of mechanical model of U-particle by using existing achievements of classical electromagnetism, definition and unit of classical electromagnetism should be used uniformly, therefore, one electron in mechanical model of U-particle is equivalent to $1.602 * 10^{-19}\ C$ in classical electromagnetism.

U-16: The gradient field of rotational kinetic energy of U-particle constitutes electrostatic field. Suppose that electric charge of proton is positive and that of electron is negative, $q$ is electric charge of protons plus that of electrons, absolute value of the electric charge carried by one electron is $Q_e$, then electrostatic field intensity generated by $q$ at a distance of $r$ is

$$\boldsymbol{E}_s(r) = \frac{4\pi R^3 \rho_N}{9Q_e} * \boldsymbol{\nabla} E_R(r) = \frac{2\pi R^4 K_U \rho C_U^2}{9Q_e^2} * \frac{q}{r^2} * \boldsymbol{e}_r \qquad (10)$$

$$\boldsymbol{\nabla} \times \boldsymbol{E}_s = 0 \qquad (11)$$

Assuming that the gradient of rotational kinetic energy of Up is $\boldsymbol{\nabla} E_{Rp}(r)$ and that of Ue is $\boldsymbol{\nabla} E_{Re}(r)$, then the electrostatic field intensity is proportional to $\boldsymbol{\nabla} E_{Rp}(r) + \boldsymbol{\nabla} E_{Re}(r)$.

The negative value of the gradient of electrostatic potential $\varphi_s$ is electrostatic field intensity $\boldsymbol{E}_s$, $\boldsymbol{E}_s = -\boldsymbol{\nabla}\varphi_s$. The electrostatic potential $\varphi_s(r)$ generated by electric charge $q$ at a point with distance $r$ from it is

$$\varphi_s(r) = \frac{2\pi R^4 K_U \rho C_U^2}{9Q_e^2} * \frac{q}{r} = -\frac{1}{3Q_e} * \left(\rho_{ER}(r) * \frac{4\pi R^3}{3}\right) = -\frac{E_h(r)}{3Q_e} \qquad (12)$$

$\rho_{ER}(r) = E_R(r) * \rho_N$, where $\rho_{ER}(r)$ is rotational kinetic energy density of U-particle, and its symbol is the same as that of $E_R(r)$. Hidden energy $E_h(r)$ at a certain point is equal to rotational kinetic energy density $\rho_{ER}(r)$ of U-particle at the point times the volume of a sphere with radius R.

Explanation: The definition of electric field intensity in classical electromagnetism is the force exerted on unit positive charge in electric field. According to equation (4), the repulsive force of one proton to another is

$$F_e = \frac{2\pi R^4 K_U \rho C_U^2}{9} * \frac{1}{r^2}$$

The number of protons equivalent to electric charge $q$ is $N = q/Q_e$ and the number of protons of per unit positive charge is $M = 1/Q_e$, according to the superposition principle, the repulsion force exerted on unit positive charge which is generated by electric charge $q$ is

$$F_E = N * M * F_e = \frac{q}{Q_e} * \frac{1}{Q_e} * F_e = \frac{2\pi R^4 K_U \rho C_U^2}{9Q_e^2} * \frac{q}{r^2}$$

Electrostatic field intensity generated by electric charge $q$ is

$$E_s = \frac{2\pi R^4 K_U \rho C_U^2}{9Q_e^2} * \frac{q}{r^2}$$

U-particle around a proton is Ue, so take its rotational kinetic energy as a negative value. For a single proton, according to equation (1), $E_R(r) = -K_U E_U * R/r$, for $N = q/Q_e$ protons, using the superposition principle

$$E_R(r) = -K_U E_U * \frac{R}{r} * \frac{q}{Q_e}$$

$$\left(\boldsymbol{\nabla} E_R(r)\right)_r = K_U E_U * \frac{R}{r^2} * \frac{q}{Q_e} = \frac{K_U M_U C_U^2}{2} * \frac{R}{r^2} * \frac{q}{Q_e} = \frac{K_U M_U C_U^2 R q}{2r^2 Q_e}$$

$$E_s = \frac{2\pi R^4 K_U \rho C_U^2}{9Q_e^2} * \frac{q}{r^2} = \frac{4\pi R^3 \rho_N}{9Q_e} * \frac{K_U M_U C_U^2 R q}{2r^2 Q_e} = \frac{4\pi R^3 \rho_N}{9Q_e} * \left(\boldsymbol{\nabla} E_R(r)\right)_r$$

$$\boldsymbol{E}_s(r) = \frac{4\pi R^3 \rho_N}{9Q_e} * \boldsymbol{\nabla} E_R(r)$$

The calculation method of electrostatic field intensity $\boldsymbol{E}_s$ in the above equation refers to the definition of electrostatic field intensity $\boldsymbol{E}_s$ in classical electromagnetism, but it can define the electrostatic field intensity $\boldsymbol{E}_s$ at a certain point. Therefore, this paper takes it as the accurate definition of electrostatic field intensity $\boldsymbol{E}_s$ under the mechanical model of U-particle. Because curl of gradient is always zero, the electrostatic field is irrotational field.

$$\boldsymbol{\nabla} \times \boldsymbol{E}_s = 0$$

After the rotational kinetic energy of U-particle is given a positive or negative sign, for the electrostatic force on an electron/proton, the effect of Up's rotational kinetic energy gradient $\boldsymbol{\nabla} E_{Rp}(r)$ is the same as that of Ue's rotational kinetic energy gradient $\boldsymbol{\nabla} E_{Re}(r)$, that is, the electrostatic field intensity is proportional to $\boldsymbol{\nabla} E_{Rp}(r) + \boldsymbol{\nabla} E_{Re}(r)$.

Classical electromagnetism defines the electrostatic potential of a point as: the work done by electric field force when a unit positive charge travels from the point to infinity through any path. Referring to the definition of classical electromagnetism, under the mechanical model of U-particle, the electrostatic potential of electric charge $q$ at a point with distance $r$ from it is defined as follows

$$\begin{aligned}
\varphi_s(r) &= \int_r^\infty E_s(x)dx = \int_r^\infty \frac{2\pi R^4 K_U \rho C_U^2}{9Q_e^2} * \frac{q}{x^2}dx = \frac{2\pi R^4 K_U \rho C_U^2}{9Q_e^2} * \frac{q}{r} \\
&= \frac{4\pi R^3}{9Q_e} * \left(\rho_N * \frac{K_U M_U C_U^2}{2} * \frac{R}{r} * \frac{q}{Q_e}\right) = \frac{4\pi R^3}{9Q_e} * \rho_N * \left(K_U E_U * \frac{R}{r} * \frac{q}{Q_e}\right) \\
&= \frac{4\pi R^3}{9Q_e} * \rho_N * \left(-E_R(r)\right) = -\frac{1}{3Q_e} * \left(\rho_N E_R(r) * \frac{4\pi R^3}{3}\right) \\
&= -\frac{1}{3Q_e} * \left(\rho_{ER}(r) * \frac{4\pi R^3}{3}\right)
\end{aligned}$$

The positive and negative sign of $\varphi_s(r)$ is the same as the sign of electric charge $q$. For the mechanical model of U-particle, the rotational kinetic energy of U-particle that is infinitely far away from electrons/protons is zero, so the electrostatic potential is zero. Therefore zero potential has a clear physical meaning.

Hidden energy $E_h(r)$ at a certain point is equal to rotational kinetic energy density $\rho_{ER}(r)$ of U-particle at the point times the volume of a sphere with radius R, therefore, the above equation can also be written as $\varphi_s(r) = -E_h(r)/3Q_e$ .

U-17: Infinite long straight wire in the direction of Z-axis, $\lambda$ electrons per unit length move in the wire at constant speed $v_i$, distance between point A and the wire is $r$, then the curl of the macro velocity $\boldsymbol{v}_U$ of Up at point A is

$$\boldsymbol{\nabla} \times \boldsymbol{v}_U = \frac{2K_U R \lambda v_i}{r} * \boldsymbol{e}_\varphi \qquad (13)$$

An electron/proton will rotate on its axis at point A. A proton rotates in the opposite direction to an electron at point A. The angular velocity of a proton is

$$\boldsymbol{\omega} = \frac{\boldsymbol{\nabla} \times \boldsymbol{v}_U}{2} = \frac{K_U R \lambda v_i}{r} * \boldsymbol{e}_\varphi \qquad (14)$$

The relationship between direction of electron angular velocity at point A and direction of current is right-handed helix. Hold the long straight wire with right hand, point four fingers in direction of angular velocity of the electron rotation, and direction of thumb is direction of the current. Velocity curl of Ue and that of Up cause electrons or protons to rotate in the opposite direction.

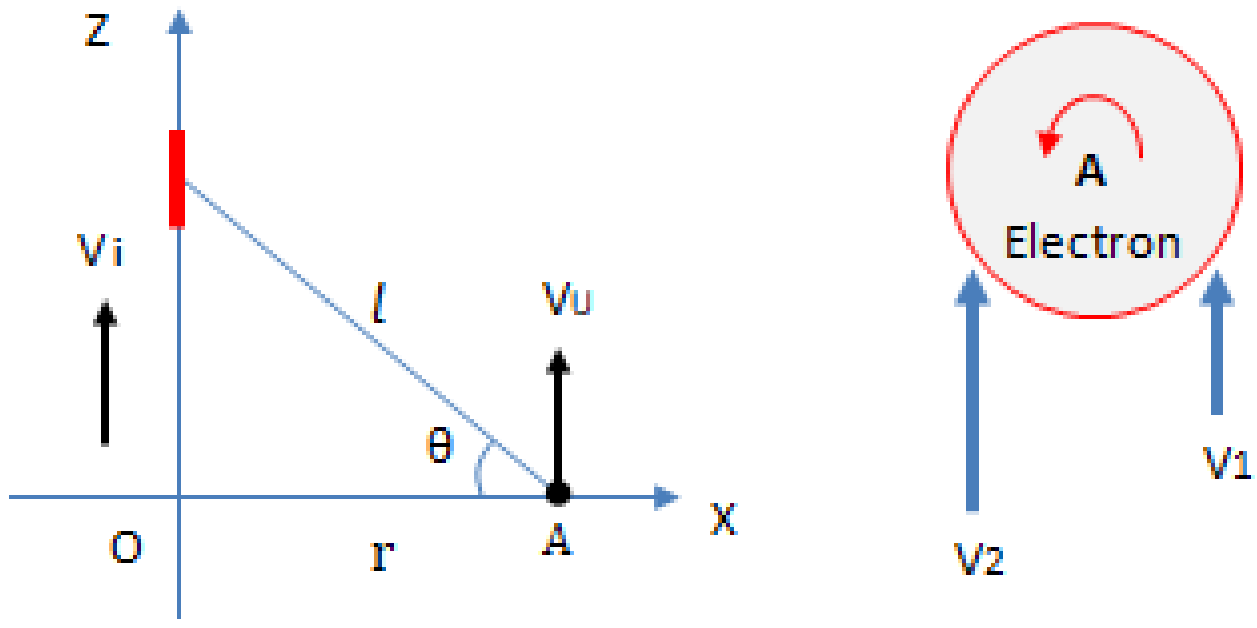

**Fig-A19** The velocity curl of Up and electron rotation

Explanation: As shown in Fig-A19, the infinite long straight wire overlaps with Z-axis, and direction of electron movement in the wire is the positive direction of Z-axis. Take a small length of wire $dz$ on Z-axis, according to U-8, the movement of a single electron with velocity of $v_i$ in $dz$ causes the macro velocity of Up at point A to be $v_{Ue} = K_U R v_i / l$. Since $\boldsymbol{v}_{Ue}$ has only a component in $\boldsymbol{e}_z$ direction, curl of $\boldsymbol{v}_{Ue}$ is calculated with cylindrical coordinates. According to appendix C "Three dimensional coordinate system and simplified operation of Hamilton operator", curl of $\boldsymbol{v}_{Ue}$ has only a component in direction of $\boldsymbol{e}_\varphi$, and the magnitude is

$$(\boldsymbol{\nabla} \times \boldsymbol{v}_{Ue})_\varphi = -\frac{\partial}{\partial r}\left(\frac{K_U R v_i}{l}\right) = \frac{K_U R v_i}{l^2} * \frac{\partial l}{\partial r} = \frac{K_U R v_i}{l^2} * \frac{\partial}{\partial r}\left(\sqrt{r^2 + z^2}\right) = \frac{K_U R v_i \cos\theta}{l^2}$$

Therefore, movement of a single electron with velocity of $v_i$ in $dz$ causes the velocity curl of Up at point A to be

$$\boldsymbol{\nabla} \times \boldsymbol{v}_{Ue} = \frac{K_U R v_i \cos\theta}{l^2} * \boldsymbol{e}_\varphi$$

There are $\lambda dz$ moving electrons in $dz$ length, according to the superposition principle, their movement cause the velocity curl component in direction of $\boldsymbol{e}_\varphi$ of Up at point A to be

$$\lambda dz * (\boldsymbol{\nabla} \times \boldsymbol{v}_{Ue})_\varphi = \lambda d(r \tan\theta) * \frac{K_U R v_i \cos\theta}{l^2} = \frac{K_U R \lambda v_i \cos\theta}{r} d\theta$$

All moving electrons in the infinite wire cause the velocity curl component in direction of $\boldsymbol{e}_\varphi$ of Up at point A to be

$$(\boldsymbol{\nabla} \times \boldsymbol{v}_U)_\varphi = \int \lambda dz * (\boldsymbol{\nabla} \times \boldsymbol{v}_{Ue})_\varphi = \int_{-\pi/2}^{\pi/2} \frac{K_U R \lambda v_i \cos\theta}{r} d\theta = \frac{2K_U R \lambda v_i}{r}$$

So

$$\boldsymbol{\nabla} \times \boldsymbol{v}_U = \frac{2K_U R \lambda v_i}{r} * \boldsymbol{e}_\varphi$$

In Fig-A19, the direction of velocity curl of Up at point A is perpendicular to the XZ plane inward. For the electron at point A, the velocity $V_2$ of Up on the left-side is bigger than $V_1$ on the right-side, referring to analysis of Fig-A18, upward force on the left side of the electron is smaller than upward force on the right side, so the electron will rotate anticlockwise, and its angular velocity direction is opposite to the direction of velocity curl of Up at point A. Hold the long straight wire with right hand, point four fingers in direction of angular velocity of the electron rotation, and direction of thumb is direction of the current. Referring to analysis of Fig-A18, if the electron at point A is replaced by a proton, the proton will rotate clockwise. In the same way, velocity curl of Ue and that of Up cause electrons or protons to rotate in the opposite direction.

Mathematically, curl of the linear velocity of a rigid body is equal to twice the angular velocity, therefore, when angular velocity of the electron at point A is 1/2 of velocity curl of Up at point A, the electron is in equilibrium state. The angular velocity of an electron at point A is

$$\omega = \frac{(\boldsymbol{\nabla} \times \boldsymbol{v}_U)_\varphi}{2} = \frac{K_U R \lambda v_i}{r}$$

U-18: The curl field of macro velocity of U-particle constitutes magnetic field, magnetic induction intensity $\boldsymbol{B}$ of the magnetic field generated by $\boldsymbol{\nabla} \times \boldsymbol{v}_U$ of macro velocity of Up is

$$\boldsymbol{B} = -\frac{2\pi R^3 \rho}{9Q_e} * (\boldsymbol{\nabla} \times \boldsymbol{v}_U) \qquad (15)$$

The curl of magnetic vector potential $\boldsymbol{A}$ is magnetic induction intensity $\boldsymbol{B}$, $\boldsymbol{B} = \boldsymbol{\nabla} \times \boldsymbol{A}$ and

$$\boldsymbol{A} = -\frac{\boldsymbol{P}_{em}}{6Q_e} = -\frac{1}{6Q_e} * \left(\boldsymbol{\rho}_{PU} * \frac{4\pi R^3}{3}\right) \qquad (16)$$

Electrical momentum $\boldsymbol{P}_{em}(r,t)$ at a certain point is equal to macro momentum density $\boldsymbol{\rho}_{PU}$ of Up at the point times the volume of a sphere with radius R.

Infinite long straight wire, λ electrons per unit length move in the wire at constant speed $v_i$ to the positive direction of Z-axis, current in the wire is $I$, distance between point A and the wire is $r$, then the magnetic induction intensity at point A is

$$\boldsymbol{B} = -\frac{4\pi R^4 K_U \rho \lambda v_i}{9Q_e r} * \boldsymbol{e}_\varphi = \frac{4\pi R^4 K_U \rho}{9Q_e^2} * \frac{I}{r} * \boldsymbol{e}_\varphi \qquad (17)$$

The relationship between direction of magnetic induction intensity and direction of current is right-handed helix.

Explanation: In Fig-A19, angular velocity $\boldsymbol{\omega}$ of the electron rotation at point A is perpendicular to the XZ plane outward, if the electron moves upward at speed $v$, according to equation (8), the electron is pushed by U-particle and the direction of $\boldsymbol{F}_B$ is to the left. When calculating the ideal Lorentz force in U-14, it is assumed that the states of U-particles around the electron are the same, but this assumption is impossible in the actual physical environment, because it is the different macro motion states of U-particles around the electron that leads to the angular velocity $\boldsymbol{\omega}$ of the electron. Although U-particle has a constant macro velocity in the Z-axis direction, the macro motion velocity of U-particle in direction of $\boldsymbol{F}_B$ is constantly zero, so there will be no additional momentum change in direction of $\boldsymbol{F}_B$, and $\boldsymbol{F}_B$ can still be approximately calculated by equation (8). Magnitude of $\boldsymbol{F}_B$ is

$$F_B = \frac{4\pi R^3 \rho}{9} * v * \omega$$

The definition of magnetic induction intensity $\boldsymbol{B}$ in classical electromagnetism comes from Lorentz force $\boldsymbol{F}_B = q * \boldsymbol{v} \times \boldsymbol{B}$. The electric charge of an electron is $q = Q_e$, therefore, the magnitude of magnetic induction intensity $\boldsymbol{B}$ is

$$B = \frac{F_B}{qv} = \frac{4\pi R^3 \rho}{9} * v * \omega * \frac{1}{Q_e v} = \frac{4\pi R^3 \rho}{9 Q_e} * \omega$$

According to equation (14)

$$\omega = \frac{(\boldsymbol{\nabla} \times \boldsymbol{v}_U)_\varphi}{2} = \frac{K_U R \lambda v_i}{r}$$

$$B = \frac{2\pi R^3 \rho}{9 Q_e} * (\boldsymbol{\nabla} \times \boldsymbol{v}_U)_\varphi = \frac{4\pi R^3 \rho}{9 Q_e} * \frac{(\boldsymbol{\nabla} \times \boldsymbol{v}_U)_\varphi}{2} = \frac{4\pi R^3 \rho}{9 Q_e} * \frac{K_U R \lambda v_i}{r} = \frac{4\pi R^4 K_U \rho \lambda v_i}{9 Q_e r}$$

The wire with current $I$, $I = \lambda v_i * Q_e$. Magnitude of magnetic induction intensity of the long straight wire is

$$B = \frac{4\pi R^4 K_U \rho}{9 Q_e^2} * \frac{\lambda v_i * Q_e}{r} = \frac{4\pi R^4 K_U \rho}{9 Q_e^2} * \frac{I}{r}$$

In Fig-A19, direction of $\boldsymbol{\nabla} \times \boldsymbol{v}_U$ at point A is perpendicular to the XZ plane inward. Angular velocity of the electron rotation at point A is perpendicular to the XZ plane outward, if the electron at point A is replaced by a proton, angular velocity of the proton rotation is perpendicular to the XZ plane inward, according to equation (8), if the proton moves downward, force direction on the proton is to the left. According to the definition of Lorentz force $\boldsymbol{F}_B = q * \boldsymbol{v} \times \boldsymbol{B}$ in classical electromagnetism, if the proton moves downward and the force direction is to the left, then the direction of B is perpendicular to the XZ plane outward, therefore, the direction of $\boldsymbol{B}$ is opposite to that of $\boldsymbol{\nabla} \times \boldsymbol{v}_U$.

$$\boldsymbol{B} = -\frac{2\pi R^3 \rho}{9 Q_e} * (\boldsymbol{\nabla} \times \boldsymbol{v}_U)$$

The calculation method of magnetic induction intensity $\boldsymbol{B}$ in the above equation refers to the definition of magnetic induction intensity $\boldsymbol{B}$ in classical electromagnetism, but it can define the magnetic induction intensity $\boldsymbol{B}$ at a certain point. Therefore, this paper takes it as the accurate definition of magnetic induction intensity $\boldsymbol{B}$ under the mechanical model of U-particle, which does not need to be based on Lorentz force. This definition is still used when the current changes. The relationship between direction of magnetic induction intensity and direction of current is right-handed helix. Hold the long straight wire with right hand, point four fingers in direction of magnetic induction intensity, and direction of thumb is the direction of the current. The direction of magnetic induction intensity is the same as that of angular velocity of the electron rotation.

In electromagnetics, the curl of magnetic vector potential $\boldsymbol{A}$ is magnetic induction intensity $\boldsymbol{B}$, $\boldsymbol{B} = \boldsymbol{\nabla} \times \boldsymbol{A}$, so

$$\boldsymbol{A} = -\frac{2\pi R^3 \rho}{9 Q_e} * \boldsymbol{v}_U = -\left(\rho \boldsymbol{v}_U * \frac{4\pi R^3}{3}\right) * \frac{1}{6 Q_e} = -\left(\boldsymbol{\rho}_{PU} * \frac{4\pi R^3}{3}\right) * \frac{1}{6 Q_e} = -\boldsymbol{P}_{em} * \frac{1}{6 Q_e}$$

Electrical momentum $\boldsymbol{P}_{em}(r,t)$ at a certain point is equal to macro momentum density $\boldsymbol{\rho}_{PU}$ of Up at the point times the volume of a sphere with radius R, it is similar to the electromagnetic momentum assumed by Maxwell.

U-19: The permittivity and permeability of vacuum are

$$\varepsilon_0 = \frac{9Q_e^2}{8\pi^2 R^4 K_U \rho C_U^2} \qquad (18)$$

$$\mu_0 = \frac{8\pi^2 R^4 K_U \rho}{9Q_e^2} \qquad (19)$$

Explanation: In classical electromagnetism, when electric charge of a charged object is $q$ and distance from the charged object is $r$, electrostatic field intensity is

$$E_s = \frac{1}{4\pi\varepsilon_0} * \frac{q}{r^2}$$

According equation (10), when electric charge of a charged object is $q$ and distance from the charged object is $r$, electrostatic field intensity is

$$E_s = \frac{2\pi R^4 K_U \rho C_U^2}{9Q_e^2} * \frac{q}{r^2}$$

Comparing the two equations, we can see that the permittivity of vacuum is

$$\varepsilon_0 = \frac{9Q_e^2}{8\pi^2 R^4 K_U \rho C_U^2}$$

The mathematical expression of vacuum permittivity can be derived by the mechanical model base on U-particle, and in classical electromagnetism, the vacuum permittivity constant is only a value measured by experiment.

In classical electromagnetism, when current in long straight wire is $I$ and distance from the wire is $r$, magnetic induction intensity is

$$B = \frac{\mu_0}{2\pi} * \frac{I}{r}$$

According equation (17), when current in long straight wire is $I$ and distance from the wire is $r$, magnetic induction intensity is

$$B = \frac{4\pi R^4 K_U \rho}{9Q_e^2} * \frac{I}{r}$$

Comparing the two equations, we can see that the permeability of vacuum is

$$\mu_0 = \frac{8\pi^2 R^4 K_U \rho}{9Q_e^2}$$

U-20: The maximum speed $C_U$ of U-particle translational motion equals the speed of light. Speed of electromagnetism propagation equals the speed of light, and the speed of light $C_U$ is a physical quantity with a constant value. Field constant is $K_U R^4 \rho = 3.68 * 10^{-45}\ Kg.m.$ There is no gravitation between two U-particles, and there is only inertial mass but no gravitational mass in U-particle.

Explanation: Based on the definition of current unit Ampere in classical electromagnetism, the permeability of vacuum can be determined as $\mu_0 = 4\pi * 10^{-7}\ N/A^2$, according equation (18) and (19), the product of $\varepsilon_0$ and $\mu_0$ can be calculated by mechanical model of U-particle as follows

$$\varepsilon_0 * \mu_0 = \frac{9Q_e^2}{8\pi^2 R^4 K_U \rho C_U^2} * \frac{8\pi^2 R^4 K_U \rho}{9Q_e^2} = \frac{1}{C_U^2} \qquad (20)$$

$$C_U = \sqrt{\frac{1}{\varepsilon_0 * \mu_0}} = \sqrt{\frac{1}{8.854 * 10^{-12} * 4\pi * 10^{-7}}} \approx 3.0 * 10^8\ m/s$$

That is to say, $\varepsilon_0 * \mu_0 * C_U^2 = 1$ can be confirmed by using the mathematical expressions of $\varepsilon_0$ and $\mu_0$ derived from mechanical model of U-particle. Using the value of $\mu_0$ determined by classical electromagnetism and the value of $\varepsilon_0$ measured by experiment, the maximum speed of U-particle translational motion can be directly calculated, it is equal to the speed of light.

According to U-7, rotational kinetic energy of U-particle is very small when distance between U-particle and centre of an electron/proton is 1 nm, that is to say, translational motion speed of U-particle is close to $C_U$, so it can be approximately considered that translational motion speed of U-particle is $C_U$. Therefore, the speed of propagation of any physical properties of U-particle, including electric field and magnetic field, is $C_U$. According to equation (19)

$$\mu_0 = \frac{8\pi^2 R^4 K_U \rho}{9Q_e^2}$$

$$K_U R^4 \rho = \frac{9Q_e^2 \mu_0}{8\pi^2} = \frac{9 * (1.602 * 10^{-19})^2 * 4\pi * 10^{-7}}{8\pi^2} \approx 3.68 * 10^{-45}$$

$K_U R^4 \rho$ can be called the field constant

$$K_U R^4 \rho \approx 3.68 * 10^{-45}\ Kg.m$$

An electron is isotropic, but there is no ideal smooth boundary. On the sphere with centre of the electron and radius of R, collision between U-particle and the electron satisfies collision characteristics of mechanical model of U-particle. An electron cannot be regarded as a homogeneous sphere with radius R and the same internal density.

According to equation (4), when distance between two electrons is 1 meter, electrostatic repulsion between them is

$$F_e = \frac{2\pi R^4 K_U \rho C_U^2}{9} * \frac{1}{L^2} = \frac{2\pi * 3.68 * 10^{-45} * 9 * 10^{16}}{9} \approx 2.3 * 10^{-28}\ N$$

According to U-11, component force on half sphere of an electron is

$$hF_e \gg F_e * 7.5 * 10^{31} = 2.3 * 10^{-28} * 7.5 * 10^{31} \approx 1.73 * 10^4\ N$$

To make an analogy, the electron is like a shrimp at the bottom of the sea 8000 meters deep, although the pressure there is huge, but the shrimp only endure differential pressure, so the shrimp can move freely without being crushed to death. The pressure generated by U-particle around us is much bigger than this, is it exciting? Don't worry, U-particle only works on electrons/protons, it has no interest in your huge body ☺. The huge pressure caused by U-particle ensures the stability of structure of an electron/proton.

The physical properties of microscopic particles may seem extreme. For example, in a cathode-ray tube, an acceleration voltage about 2000 volts can accelerate electrons to $3 * 10^7\ m/s$. The mass of an electron is $M_e = 9.10956 * 10^{-31} Kg$, assuming the radius of electron is $R = 1.0 * 10^{-16}\ m$, then the density of electron is

$$\rho_e = \frac{M_e}{4\pi R^3/3} = \frac{3 * 9.10956 * 10^{-31}}{4\pi * (1.0 * 10^{-16})^3} \approx 2.2 * 10^{17} Kg/m^3$$

The speed and density of electron are much higher than that of macroscopic object, so it is normal that the density of U-particle is much higher than that of electron. The density of U-particle in space can be estimated as follows by field constant

$$K_U \rho = \frac{3.68 * 10^{-45}}{(1.0 * 10^{-16})^4} \approx 3.68 * 10^{19}\ Kg/m^3$$

According to U-7, $K_U \ll 1$, so $\rho \gg 3.68 * 10^{19}\ Kg/m^3$. The sum of translational kinetic energy of U-particle in the unit space is $E = \rho C_U^2/2$. If the speed of light $C_U$ accumulates a change rate of $10^{-10}$ within 1 million years, that is, the change rate of the speed of light per second is

$$\frac{dC_U}{dt} = \frac{10^{-10}}{10^6 * 365 * 24 * 3600} \approx 3.17 * 10^{-24}$$

then

$$\frac{dE}{dt} = \rho C_U * \frac{dC_U}{dt} \gg 3.68 * 10^{19} * 3 * 10^8 * 3.17 * 10^{-24} \approx 3500\ W$$

It is impossible for any cubic meter of space to continuously generate 3500 watts of power, which would result in a huge disaster. Therefore, the change rate of the speed of light per second is much smaller than $3.17 * 10^{-24}$, and the speed of light $C_U$ can be approximated as a physical quantity with a constant value.

Although density of U-particle in space is huge, it does not affect movement of normal objects, including movement of our bodies. The reason is that U-particle only works on electron/proton, according to the analysis in U-13, for a stationary electron/proton, if acceleration of Up and Ue in surrounding environment are the same, the direction of the force caused by accelerating Up is opposite to that of the force caused by accelerating Ue, so resultant force of ideal induced forces exerted on the stationary electron/proton is zero. Similarly, if the macro velocities of Up and Ue in surrounding environment are zero, then the uniformly moving electron/proton is subjected to zero resultant force from U-partilce. In addition, although density of U-particle in space is huge, it can be seen from the cause of gravitation that gravitation cannot be formed between two U-particles with inertial mass. Therefore, there is only inertial mass but no gravitational mass in U-particle. According to equation (12), the electrostatic potential is proportional to the rotational kinetic energy density of the U-particles. Similarly, the gravitational potential is proportional to the rotational kinetic energy density of TB. Therefore, for macroscopic objects, gravitational mass and inertial mass are linearly related.

## 5 Approximately stationary point

Suppose that there is an isolated electron, and there is no other electron/proton within infinite distance from this electron, then, the isolated electron takes itself as reference point, the electron is in stationary state defined by U-1.

Suppose that an electrically neutral ball with the centre of O, and there are no other matter in infinite distance from the ball. Taking point O as reference point and taking arbitrary point P in space, length of OP is a finite value, then velocity of macro motion of U-particle at point P is zero, and rotational kinetic energy of U-particle at point P remains unchanged. According to U-1, velocity of macro motion of U-particle at point P is zero, and translational kinetic energy density of U-particle at point P remains unchanged, so point O is a stationary point.

Due to the mutual motion between celestial bodies, the ideal stationary point does not exist, and we can only choose an approximately stationary point. Assuming that radius of ball A is $R_a$ and mass is $m_a$, radius of ball B is $R_b$ and mass is $m_b$, and point P on the surface of ball A is on line AB, $AB = L$. According to U-12 and the superposition principle, rotational kinetic energy of U-particle at point P is

$$E_R \propto \left(\frac{m_a}{R_a^2} + \frac{m_b}{(L - R_a)^2}\right)$$

For example, mass of the Earth is $m_a \approx 5.97 * 10^{24} Kg$ and radius is $R_a \approx 6371 Km$, mass of the Sun is $m_b \approx 1.989 * 10^{30} Kg$, and distance between the Earth and the Sun is $L \approx 1.5 * 10^8 Km$, so

$$\frac{m_a}{R_a^2} : \frac{m_b}{(L - R_a)^2} \approx \frac{5.97 * 10^{24}}{6371^2} * \frac{(1.5 * 10^8 - 6371)^2}{1.989 * 10^{30}} \approx 1664:1$$

On the surface of the Earth, rotational kinetic energy of U-particle is mainly influenced by the Earth, with little influence from the Sun. Therefore, in most cases, any point on the earth can be regarded as an approximately stationary point to calculate U-particle.

## 6 Changing electric field and changing magnetic field

In the process of calculating ideal induced force, angular velocity of the stationary electron/proton is required to be zero, which is impossible in actual physical environment, because the curl of U-particles generated by moving electrons in wire is not zero, this will cause angular velocity of the stationary electron/proton to be non-zero. In the process of calculating ideal Lorentz force, the states of U-particle around the moving electron/proton are required to be the same, which is impossible in actual physical environment, because angular velocity of the electron/proton can be generated only when $\boldsymbol{v}_U(r,t)$ varies with the parameter $r$.

U-21: Infinite long straight wire, the current is zero before $t = 0$, and from $t = 0$, λ electrons per unit length move in the wire in the Z-axis direction with constant acceleration $a$, speed of electrons in the wire is $v_i(t) = at$ at the moment of $t$. Suppose that the distance between point A and the wire is $r$, if $r < C_U t$, then the curl of the macro velocity $\boldsymbol{v}_U(r,t)$ of Up at point A is

$$\boldsymbol{\nabla} \times \boldsymbol{v}_U(r,t) = \frac{2K_U R\lambda a\sqrt{C_U^2 t^2 - r^2}}{C_U r} * \boldsymbol{e}_\varphi \qquad (21)$$

An electron/proton will rotate on its axis at point A. A proton rotates in the opposite direction to an electron at point A. The relationship between direction of electron angular velocity at point A and direction of current is right-handed helix. A proton rotates in the opposite direction to the electron at point A.

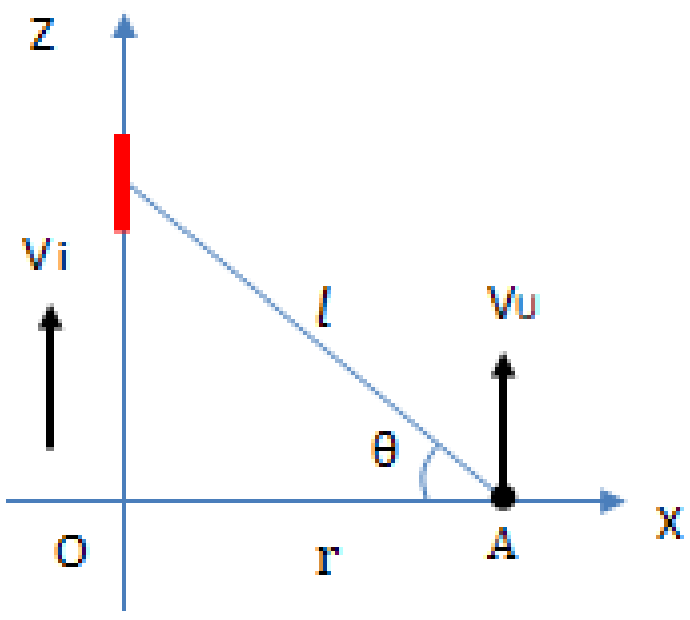

**Fig-A20** Curl generated by accelerating U-particles

Explanation: As shown in Fig-A20, the infinite long straight wire overlaps with Z-axis. Because the distance $l$ between each segment of wire $dz$ and point A is different, according to U-20, after a time interval of $l/C_U$, movement of electrons in the wire $dz$ can affect motion of U-particle on point A, therefore, motion of Up on point A at the moment of $t$ is caused by movement of electrons in the wire $dz$ before the moment of $t$. While $t > l/C_U$, according to equation (3), the movement of a single electron in $dz$ at the moment of $(t - l/C_U)$ causes that the macro velocity of Up at point A at the moment of $t$ is

$$v_{Ue}(r,t) = \frac{K_U R * v_i(t - l/C_U)}{l} = \frac{K_U R[\,v_i(t) - a * l/C_U\,]}{l} = \frac{K_U R v_i(t)}{l} - \frac{K_U R a}{C_U}$$

$$= \frac{K_U R a t}{\sqrt{z^2 + r^2}} - \frac{K_U R a}{C_U}$$

There are $\lambda dz$ moving electrons in the $dz$ length. The macro speed of Up at point A $v_U(r,t)$ can be calculated by integrating. The integration space must ensure that $l < C_U t$, that is, $|z| < \sqrt{C_U^2 t^2 - r^2}$, otherwise influence of the motion of electrons on $dz$ wire has not yet propagated to point A. So

$$v_U(r,t) = \int_{-\sqrt{C_U^2 t^2 - r^2}}^{\sqrt{C_U^2 t^2 - r^2}} \lambda dz * v_{Ue}(r,t) = \int_{-\sqrt{C_U^2 t^2 - r^2}}^{\sqrt{C_U^2 t^2 - r^2}} \left( \frac{K_U R \lambda a t}{\sqrt{z^2 + r^2}} - \frac{K_U R \lambda a}{C_U} \right) dz$$

The integration result is

$$v_U(r,t) = \frac{-2K_U R \lambda a}{C_U} \left( \sqrt{C_U^2 t^2 - r^2} + C_U t \ln \frac{C_U t - \sqrt{C_U^2 t^2 - r^2}}{r} \right) \qquad (22)$$

So

$$\boldsymbol{\nabla} \times \boldsymbol{v}_U(r,t) = \frac{2K_U R \lambda a \sqrt{C_U^2 t^2 - r^2}}{C_U r} * \boldsymbol{e}_\varphi$$

For detailed calculation, please refer to appendix F "Calculation of partial derivatives related to induced electric field and displacement current".

According to U-17, the relationship between direction of electron angular velocity at point A and direction of current is right-handed helix. A proton rotates in the opposite direction to the electron at point A.

U-22: A static electron/proton will be pushed by accelerating U-particles, the pushing force is the induced force, which is proportional to the first derivative of the macro velocity of U-particle with respect to time. Induced electric field intensity $\boldsymbol{E}_i$ is defined as follows. The induced force is the result of induced electric field acting on stationary electron/proton.

$$\boldsymbol{E}_i = \frac{2\pi R^3\rho}{9Q_e} * \frac{\partial \boldsymbol{v}_U}{\partial t} = \frac{1}{6Q_e} * \frac{\partial \boldsymbol{P}_{em}}{\partial t} \qquad (23)$$

Relationship between induced electric field intensity and magnetic induction intensity at a stationary point satisfies

$$\boldsymbol{\nabla} \times \boldsymbol{E}_i = -\frac{\partial \boldsymbol{B}}{\partial t} \qquad (24)$$

$$\boldsymbol{\nabla} \cdot \boldsymbol{E}_i = 0 \qquad (25)$$

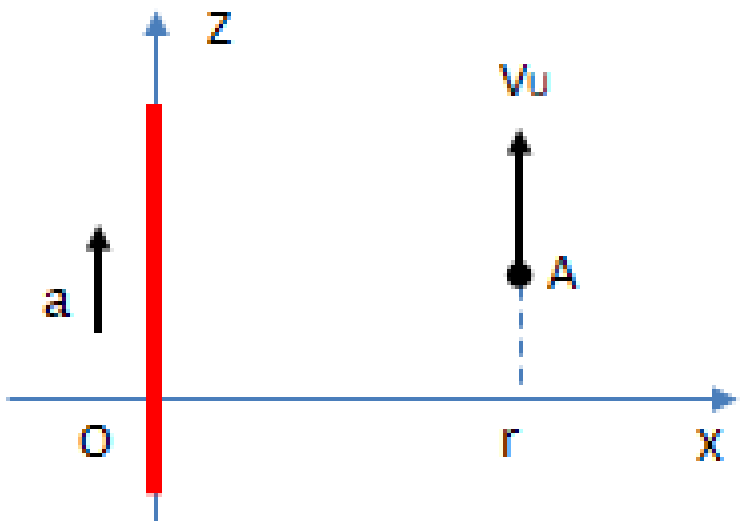

**Fig-A21** Accelerated U-particle generates induced electric field

Explanation: As shown in Fig-A21, the infinite long straight wire overlaps with Z-axis, the current is zero before $t = 0$, and from $t = 0$, $\lambda$ electrons per unit length move in the wire in the Z-axis direction with constant acceleration $a$, speed of electrons in the wire is $v_i(t) = at$ at the moment of $t$. There is a static electron at point A. Distance between point A and the wire is $r$, macro velocity of Up at point A is $\boldsymbol{v}_U(r,t)$, according to equation (21),

$$\boldsymbol{\nabla} \times \boldsymbol{v}_U(r,t) = \frac{2K_U R\lambda a\sqrt{C_U^2 t^2 - r^2}}{C_U r} * \boldsymbol{e}_\varphi$$

According to equation (15)

$$\boldsymbol{B}(r,t) = -\frac{2\pi R^3\rho}{9Q_e} * \boldsymbol{\nabla} \times \boldsymbol{v}_U(r,t) = -\frac{2\pi R^3\rho}{9Q_e} * \frac{2K_U R\lambda a\sqrt{C_U^2 t^2 - r^2}}{C_U r} * \boldsymbol{e}_\varphi$$

$$= -\frac{4\pi R^4 K_U \rho\lambda a\sqrt{C_U^2 t^2 - r^2}}{9Q_e C_U r} * \boldsymbol{e}_\varphi$$

$$\frac{\partial \boldsymbol{B}(r,t)}{\partial t} = -\frac{4\pi R^4 K_U \rho\lambda a}{9Q_e r} * \frac{C_U t}{\sqrt{C_U^2 t^2 - r^2}} * \boldsymbol{e}_\varphi$$

In the process of calculating ideal induced force, angular velocity $\omega$ of the electron rotating on its axis is required to be zero, but $\omega$ is not zero here. According to equation (7)

$$\boldsymbol{F}_i = -\frac{2\pi R^3\rho}{9} * \frac{d\boldsymbol{v}_U(t)}{dt}$$

The above formula requires that macro velocity v of Up is the same in the space occupied by the electron, that is, $\boldsymbol{v}_U(r,t)$ will not change due to the change of variable $r$. Because the space occupied by the electron is very small, the force on the electron can be approximately calculated by $\boldsymbol{v}_U(r,t)$ at the center of the electron. Let

$$\boldsymbol{F}_i(r,t) = -\frac{2\pi R^3\rho}{9} * \frac{\partial \boldsymbol{v}_U(r,t)}{\partial t}$$

Induced electric field intensity $\boldsymbol{E}_i(r,t) = \boldsymbol{F}_i(r,t)/(-Q_e)$, so

$$\boldsymbol{E}_i(r,t) = \frac{2\pi R^3\rho}{9Q_e} * \frac{\partial \boldsymbol{v}_U(r,t)}{\partial t} = \frac{1}{6Q_e} * \frac{4\pi R^3\rho}{3} * \frac{\partial \boldsymbol{v}_U(r,t)}{\partial t} = \frac{1}{6Q_e} * \frac{\partial \boldsymbol{P}_{em}(r,t)}{\partial t}$$

The calculation method of induced electric field intensity $\boldsymbol{E}_i$ in the above equation refers to the definition of induced electric field intensity $\boldsymbol{E}_i$ in classical electromagnetism, but it can define the induced electric field intensity $\boldsymbol{E}_i$ at a certain point. Therefore, this paper takes it as the accurate definition of induced electric field intensity $\boldsymbol{E}_i$ under the mechanical model of U-particle. The partial derivative of equation (22) with respect to time $t$ is

$$\frac{\partial v_U(r,t)}{\partial t}=-2K_U R\lambda a\, ln\frac{C_U t-\sqrt{C_U^2 t^2-r^2}}{r}$$

For detailed calculation, please refer to appendix F "Calculation of partial derivatives related to induced electric field and displacement current". So

$$\boldsymbol{E}_i(r,t)=\frac{2\pi R^3\rho}{9Q_e}*\frac{\partial \boldsymbol{v}_U(r,t)}{\partial t}=-\frac{4\pi R^4 K_U\rho\lambda a}{9Q_e}ln\frac{C_U t-\sqrt{C_U^2 t^2-r^2}}{r}*\boldsymbol{e}_z \tag{26}$$

Since $\boldsymbol{E}_i(r,t)$ has only a component in $\boldsymbol{e}_z$ direction, in cylindrical coordinate system, $\boldsymbol{\nabla}\times\boldsymbol{E}_i(r,t)$ has only a component in direction of $\boldsymbol{e}_\varphi$, so

$$\boldsymbol{\nabla}\times\boldsymbol{E}_i(r,t)=-\frac{\partial E_i(r,t)}{\partial r}*\boldsymbol{e}_\varphi=\frac{4\pi R^4 K_U\rho\lambda a}{9Q_e}*\frac{C_U t}{r\sqrt{C_U^2 t^2-r^2}}*\boldsymbol{e}_\varphi$$

For detailed calculation, please refer to appendix F "Calculation of partial derivatives related to induced electric field and displacement current". Compare with

$$\frac{\partial \boldsymbol{B}(r,t)}{\partial t}=-\frac{4\pi R^4 K_U\rho\lambda a}{9Q_e r}*\frac{C_U t}{\sqrt{C_U^2 t^2-r^2}}*\boldsymbol{e}_\varphi$$

So

$$\boldsymbol{\nabla}\times\boldsymbol{E}_i=-\frac{\partial \boldsymbol{B}}{\partial t}$$

$$\boldsymbol{\nabla}\cdot\boldsymbol{E}_i(r,t)=\frac{\partial E_i(r,t)}{\partial z}=\frac{\partial}{\partial z}\left(-\frac{4\pi R^4 K_U\rho\lambda a}{9Q_e}ln\frac{C_U t-\sqrt{C_U^2 t^2-r^2}}{r}\right)=0$$

**Fig-A22** Induced electric field and direction of force on electron/proton

Fig-A22 shows a circle with infinite radius, and the circle in the middle represents an infinite long straight wire, horizontal line intersects with three circles at P1, P2 and P3 respectively. Compared with Fig-A21, current at point P2 is downward, according to analysis of Fig-A15, the proton at point P1 or P3 moves upward, according to analysis of Fig-A16, the electron at point P1 or P3 moves downward, that is, induced electric field blocks the growth of current in the middle circle. The negative sign of $\boldsymbol{\nabla}\times\boldsymbol{E}_i=-\partial\boldsymbol{B}/\partial t$ coincides with this "blocking" effect.

When current of infinite long straight wire increases, according to Lenz's law of classical electromagnetism, due to the increase of magnetic flux through the middle circle, direction of magnetic induction intensity generated by induced electric field is perpendicular to paper inward, therefore, direction of current generated by induced electric field is opposite to that of the infinite long straight wire, this is consistent with mechanical model of U-particle.

Induced electric field is generated by the accelerated motion of U-particle, not by the change of velocity curl of U-particle. Both induced electric field and changing magnetic field are generated by moving electrons; there is a concomitant relationship between induced electric field and changing magnetic field not a causal relationship. According to classical electromagnetism, electric field and magnetic field are in same phase in electromagnetic wave, which is manifestation of non causal relationship but concomitant relationship between changing electric field and changing magnetic field.

Mathematically, $\boldsymbol{\nabla} \times \boldsymbol{E}_i = -\partial \boldsymbol{B}/\partial t$ means that although the derivation order of electrical momentum $P_{em}(r,t)$ for time and space is different, the two second-order partial derivatives are equal, that is

$$\frac{\partial}{\partial r}\left(\frac{\partial P_{em}(r,t)}{\partial t}\right) = \frac{\partial}{\partial t}\left(\frac{\partial P_{em}(r,t)}{\partial r}\right)$$

According to equation (23)

$$\boldsymbol{E}_i(r,t) = \frac{1}{6Q_e} * \frac{\partial \boldsymbol{P}_{em}(r,t)}{\partial t}$$

So

$$\boldsymbol{\nabla} \times \boldsymbol{E}_i = -\frac{\partial E_i(r,t)}{\partial r} * \boldsymbol{e}_\varphi = -\frac{\partial}{\partial r}\left(\frac{1}{6Q_e} * \frac{\partial P_{em}(r,t)}{\partial t}\right) * \boldsymbol{e}_\varphi = -\frac{1}{6Q_e} * \frac{\partial}{\partial r}\left(\frac{\partial P_{em}(r,t)}{\partial t}\right) * \boldsymbol{e}_\varphi$$

According to equation (16)

$$\boldsymbol{A} = -\frac{\boldsymbol{P}_{em}}{6Q_e}$$

So

$$\boldsymbol{B} = \boldsymbol{\nabla} \times \boldsymbol{A} = \frac{1}{6Q_e} * \frac{\partial P_{em}(r,t)}{\partial r} * \boldsymbol{e}_\varphi$$

$$\frac{\partial \boldsymbol{B}}{\partial t} = \frac{1}{6Q_e} * \frac{\partial}{\partial t}\left(\frac{\partial P_{em}(r,t)}{\partial r}\right) * \boldsymbol{e}_\varphi$$

U-23: Electrons move in a wire at a constant speed before $t = 0$, after $t > 0$, electrons move with constant acceleration. Relationship between electric field intensity $\boldsymbol{E}$, magnetic induction intensity $\boldsymbol{B}$ and conduction current density $\boldsymbol{J}_c$ satisfies

$$\boldsymbol{\nabla} \times \boldsymbol{B} = \mu_0\left(\boldsymbol{J}_c + \varepsilon_0 * \frac{\partial \boldsymbol{E}}{\partial t}\right) = \mu_0\left(\boldsymbol{J}_c + \frac{Q_e}{4\pi R K_U C_U^2} * \frac{\partial^2 \boldsymbol{v}_U}{\partial t^2}\right) \quad (27)$$

Displacement current density $\boldsymbol{J}_d$ is defined as follows

$$\boldsymbol{J}_d(r,t) = \varepsilon_0 * \frac{\partial \boldsymbol{E}_i(r,t)}{\partial t} = \frac{Q_e}{4\pi R K_U C_U^2} * \frac{\partial^2 \boldsymbol{v}_U(r,t)}{\partial t^2} \quad (28)$$

It is proportional to the first derivative of induced electric field with respect to time, and also proportional to the second derivative of macro velocity of U-particle with respect to time.

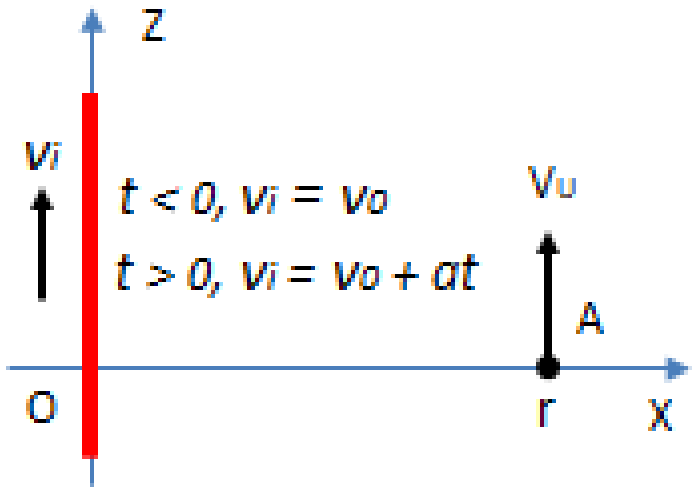

**Fig-A23** Changing conduction current generates displacement current

Explanation: As shown in Fig-A23, the infinite long straight wire overlaps with Z-axis, λ electrons per unit length move upward in the wire at a constant speed of $v_0$ before $t = 0$, after $t=0$, electrons move upward with constant acceleration $a$, speed of electrons moving in the wire is $v_i(t)=v_0+at$. Distance between point A and the wire is $r$. Suppose that the wire passes through surface $\Sigma$ bounded by closed curve $\Gamma$. The current $I(t)$ is divided into constant current $I_1(t)$ and variable current $I_2(t)$, $I_1(t)=-\lambda Q_e v_0$, $I_2(t)=-\lambda Q_e at$, $I(t)=I_1(t)+I_2(t)=-\lambda Q_e v_i(t)$. The constant current and the variable current are calculated respectively.

For the constant current part, according to equation (17)

$$\boldsymbol{B}_1=-\frac{4\pi R^4\rho K_U\lambda v_0}{9Q_e r}*\boldsymbol{e}_\varphi$$

$$\oint_\Gamma \boldsymbol{B}_1(r,t)\cdot d\boldsymbol{l}=\oint_\Gamma -\frac{4\pi R^4\rho K_U\lambda v_0}{9Q_e r}*rd\varphi=-\frac{8\pi^2R^4K_U\rho\lambda v_0}{9Q_e}=\frac{8\pi^2R^4K_U\rho I_1(t)}{9Q_e^2}=\mu_0 I_1(t)$$

According to Stokes mathematical theorem

$$\iint_\Sigma [\boldsymbol{\nabla}\times\boldsymbol{B}_1(r,t)]\cdot d\boldsymbol{S}=\oint_\Gamma \boldsymbol{B}_1(r,t)\cdot d\boldsymbol{l}$$

Divide both sides of the above equation by area $\Sigma$ and take the limit that $\Sigma$ tends to zero, when $\Sigma$ tends to zero, $r$ also tends to zero

$$\left(\boldsymbol{\nabla}\times\boldsymbol{B}_1(0,t)\right)_{\boldsymbol{z}}=\lim_{\Sigma\to 0}\frac{1}{\Sigma}*\oint_\Gamma \boldsymbol{B}_1(r,t)\cdot d\boldsymbol{l}=\lim_{\Sigma\to 0}\frac{1}{\Sigma}*\mu_0 I_1(t)$$

While $r>0$

$$\boldsymbol{\nabla}\times\boldsymbol{B}_1(r,t)=\frac{1}{r}*\frac{\partial}{\partial r}\left[r*\left(-\frac{4\pi R^4K_U\rho\lambda v_0}{9Q_e r}\right)\right]*\boldsymbol{e}_z=0$$

For the variable current part, according to U-22

$$\boldsymbol{B}_2(r,t)=-\frac{4\pi R^4K_U\rho\lambda a\sqrt{C_U^2t^2-r^2}}{9Q_eC_Ur}*\boldsymbol{e}_\varphi$$

$$\oint_\Gamma \boldsymbol{B}_2(r,t)\cdot d\boldsymbol{l}=\oint_\Gamma -\frac{4\pi R^4K_U\rho\lambda a\sqrt{C_U^2t^2-r^2}}{9Q_eC_Ur}*rd\varphi=-\frac{4\pi R^4K_U\rho\lambda at}{9Q_e}\oint_\Gamma\sqrt{1-\frac{r^2}{C_U^2t^2}}d\varphi$$

According to Stokes mathematical theorem

$$\iint_\Sigma [\boldsymbol{\nabla}\times\boldsymbol{B}_2(r,t)]\cdot d\boldsymbol{S}=\oint_\Gamma \boldsymbol{B}_2(r,t)\cdot d\boldsymbol{l}$$

Divide both sides of the above equation by $\Sigma$ and take the limit that $\Sigma$ tends to zero, when $\Sigma$ tends to zero, $r$ also tends to zero

$$\begin{aligned}\left(\boldsymbol{\nabla}\times\boldsymbol{B}_2(0,t)\right)_{\boldsymbol{z}}&=\lim_{\Sigma\to 0}\frac{1}{\Sigma}*\oint_\Gamma \boldsymbol{B}_2(r,t)\cdot d\boldsymbol{l}=\lim_{\Sigma\to 0}\frac{1}{\Sigma}*\left(-\frac{4\pi R^4K_U\rho\lambda at}{9Q_e}\oint_\Gamma d\varphi\right)\\&=\lim_{\Sigma\to 0}\frac{1}{\Sigma}*\left(-\frac{8\pi^2R^4K_U\rho\lambda at}{9Q_e}\right)=\lim_{\Sigma\to 0}\frac{1}{\Sigma}*\left(\frac{8\pi^2R^4K_U\rho I_2(t)}{9Q_e^2}\right)=\lim_{\Sigma\to 0}\frac{1}{\Sigma}*\mu_0 I_2(t)\end{aligned}$$

While $r>0$

$$\boldsymbol{\nabla}\times\boldsymbol{B}_2(r,t)=\frac{1}{r}*\frac{\partial}{\partial r}\left[r*\left(-\frac{4\pi R^4K_U\rho\lambda a\sqrt{C_U^2t^2-r^2}}{9Q_eC_Ur}\right)\right]*\boldsymbol{e}_z=\frac{4\pi R^4K_U\rho\lambda a}{9Q_eC_U\sqrt{C_U^2t^2-r^2}}*\boldsymbol{e}_z$$

According to equation (26)

$$\boldsymbol{E}_i(r,t)=\frac{2\pi R^3\rho}{9Q_e}*\frac{\partial\boldsymbol{v}_U(r,t)}{\partial t}=-\frac{4\pi R^4K_U\rho\lambda a}{9Q_e}ln\frac{C_Ut-\sqrt{C_U^2t^2-r^2}}{r}*\boldsymbol{e}_z$$

So

$$\frac{\partial \boldsymbol{E}_i(r,t)}{\partial t}=\frac{2\pi R^3\rho}{9Q_e}*\frac{\partial^2 \boldsymbol{v}_U(r,t)}{\partial t^2}=\frac{4\pi R^4 K_U\rho\lambda a}{9Q_e}*\frac{C_U}{\sqrt{C_U^2t^2-r^2}}*\boldsymbol{e}_z$$

For detailed calculation, please refer to appendix F "Calculation of partial derivatives related to induced electric field and displacement current". Using the superposition principle, the electromagnetic effects of the constant current $I_1(t)$ and the variable current $I_2(t)$ are added. The constant current $I_1(t)$ does not generate induced electric field, and the change rate of electrostatic field to time is zero, so

$$\frac{\partial \boldsymbol{E}(r,t)}{\partial t}=\frac{\partial \boldsymbol{E}_s(r,t)}{\partial t}+\frac{\partial \boldsymbol{E}_i(r,t)}{\partial t}=\frac{\partial \boldsymbol{E}_i(r,t)}{\partial t}=\frac{4\pi R^4 K_U\rho\lambda a}{9Q_e}*\frac{C_U}{\sqrt{C_U^2t^2-r^2}}*\boldsymbol{e}_z$$

While $r=0$

$$\boldsymbol{\nabla}\times\boldsymbol{B}(0,t)=\boldsymbol{\nabla}\times\boldsymbol{B}_1(0,t)+\boldsymbol{\nabla}\times\boldsymbol{B}_2(0,t)=\lim_{\Sigma\to 0}\frac{1}{\Sigma}*\mu_0 I_1(t)*\boldsymbol{e}_z+\lim_{\Sigma\to 0}\frac{1}{\Sigma}*\mu_0 I_2(t)*\boldsymbol{e}_z$$

$$=\lim_{\Sigma\to 0}\frac{1}{\Sigma}*\mu_0 I(t)*\boldsymbol{e}_z=\mu_0\boldsymbol{J}_c(t)$$

While $0<r<C_Ut$

$$\boldsymbol{\nabla}\times\boldsymbol{B}(r,t)=\boldsymbol{\nabla}\times\boldsymbol{B}_1(r,t)+\boldsymbol{\nabla}\times\boldsymbol{B}_2(r,t)=\frac{4\pi R^4 K_U\rho\lambda a}{9Q_e C_U\sqrt{C_U^2t^2-r^2}}*\boldsymbol{e}_z$$

So

$$\boldsymbol{\nabla}\times\boldsymbol{B}(r,t)=\frac{1}{C_U^2}*\frac{\partial \boldsymbol{E}(r,t)}{\partial t}$$

Therefore, while $0\le r<C_Ut$

$$\boldsymbol{\nabla}\times\boldsymbol{B}(r,t)=\mu_0\boldsymbol{J}_c(t)+\frac{1}{C_U^2}*\frac{\partial \boldsymbol{E}(r,t)}{\partial t}=\mu_0\left(\boldsymbol{J}_c(t)+\varepsilon_0*\frac{\partial \boldsymbol{E}(r,t)}{\partial t}\right)$$

$$=\mu_0\left(\boldsymbol{J}_c(t)+\varepsilon_0*\frac{\partial \boldsymbol{E}_i(r,t)}{\partial t}\right)$$

$$=\mu_0\left(\boldsymbol{J}_c(t)+\frac{9Q_e^2}{8\pi^2R^4K_U\rho C_U^2}*\frac{2\pi R^3\rho}{9Q_e}*\frac{\partial^2 \boldsymbol{v}_U(r,t)}{\partial t^2}\right)$$

$$=\mu_0\left(\boldsymbol{J}_c(t)+\frac{Q_e}{4\pi RK_UC_U^2}*\frac{\partial^2 \boldsymbol{v}_U(r,t)}{\partial t^2}\right)$$

Many teaching materials use charge-discharge model of parallel plate capacitor to illustrate displacement current, in fact, the key factor causing displacement current is the second derivative of the macro velocity of U-particle with respect to time. During the charging and discharging process of the parallel plate capacitor, moving speed of electron gradually decreases to zero after it reaching the parallel plate. The change of moving speed of electron leads to the change of macro speed of U-particle. Displacement current density $\boldsymbol{J}_d$ is

$$\boldsymbol{J}_d(r,t)=\varepsilon_0*\frac{\partial \boldsymbol{E}_i(r,t)}{\partial t}=\frac{Q_e}{4\pi RK_UC_U^2}*\frac{\partial^2 \boldsymbol{v}_U(r,t)}{\partial t^2}$$

The calculation method of displacement current density $\boldsymbol{J}_d$ in the above equation refers to the definition of displacement current density $\boldsymbol{J}_d$ in classical electromagnetism, therefore, this paper takes it as the accurate definition of displacement current density $\boldsymbol{J}_d$ under the mechanical model of U-particle. $\boldsymbol{J}_d$ is proportional to the first-order derivative of induced electric field intensity with respect to time, or $\boldsymbol{J}_d$ is proportional to the second-order derivative of macro velocity of U-particle with respect to time.

U-24: According to mechanical model of U-particle, Maxwell's equations of electromagnetic field can be derived as follows

$$\begin{cases} \boldsymbol{\nabla}\cdot\boldsymbol{E}=\dfrac{\rho_q}{\varepsilon_0} \\ \boldsymbol{\nabla}\cdot\boldsymbol{B}=0 \\ \boldsymbol{\nabla}\times\boldsymbol{E}=-\dfrac{\partial\boldsymbol{B}}{\partial t} \\ \boldsymbol{\nabla}\times\boldsymbol{B}=\mu_0\left(\boldsymbol{J}_c+\varepsilon_0*\dfrac{\partial\boldsymbol{E}}{\partial t}\right) \end{cases} \tag{29}$$

$\rho_q$ is volume density of electric charge.

Explanation: It is completely consistent with the results of classical electromagnetism. Most of its derivation process has been in previous chapters, which is summarized as follows. According to equation (10)

$$E_s=\frac{2\pi R^4K_U\rho C_U^2}{9Q_e^2}*\frac{q}{r^2}$$

According to equation (18)

$$\varepsilon_0=\frac{9Q_e^2}{8\pi^2R^4K_U\rho C_U^2}$$

Therefore, for the electrostatic field

$$E_s=\frac{1}{4\pi\varepsilon_0}*\frac{q}{r^2}$$

Suppose that electric charge $q$ is surrounded by a closed surface $\Sigma$, volume inside the curved surface $\Sigma$ is $\Omega$, then

$$\oiint_{\Sigma}\boldsymbol{E}_s\cdot d\boldsymbol{S}=\oiint_{\Sigma}\frac{1}{4\pi\varepsilon_0}*\frac{q}{r^2}*r^2\sin\theta\, d\theta d\varphi=\frac{q}{\varepsilon_0}$$

According to Gauss mathematical theorem

$$\iiint_{\Omega}(\boldsymbol{\nabla}\cdot\boldsymbol{E}_s)*d\Omega=\oiint_{\Sigma}\boldsymbol{E}_s\cdot d\boldsymbol{S}$$

According to equation (25), the divergence of induced electric field intensity is $\boldsymbol{\nabla}\cdot\boldsymbol{E}_i=0$. Total electric field intensity is $\boldsymbol{E}=\boldsymbol{E}_s+\boldsymbol{E}_i$, and $\boldsymbol{\nabla}\cdot\boldsymbol{E}=\boldsymbol{\nabla}\cdot\boldsymbol{E}_s+\boldsymbol{\nabla}\cdot\boldsymbol{E}_i=\boldsymbol{\nabla}\cdot\boldsymbol{E}_s$, so

$$\iiint_{\Omega}(\boldsymbol{\nabla}\cdot\boldsymbol{E})*d\Omega=\iiint_{\Omega}(\boldsymbol{\nabla}\cdot\boldsymbol{E}_s)*d\Omega=\oiint_{\Sigma}\boldsymbol{E}_s\cdot d\boldsymbol{S}=\frac{q}{\varepsilon_0}$$

Divide both sides of the above equation by $\Omega$ and take the limit that $\Omega$ tends to zero

$$\boldsymbol{\nabla}\cdot\boldsymbol{E}=\lim_{\Omega\to 0}\frac{q}{\Omega}*\frac{1}{\varepsilon_0}=\frac{\rho_q}{\varepsilon_0}$$

According to equation (15)

$$\boldsymbol{B}=-\frac{2\pi R^3\rho}{9Q_e}*(\boldsymbol{\nabla}\times\boldsymbol{v}_U)$$

Since divergence of curl is always zero, hence

$$\boldsymbol{\nabla}\cdot\boldsymbol{B}=0$$

According to equation (11) $\boldsymbol{\nabla}\times\boldsymbol{E}_s=0$, according to equation (24) induced electric field intensity $\boldsymbol{\nabla}\times\boldsymbol{E}_i=-\partial\boldsymbol{B}/\partial t$, since $\boldsymbol{\nabla}\times\boldsymbol{E}=\boldsymbol{\nabla}\times(\boldsymbol{E}_s+\boldsymbol{E}_i)=\boldsymbol{\nabla}\times\boldsymbol{E}_i$, hence $\boldsymbol{\nabla}\times\boldsymbol{E}$ is equal to the negative value of the change rate of magnetic induction intensity, so

$$\boldsymbol{\nabla} \times \boldsymbol{E} = -\frac{\partial \boldsymbol{B}}{\partial t}$$

According to equation (27)

$$\boldsymbol{\nabla} \times \boldsymbol{B} = \mu_0 \left( \boldsymbol{J}_c + \varepsilon_0 * \frac{\partial \boldsymbol{E}}{\partial t} \right)$$

U-25: The force acting on a moving electron in magnetic field is the sum of induced force and Lorentz force.

Explanation: In Fig-A23, if there is an electron at point A moving to the right at speed $\boldsymbol{v}$, that is, $dr/dt = v$, according to equation (8), the Lorentz force acting on the moving electron is

$$\boldsymbol{F}_B = -\frac{4\pi R^3 \rho}{9} * (\boldsymbol{v} \times \boldsymbol{\omega}) = -\frac{4\pi R^3 \rho}{9} * \left( \boldsymbol{v} \times \frac{\boldsymbol{\nabla} \times \boldsymbol{v}_U}{2} \right) = -\frac{2\pi R^3 \rho}{9} * \left[ \frac{\partial \boldsymbol{v}_U(r,t)}{\partial r} * \frac{dr}{dt} \right]$$

According to eqation (23)

$$\boldsymbol{E}_i = \frac{2\pi R^3 \rho}{9 Q_e} * \frac{\partial \boldsymbol{v}_U(r,t)}{\partial t}$$

So, the induced force on the moving electron is

$$\boldsymbol{F}_i = -\frac{2\pi R^3 \rho}{9} * \frac{\partial \boldsymbol{v}_U(r,t)}{\partial t}$$

Since

$$\frac{d\boldsymbol{v}_U(r,t)}{dt} = \frac{\partial \boldsymbol{v}_U(r,t)}{\partial t} + \frac{\partial \boldsymbol{v}_U(r,t)}{\partial r} * \frac{dr}{dt} = -\frac{9}{2\pi R^3 \rho} (\boldsymbol{F}_i + \boldsymbol{F}_B)$$

So, non-electrostatic force acting on the moving electron in magnetic field is the sum of induced force and Lorentz force. In electromagnetism, these two forces are often confused by students. Feynman understood them as two independent forces, and Einstein regarded them as evidence of Special Theory of Relativity. However, in the mechanical model of U-particle, their relationship is so simple.

## 7 Summary

So far, in stationary frame of reference, Maxwell's equations of classical electromagnetism are derived by the mechanical model based on U-particle, physical laws used are conservation of momentum, conservation of kinetic energy and Newton's three laws of motion.

The electron/proton transforms translational kinetic energy of U-particle into rotational kinetic energy and acts as energy converter. This is manifestation of energy conservation, and also source of power in the world.

The electron/proton is exerted three kinds of forces generated by U-particle: (1) electrostatic force is generated by gradient of rotational kinetic energy of U-particle, electrostatic force constitutes electrostatic field; (2) the induced force on static electrons/protons generated by accelerating or decelerating motion of U-particle constitutes induced electric field; (3) Lorentz force is generated by Magnus effect of rotating electron/proton with translational velocity in U-particle environment. In addition, moving electron/proton causes macro velocity of U-particle, and curl of the macro velocity of U-particle constitutes magnetic field. Electrostatic potential is proportional to the rotational kinetic energy of U-particle, and magnetic vector potential is proportional to the macro momentum of U-particle. By replacing potential energy with rotational kinetic energy, the law of conservation of energy is simplified to the law of conservation of kinetic energy, and analytical mechanics can be more perfect without relying on potentials with action at a distance.

Electromagnetic field is physical properties generated by motions of U-particle; it is field in mathematical meaning. The statement that "electromagnetic field is a kind of matter" is not accurate. The reason is the same as that we can't say "dance is a kind of matter". Dance is only beautiful movement displayed by dancers. In order to avoid abusing definition of "matter", any object with zero inertial mass should not be defined as "matter". The statement that "changing electric field generates magnetic field" or "changing magnetic field generates electric field" is not accurate. The change of electric field and magnetic field are the result of electron/proton movement, and there is no causal relationship between them, but concomitant.

Ampere force is macro manifestation of Lorentz force. If two parallel straight wires A and B are applied with same direction current, movement of electrons in wire A causes velocity curl of U-particle at position of wire B, and electrons moving in wire B are pushed towards wire A by Lorentz force, as a result, two straight wires attract each other. As shown in Fig-A24, replace the two straight wires with two coils, number of turns of the two coils on cylinder is 1. When current in same direction is applied, the two coils attract each other, and when reverse current is applied, the two coils repel each other. Therefore, children can intuitively understand attraction or repulsion between magnets. Since clockwise rotation on front of paper turns into anticlockwise rotation on reverse side of the paper, children can intuitively understand why there is no "magnetic monopole".

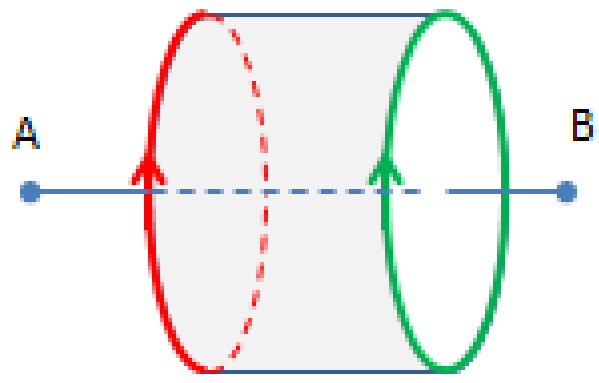

**Fig-A24** Attraction and repulsion of current-carrying coils

Whether it's electrostatic attraction, magnetic attraction, or gravitation, "attraction" is misleading, it mislead people to understand "attraction" as some kind of magic pulling force. It can be seen from this paper that the "attraction" comes from differential pushing-force of U-particle or TB on electrons/protons.

The essence of magnetic field is curl of the macro motion of U-particle, and electrons/protons will rotate on its axis in magnetic field, macro motion of the U-particle can store mechanical energy, this energy is magnetic field energy in classical electromagnetism.

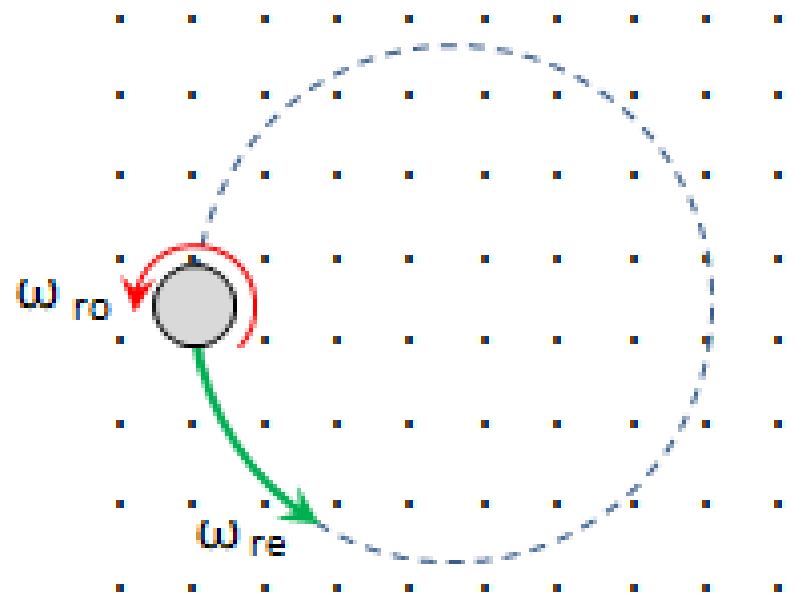

**Fig-A25** Revolution and rotation of an electron

As shown in Fig-A25, an electron makes a uniform circular motion under the action of Lorentz force in a constant magnetic field. If this circular motion is called revolution, the ratio of the angular velocity of the electron's rotation in the magnetic field to the angular velocity of revolution can be calculated. Suppose angular velocity of the electron revolution is $\omega_{re}$, angular velocity of the electron rotation is $\omega_{ro}$, inertial mass of an electron is $M_e$, mass of U-particle contained in a virtual sphere with the same radius as electron is $M_V$, magnetic induction intensity of the constant magnetic field is $B$. Radius of the uniform circular motion of the electron is $r$, period is $T$, and linear speed is $v$. Because centripetal force of an electron moving in a uniform circular motion is equal to Lorentz force,

$$\frac{M_e * v^2}{r} = Q_e v B$$

$$\frac{v}{r} = \frac{Q_e}{M_e} * B$$

$$\omega_{re} = \frac{2\pi}{T} = 2\pi * \frac{v}{2\pi r} = \frac{Q_e}{M_e} * B$$

According to equation (15)

$$B = \frac{2\pi R^3 \rho}{9Q_e} * 2\omega_{ro} = \frac{4\pi R^3 \rho}{9Q_e} * \omega_{ro}$$

$$\omega_{re} = \frac{Q_e}{M_e} * \frac{4\pi R^3 \rho}{9Q_e} * \omega_{ro} = \frac{4\pi R^3 \rho}{3} * \frac{\omega_{ro}}{3M_e} = \frac{M_V \omega_{ro}}{3M_e}$$

So

$$\frac{\omega_{ro}}{\omega_{re}} = \frac{3M_e}{M_V}$$

There is no electron or proton or conduction current in vacuum. Maxwell obtained the wave equation of electromagnetic field through the following mathematical calculation.

$$\begin{cases} \boldsymbol{\nabla} \cdot \boldsymbol{E} = 0 \\ \boldsymbol{\nabla} \cdot \boldsymbol{B} = 0 \\ \boldsymbol{\nabla} \times \boldsymbol{E} = -\dfrac{\partial \boldsymbol{B}}{\partial t} \\ \boldsymbol{\nabla} \times \boldsymbol{B} = \varepsilon_0 \mu_0 * \dfrac{\partial \boldsymbol{E}}{\partial t} \end{cases}$$

Take curl on both sides of the third equation

$$\boldsymbol{\nabla} \times (\boldsymbol{\nabla} \times \boldsymbol{E}) = \boldsymbol{\nabla}(\boldsymbol{\nabla} \cdot \boldsymbol{E}) - \boldsymbol{\nabla}^2 \boldsymbol{E} = -\boldsymbol{\nabla}^2 \boldsymbol{E}$$

$$\boldsymbol{\nabla} \times \left(-\frac{\partial \boldsymbol{B}}{\partial t}\right) = -\frac{\partial}{\partial t}(\boldsymbol{\nabla} \times \boldsymbol{B}) = -\frac{\partial}{\partial t}\left(\varepsilon_0 \mu_0 * \frac{\partial \boldsymbol{E}}{\partial t}\right) = -\varepsilon_0 \mu_0 * \frac{\partial^2 \boldsymbol{E}}{\partial t^2}$$

$$\boldsymbol{\nabla}^2 \boldsymbol{E} = \varepsilon_0 \mu_0 * \frac{\partial^2 \boldsymbol{E}}{\partial t^2}$$

It can also be derived that

$$\boldsymbol{\nabla}^2 \boldsymbol{B} = \varepsilon_0 \mu_0 * \frac{\partial^2 \boldsymbol{B}}{\partial t^2}$$

The wave equation of mechanical wave is

$$\frac{\partial^2 f}{\partial x^2} = \frac{1}{v^2} * \frac{\partial^2 f}{\partial t^2}$$

They are similar, and

$$v = \frac{1}{\sqrt{\varepsilon_0 \mu_0}} = 3.0 * 10^8\ m/s$$

That is to say, the calculated speed of electromagnetic propagation is equal to the measured speed of light, so Maxwell predicted that light is electromagnetic wave. However, mechanical wave has medium and stationary reference point of velocity, but what is the medium of electromagnetic wave? The speed of electromagnetic waves is equal to the speed of light, but what is the stationary reference point? At that time, physicists assumed that "Ether" was the medium and stationary reference point of electromagnetic wave propagation, but it has not been found so far. In this paper, U-particle can be used as the medium for Maxwell's electromagnetic wave propagation, and a new definition of stationary reference point is made in U-1, it avoids the absolute stationary point where the mathematical velocity of all U-particles is zero.

The stationary point defined by U-1 in this paper is an ideal point, which does not exist in reality. However, in many cases, the stationary point can be replaced by an approximately stationary point. It can be seen from U-13 that Up and Ue with the same acceleration cancel the induced force exerted on an electron or a proton, so the motion of an electrically neutral object will not affect the electron/proton. The geomagnetic field can be detected on the earth's surface, which indicates that the macro motion speed of U-particle is $v_U \neq 0$. However, because the geomagnetic field is very weak, it can be regarded as $v_U \approx 0$ for experiments or applications with low accuracy requirements. Although the Moon, the Sun and other celestial bodies will also affect translational kinetic energy density of U-particle on the earth's surface, the earth's influence is the most important. This is similar to the reason why we can feel the gravity of the earth, but can't feel the gravity of the Moon and the Sun. Therefore, we can regard the ground as the approximately stationary point of U-particle motion.

Electromagnetic field is physical properties generated by motions of U-particle, and its changes take U-particle as the medium of transmission, so its propagation speed is exactly the same in the same U-particle environment. Propagation speed of the light emitted by a moving car is the same as that of the light emitted by a stopping car; propagation speed of the honking sound of a moving car is equal to that of a stopping car, and the two reasons are similar, because both lights propagate in the same U-particle environment, and both honking sounds propagate in the same air environment.

In stationary frame of reference, assuming that an object moves in X-axis direction at speed $v$. For this moving object, let the maximum speed of the U-particles in the YZ plane be $V_{yz}$, then $v^2 + V_{yz}^2 = C_U^2$, the maximum value of $V_{yz}$ is reduced from $C_U$ to $\sqrt{C_U^2 - v^2}$, and the ratio is

$$\gamma = \frac{C_U}{\sqrt{C_U^2 - v^2}} = \frac{1}{\sqrt{1 - v^2/C_U^2}}$$

Since the speed of physical processes inside moving objects will slow down, so moving clock will also slow down, and the slowing down ratio is $\gamma$. This is very interesting, the mechanical model of U-particle can help us intuitively and accurately understand the physical meaning of the Lorentz transformation.

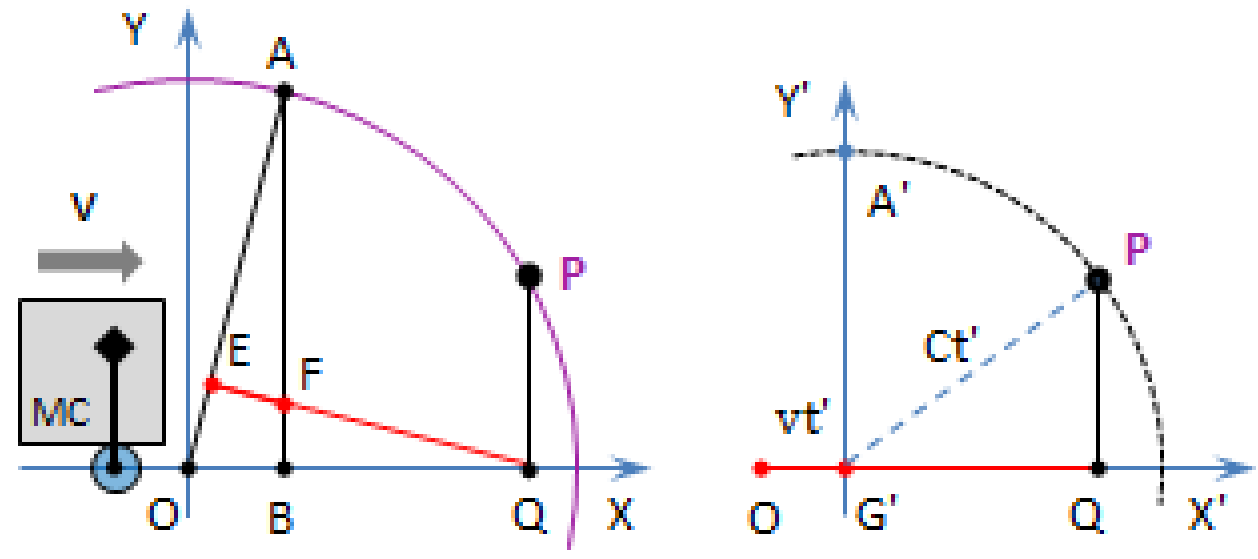

**Fig-A26** The model of Lorentz transformation

In two-dimensional space XOY shown in the left figure of Fig-A26, a car moving at a constant speed $v$ to the right passes through point O at zero time, and a flash is emitted from point O at the same time. At time $t$, the car arrives at point B, wavefront of the flash is a circle with point O as the center and radius $C_U t$. Take a point P on this circle, the projection of point P on X-axis is point Q, and the car arrives at point Q at time $\tau$. Both $t$ and $\tau$ are measured by stationary clocks on the ground, $OB = vt$, $OP = C_U t$, $OQ = v\tau$.

If a tester uses a moving clock MC to measure in the car, the time for MC to move from point O to point Q is $\tau'$. Now we need to find a lucky point G' on X-axis, the time for the moving clock MC to move from point O to point G' is $t'$, and $G'P = C_U t'$. Both $\tau'$ and $t'$ are measured by the moving clock MC. This lucky point G' can easily make the tester standing at the position of clock MC have such an illusion: the ground is moving backwards, the time for point O to leave point G' is $t'$, the time for the flash to travel from point G' to point P is also $t'$, and the speed of light is $C_U$, that is, the speed of light in the moving reference frame is the same as that in the stationary reference frame. According to this wonderful illusion, using the time measured by the moving clock MC, we can use a simple method to correctly calculate the time for light to travel to point P.

Since the moving clock MC slows down, so $\tau' = \tau\sqrt{1 - v^2/C_U^2}$. Suppose that $x = v\tau$ and $x' = v(\tau' - t')$, then $x' + vt' = x\sqrt{1 - v^2/C_U^2}$, that is

$$x = \frac{x' + vt'}{\sqrt{1 - v^2/C_U^2}}$$

This is an equation for the inverse Lorentz transformation. $x$ is the length between OQ calculated using the time $\tau$ measured by stationary clocks, abbreviated as stationary clock length, denoted as $|OQ|_S = v\tau$. $v\tau' = vt' + x'$ is the length between OQ calculated using the time $\tau'$ measured by the moving clock, abbreviated as moving clock length, denoted as $|OQ|_M = v\tau'$. Therefore $|OQ|_M = |OQ|_S\sqrt{1 - v^2/C_U^2}$.

In the left figure of Fig-A26, draw a straight line AB perpendicular to X-axis through point B, AB intersects with the circle at point A. Connect OA and draw a perpendicular line QE perpendicular to OA through point Q, with the foot perpendicular to E, QE and OB intersect at point F. It can be proved that $AF^2 - FQ^2 - QP^2 = 0$. Triangle AEF is similar to QEO, so $QE * EF = EO * AE$.

$$\begin{aligned} AF^2 - FQ^2 - QP^2 &= (AE^2 + EF^2) - FQ^2 - (PO^2 - OQ^2) \\ &= (AO - EO)^2 + (EQ - FQ)^2 - FQ^2 - AO^2 + OQ^2 \\ &= EO^2 - 2EO * AO + EQ^2 - 2EQ * FQ + OQ^2 \\ &= 2(EO^2 - EO * AO + EQ^2 - EQ * FQ) = 2(EQ * EF - OE * AE) = 0 \end{aligned}$$

Since $OP = C_U t$, $OB = vt$, so $AB = \sqrt{OA^2 - OB^2} = Ct\sqrt{1 - v^2/C_U^2}$. Triangle ABO is similar

to QEO, so $EQ = OQ * BA/OA = OQ\sqrt{1 - v^2/C_U^2}$, that is $EQ = |OQ|_M$. Previously calculated $|OQ|_M = v\tau' = vt' + x'$, so we can let $EF = vt'$ and $FQ = x'$. Triangle AEF is similar to ABO, so $AF = EF * OA/OB = vt' * C_U t/vt = C_U t'$.

The right figure of Fig-A26 is the X'O'Y' coordinate system constructed on the basis of the left figure. Take X'-axis to coincide with X-axis, and point P remains unchanged, so the projection of point P on the X'-axis is still point Q. Take $G'Q = FQ = x'$ to get point G', then take $OG' = EF = vt'$ to get point O, make Y'-axis through point G' and regard point G' as point O'. Take $A'G' = AF = C_U t'$ on the Y'-axis, and make a circle with the point G' as the circle point and $C_U t'$ as the radius. Since it has been proved that $AF^2 - FQ^2 - QP^2 = 0$, so $A'G'^2 - G'Q^2 - QP^2 = 0$, that is $C_U^2 t'^2 - x'^2 = QP^2$, it means that the point P is exactly located on the circle with point G' as the center and the radius $C_U t'$, that is, $G'P = C_U t'$.

According to the left figure of Fig-A26, $QP^2 = OP^2 - OQ^2$, that is, $QP^2 = C_U^2 t^2 - x^2$, so we get $C_U^2 t^2 - x^2 = C_U^2 t'^2 - x'^2$. Substituting the previously obtained equation for the inverse Lorentz transformation into this equation, we can get

$$C_U^2 t^2 - \frac{(x' + vt')^2}{1 - v^2/C_U^2} = C_U^2 t'^2 - x'^2$$

$$\begin{aligned}
C_U^2 t^2 - v^2 t^2 &= (C_U^2 t'^2 - x'^2)(1 - v^2/C_U^2) + (x' + vt')^2 \\
&= C_U^2 t'^2 - x'^2 - v^2 t'^2 + \frac{v^2 x'^2}{C_U^2} + x'^2 + 2vt'x' + v^2 t'^2 \\
&= C_U^2 t'^2 + \frac{v^2 x'^2}{C_U^2} + 2vt'x' = (C_U t' + vx'/C_U)^2
\end{aligned}$$

So

$$t = \frac{C_U t' + vx'/C_U}{\sqrt{C_U^2 - v^2}} = \frac{t' + vx'/C_U^2}{\sqrt{1 - v^2/C_U^2}}$$

Since the coordinates of point P on the Y-axis remain unchanged, so $y' = y$. Therefore, the inverse Lorentz transformation in two-dimensional space is

$$\begin{cases} x = \dfrac{x' + vt'}{\sqrt{1 - v^2/C_U^2}} \\ y = y' \\ t = \dfrac{t' + vx'/C_U^2}{\sqrt{1 - v^2/C_U^2}} \end{cases}$$

By the same reasoning, in regions of high electric potential, the translational kinetic energy of U-particles is smaller, and therefore clocks also slow down. If an electron/proton suddenly disintegrates for some reason, gradient of rotational kinetic energy of U-particle $\boldsymbol{\nabla} E_R(r)$ in the space around the electron/proton will suddenly change, which will lead to violent conversion between rotational kinetic energy and translational kinetic energy of U-particle. The phenomenon of an electron disintegration is disappearance of measurable inertial mass, and the phenomenon of violent conversion between rotational kinetic energy and translational kinetic energy of U-particle is output of huge energy. This may help to explain the relationship between mass and energy.

In addition, the mechanical analysis and calculation in this paper are limited to the case of stationary frame of reference and low-speed movement of electrons/protons.

## 8 Prediction

As shown in Fig-A27, cathode ray passes from left to right through an odd symmetrical magnetic field generated by two identical coils in reverse series connection, if electrons have velocity of up and down, a line pattern appears on the right phosphor screen. Fix the position of coils and move cathode-ray tube from left to right, the line pattern on the screen rotates as shown in Fig-A28. When the screen is at the **point O** in Fig-A27, inclination angle of tangent of the line pattern is maximum, which can be predicted by classical electromagnetism.

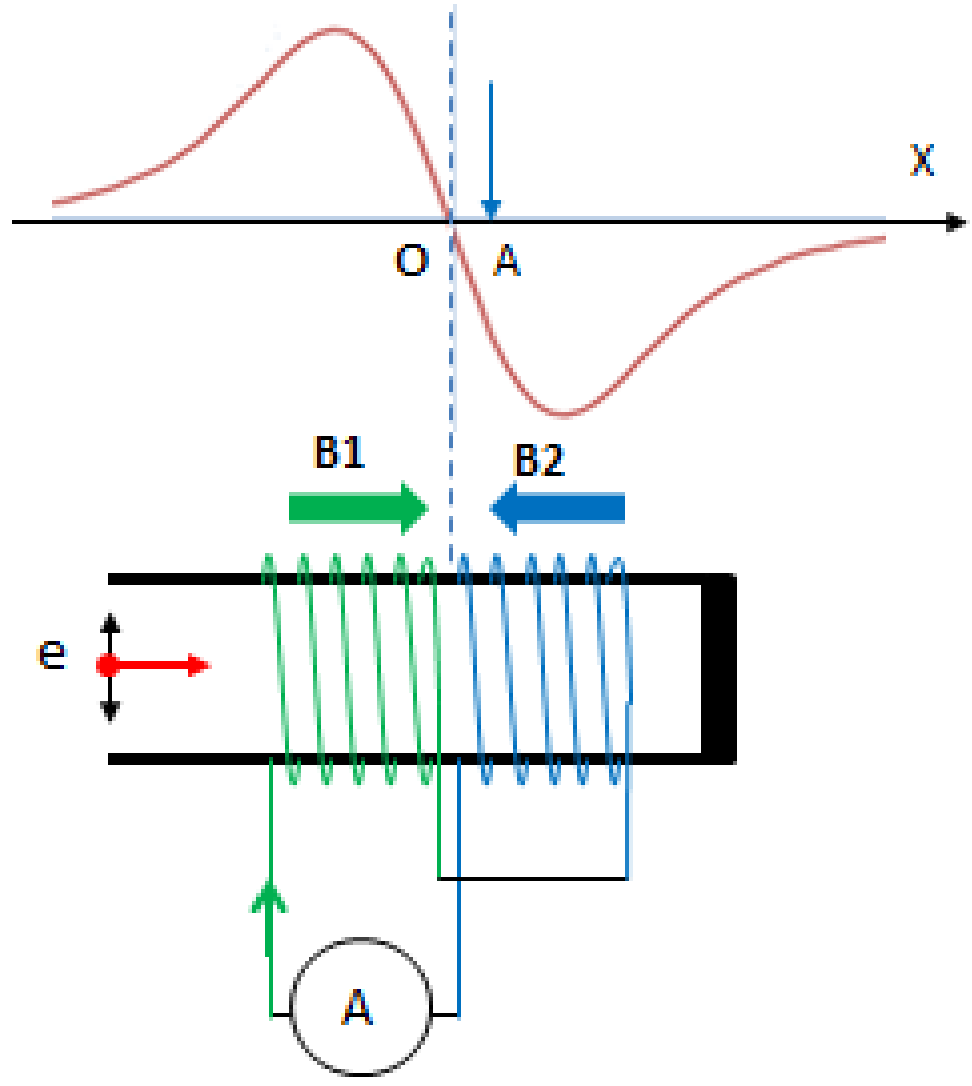

**Fig-A27** Electrons pass through odd symmetrical magnetic field

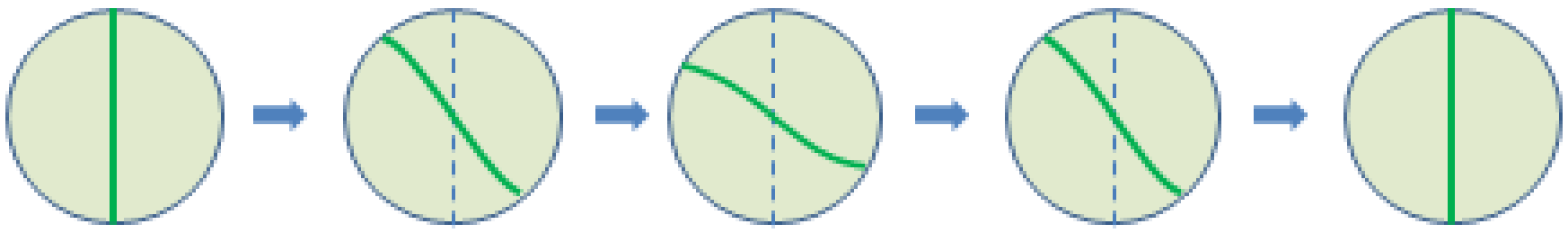

**Fig-A28** Change of line pattern on phosphor screen

According to mechanical model of U-particle in this paper, essence of Lorentz force is Magnus effect of electrons with both translational velocity and rotation on its axis in U-particle environment. In magnetic field, the electron will rotate on its axis, and the rotating electron has an inertial effect. Therefore, it can be predicted that if cathode ray passes through the odd symmetric axial magnetic field from left to right, when the screen is at **point A** in Fig-A27, inclination angle $\psi(x)$ of tangent of the line pattern will be maximum. This result, which is contradictory to classical electromagnetism, is due to the inertial effect of the electron rotating on its axis in magnetic field.

## 9 Experiment

The results of the actual measurement using cathode-ray tube are consistent with the prediction. Please refer to the paper “Anomalous deflection of an electron after passing through magnetic-field zero point” for details.

This paper is translated from Chinese into English with translation software.
Changgen Zou, April 2023@ Nanjing, China.

## Appendix B: Mathematical calculation of random collision of U-particle

First, simply calculate the random collision of ordinary particles.

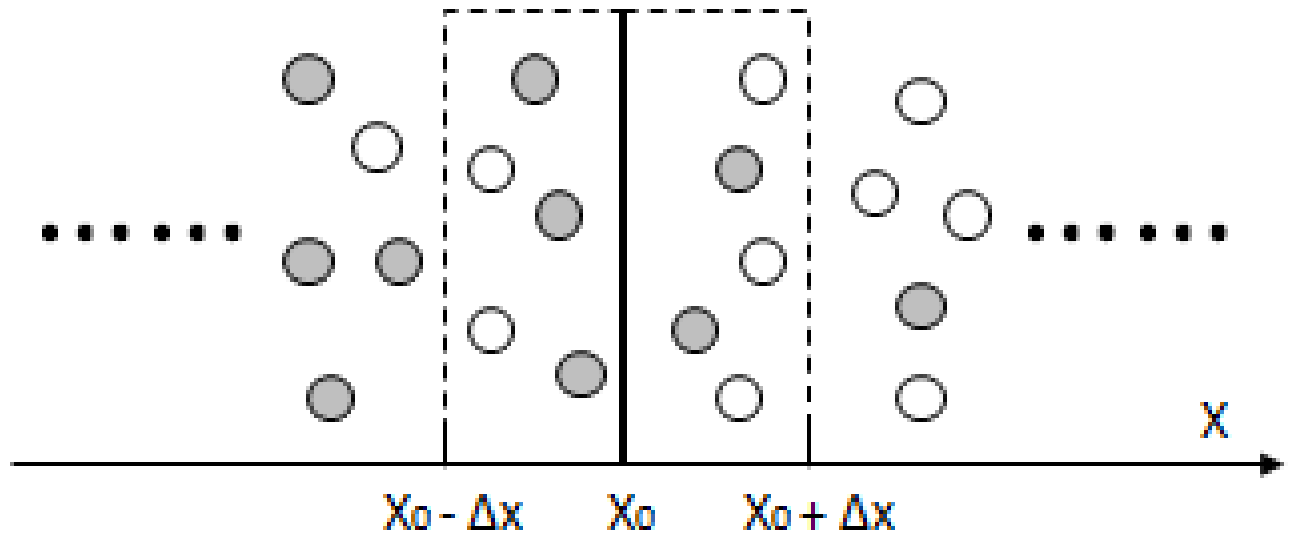

**Fig-B1** Random collision of small balls

As shown in Fig-B1, unit price of a gray ball is $x$ and unit price of a white ball is $y$. These balls have no other differences except for color and unit price. They diffuse randomly like gas molecules. Grey balls and white balls collide constantly, assuming that (1) their motion speed between two collisions is $v$ and distance is $\Delta x$, (2) the sum of quantity densities of gray and white balls is a constant $\rho_N$.

Take a virtual vertical interface at the coordinate $x_0$ on X-axis, which is a rectangle with area of $S$. Take cuboids with a thickness of $\Delta x$ on the left and right sides of the interface, and there are $\Delta x S \rho_N$ small balls in each cuboid. The opportunities of small balls moving in six directions in space are equal, so within $\Delta t = \Delta x / v$ time interval, $\Delta x S \rho_N / 6$ small balls from the left cuboid pass through the interface and enter the right cuboid, while an equal number of small balls from the right cuboid pass through the interface and enter the left cuboid.

Assuming that the proportion of gray balls in the left cuboid is $p$, and the proportion of gray balls in the right cuboid is $q$, and $p > q$, $x > y$, the price flux per unit time passing through the interface from left to right is

$$\emptyset = \frac{\{[px + (1-p)y] - [qx + (1-q)y]\} * \Delta x S \rho_N / 6}{\Delta t}$$

$$= \frac{\Delta x v}{3} * \frac{[px + (1-p)y]\rho_N - [qx + (1-q)y]\rho_N}{2\Delta x} * S$$

Where $[px + (1-p)y]\rho_N$ is the price density $\rho_L$ of the cuboid on the left side of the interface, and $[qx + (1-q)y]\rho_N$ is the price density $\rho_R$ of the cuboid on the right side of the interface, so

$$\emptyset = \frac{\Delta x v}{3} * \frac{\rho_L - \rho_R}{2\Delta x} * S = -\frac{\Delta x v}{3} * \frac{\rho_R - \rho_L}{2\Delta x} * S$$

When $\Delta x$ is very small, $(\rho_R - \rho_L)/2\Delta x$ is the price density gradient $\partial \rho / \partial x$ at $x = x_0$. Take $D = \Delta x v / 3$, then

$$\emptyset = -D * \frac{\partial \rho}{\partial x} * S$$

Therefore, for diffusion caused by random motion, the price flux passing through the interface per unit time is equal to the gradient of the price density multiplied by the area and then multiplied by a proportional coefficient. The diffusion flux per unit time is proportional to the density gradient and area, which is manifestation of mathematical theorems in physical process, it is universal. The above equation is completely consistent with Fick's first diffusion law, which also indicates that Fick's first diffusion law is suitable for diffusion motion of all microscopic particles, including U-particles. When the interface is not perpendicular to the gradient, the above eqation is $\emptyset = -D * \boldsymbol{\nabla}\rho \cdot d\boldsymbol{S}$ .

**B-1**: An isolated static electron, decreasing function of rotational kinetic energy $E_R(r)$ of Up is $E_R(r) = K_U E_U * R/r$, $r$ is the distance between the Up and centre of the electron, $K_U$ is a constant related to diffusion motion of U-particle and $K_U \ll 1$. The flux of rotational kinetic energy $E_R(r)$ per unit time is proportional to gradient of rotational kinetic energy density $\boldsymbol{\nabla}\rho_{ER}(r)$ times area.

Explanation: According to mechanical model of U-particle, a Ue becomes Up upon collision with an electron, while a Up remains Up. The Up is gradually far away from the centre of the electron after numerous collisions, and rotational kinetic energy of Up is gradually decreased to zero. This is equivalent to diffusion of rotational kinetic energy of Up from surface of the electron. Since there is no loss of rotational kinetic energy, in the stable diffusion equilibrium state, rotational kinetic energy diffused from a fixed closed surface is constant.

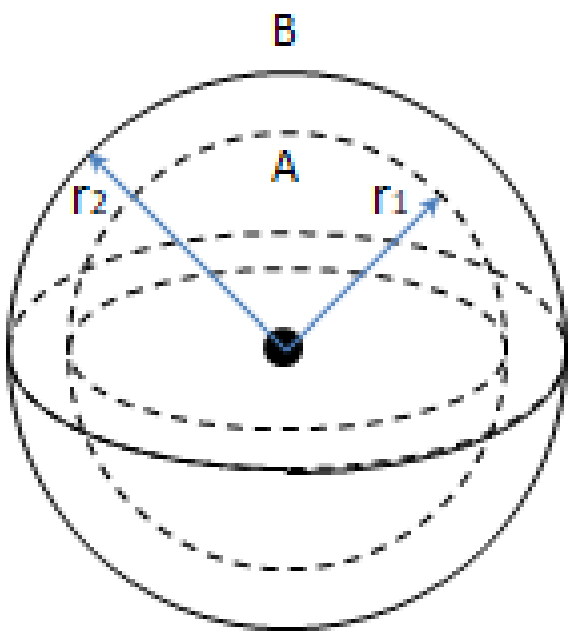

**Fig-B2** Diffusion equilibrium

As shown in Fig-B2, it is assumed that there are two spheres A and B with radii of $r_1$ and $r_2$ outside the electron, because rotational kinetic energy flowing out from surface A is equal to rotational kinetic energy flowing into surface B in the diffusion equilibrium state, according to Fick's first diffusion law, the flux of rotational kinetic energy $E_R(r)$ per unit time is proportional to gradient of rotational kinetic energy density $\boldsymbol{\nabla}\rho_{ER}(r)$ times area, so

$$\left.\frac{d\rho_{ER}(r)}{dr}\right|_{r1} * 4\pi r_1^2 = \left.\frac{d\rho_{ER}(r)}{dr}\right|_{r2} * 4\pi r_2^2 = Constant$$

$$\frac{d\rho_{ER}(r)}{dr} = \frac{Constant}{4\pi r^2}$$

Solving the equation and substituting two initial conditions, $\rho_{ER}(r) = \rho_{E0}$ when $r$ is radius $R$ of electron and $\rho_{ER}(r) = 0$ when $r$ is infinite, we can get the following results

$$\rho_{ER}(r) = \rho_{E0} * \frac{R}{r}$$

Suppose that the rotational kinetic energy diffused by a stationary electron per unit time is N times $E_U$. According to Fick's first diffusion law, the diffusion flux is equal to density gradient multiplied by diffusion area multiplied by proportional coefficient $D$. Calculate using spherical coordinates

$$N * E_U = -D * \oiint \boldsymbol{\nabla}\rho_{ER}(r) \cdot d\boldsymbol{S} = D \oiint \frac{\rho_{E0}R}{r^2} * r^2 \sin\theta \, d\theta d\varphi = D\rho_{E0}R \int_0^{\pi} \sin\theta \, d\theta \int_0^{2\pi} d\varphi$$
$$= 4\pi RD\rho_{E0}$$

So

$$\rho_{E0} = \frac{NE_U}{4\pi RD}$$

Take the scale coefficient $K_U$ as

$$K_U = \frac{N}{4\pi RD\rho_N}$$

Then

$$\rho_{E0} = \frac{N}{4\pi RD\rho_N} * E_U\rho_N = K_U E_U \rho_N$$

So

$$\rho_{ER}(r) = K_U * E_U\rho_N * \frac{R}{r} = \rho_N * E_R(r)$$

And therefore

$$E_R(r) = K_U E_U * \frac{R}{r}$$

This can be understood as follows: a stationary electron transmits rotational kinetic energy at a rate of $N * E_U$ per unit time. The rotational kinetic energy density $\rho_{ER}(r)$ of Up surrounding the electron decreases inversely with distance $r$ from center of the electron. When $r = R$, rotational kinetic energy density is $K_U E_U \rho_N$. $\rho_N$ is constant, so rotational kinetic energy $E_R(r)$ of Up decreases inversely with distance $r$ from center of the electron. When $r = R$, rotational kinetic energy is $K_U E_U$. $K_U$ must be far less than 1, otherwise $M = 1/K_U$ is not large enough, $M$ electrons can make the rotational kinetic energy density of Up close to $E_U\rho_N$. For example, if $K_U = 0.1$, then 100 electrons gathered together can make the rotational kinetic energy density of Up equal to $10E_U\rho_N$, which will cause the rotational kinetic energy of Up to be higher than $E_U$.

**B-2**: The relationship between velocity $\boldsymbol{v}_i$ of an electron moving at a low speed, macro velocity $\boldsymbol{v}_U$ of Up and distance $r$ between the Up and centre of the electron is $\boldsymbol{v}_U(r) = K_U\boldsymbol{v}_i * R/r$ .

Explanation: An electron moves at a constant velocity $\boldsymbol{v}_i$. A Ue becomes Up upon collision with the electron, while a Up remains Up. After this Up collides with other U-particles, macro momentum $\boldsymbol{P}_U$ of the Up changes. Since $v_i$ is much less than the speed V of U-particle random collision, where V is close to the speed of light, hence, $\boldsymbol{P}_U$ diffuses isotropically in space depending on the random collision of U-particle, $\boldsymbol{P}_U$ decreases gradually to zero with the increase of the distance $r$ between the Up and the electron. Take a curved surface Σ which is fixed with relative position of the electron and encloses the electron, when the electron moves uniformly, the macro momentum diffused from the closed surface Σ is a fixed value. For $v_i \ll C_U$, the macroscopic momentum transmitted by an electron per unit time is approximately $N * M_U\boldsymbol{v}_i$. Thus, $\boldsymbol{P}_U$ diffuses in the same way as $E_R$ from a stationary electron, therefore, the macro momentum density $\boldsymbol{\rho}_{PU}(r)$ of Up satisfies

$$\boldsymbol{\rho}_{PU}(r) = \boldsymbol{\rho}_{P0} * \frac{R}{r}$$

$\boldsymbol{\rho}_{P0}$ represents the value of $\boldsymbol{\rho}_{PU}(r)$ when $r$ is electron radius R. Similar to B-1, the macro momentum diffused by a moving electron per unit time is N times $M_U\boldsymbol{v}_i$, so

$$N * M_U v_i = -D * \oiint \boldsymbol{\nabla}\rho_{PU}(r)\cdot d\boldsymbol{S} = D \oiint \frac{\rho_{P0}R}{r^2} * r^2 \sin\theta \, d\theta d\varphi = D\rho_{P0}R\int_0^{\pi} \sin\theta \, d\theta \int_0^{2\pi} d\varphi$$
$$= 4\pi R D\rho_{P0}$$

So

$$\rho_{P0} = \frac{NM_U v_i}{4\pi RD} = \frac{N}{4\pi RD\rho_N} * M_U v_i \rho_N = K_U v_i \rho$$

$$\boldsymbol{\rho}_{PU}(r) = K_U \boldsymbol{v}_i \rho * \frac{R}{r} = \boldsymbol{v}_U(r) * \rho$$

So

$$\boldsymbol{v}_U(r) = K_U \boldsymbol{v}_i * \frac{R}{r}$$

Suppose that when an electron moves at a low speed and the speed changes slowly, the above equation is still approximately correct. The result of a proton is the same.

**B-3**: The mass and radius of a white ball and a grey ball are both $m$ and $r$. As shown in Fig-B3, the sphere with centre of O and radius of R is surrounded by static white balls. $R \gg r$. Grey balls are uniformly distributed in space outside the sphere O. The probability of grey balls moving in all direction is equal, but the average velocity is vector $\boldsymbol{v}$, that is, the macro velocity of grey ball is $\boldsymbol{v}$. The gap between white balls is very small, and grey ball has no chance to directly collide with the sphere O. A grey ball collides with a white ball, after each collision, the white ball enters the sphere O. The number of collisions between grey balls and white balls in unit time and unit area on surface of sphere O is a fixed value N. Under the above assumptions, white balls is randomly collided by grey balls and then enter the sphere O, the macro momentum entering the sphere O in unit time is $\emptyset$. If there is no white ball in the above assumptions, and the sphere O that grey balls random collide with is a virtual sphere O, when grey balls collide with the virtual sphere, they enter directly the virtual sphere O, the macro momentum entering the virtual sphere O in unit time is $\emptyset_V$, then $\emptyset = \emptyset_V/6$ .

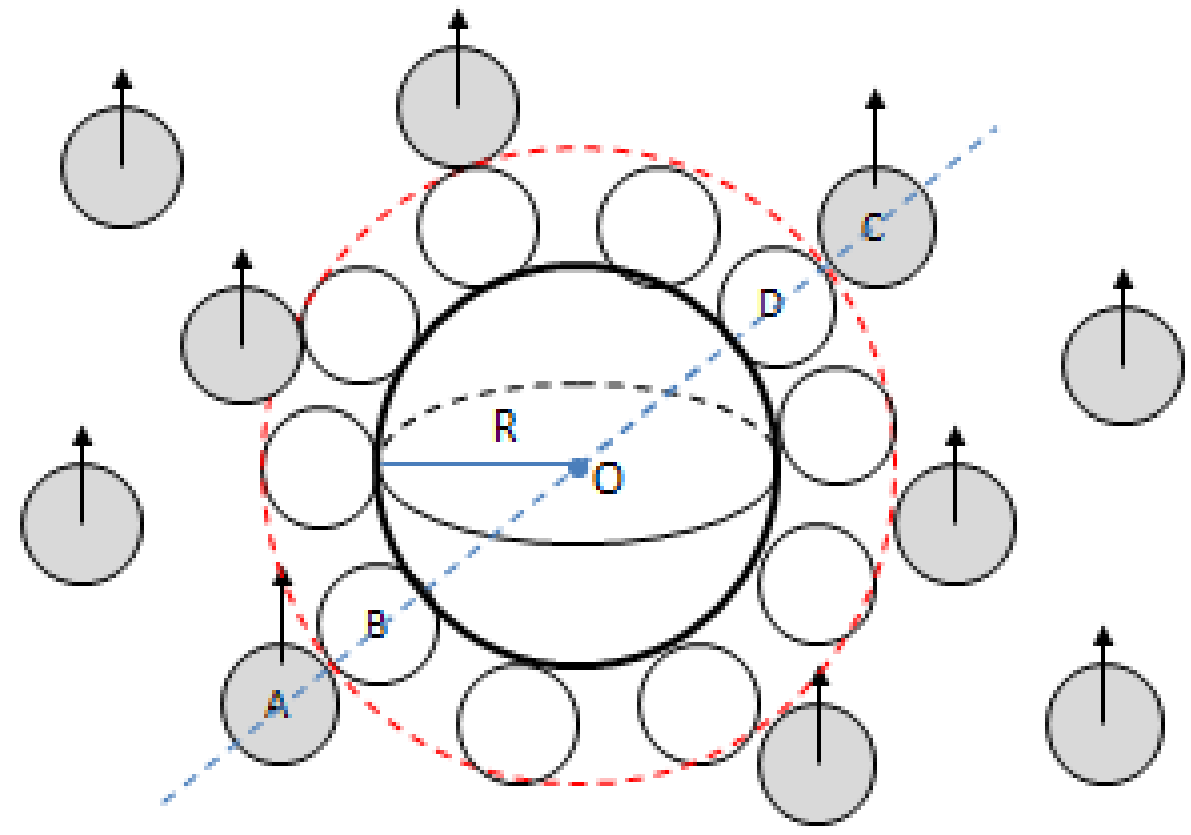

**Fig-B3** Grey and white balls outside sphere O

Explanation: As shown in Fig-B3, macro velocity $\boldsymbol{v}$ of grey balls is upward. Enlarge geometric size of white ball B to radius of $(R + 2r)$, and surface of enlarged white ball B is shown as the dotted line in Fig-B3. Since the probability of grey balls moving in all direction is equal, when $R \gg r$, component values of momentum after collision in different directions can be calculated by the white ball with the same mass but enlarged geometric size.

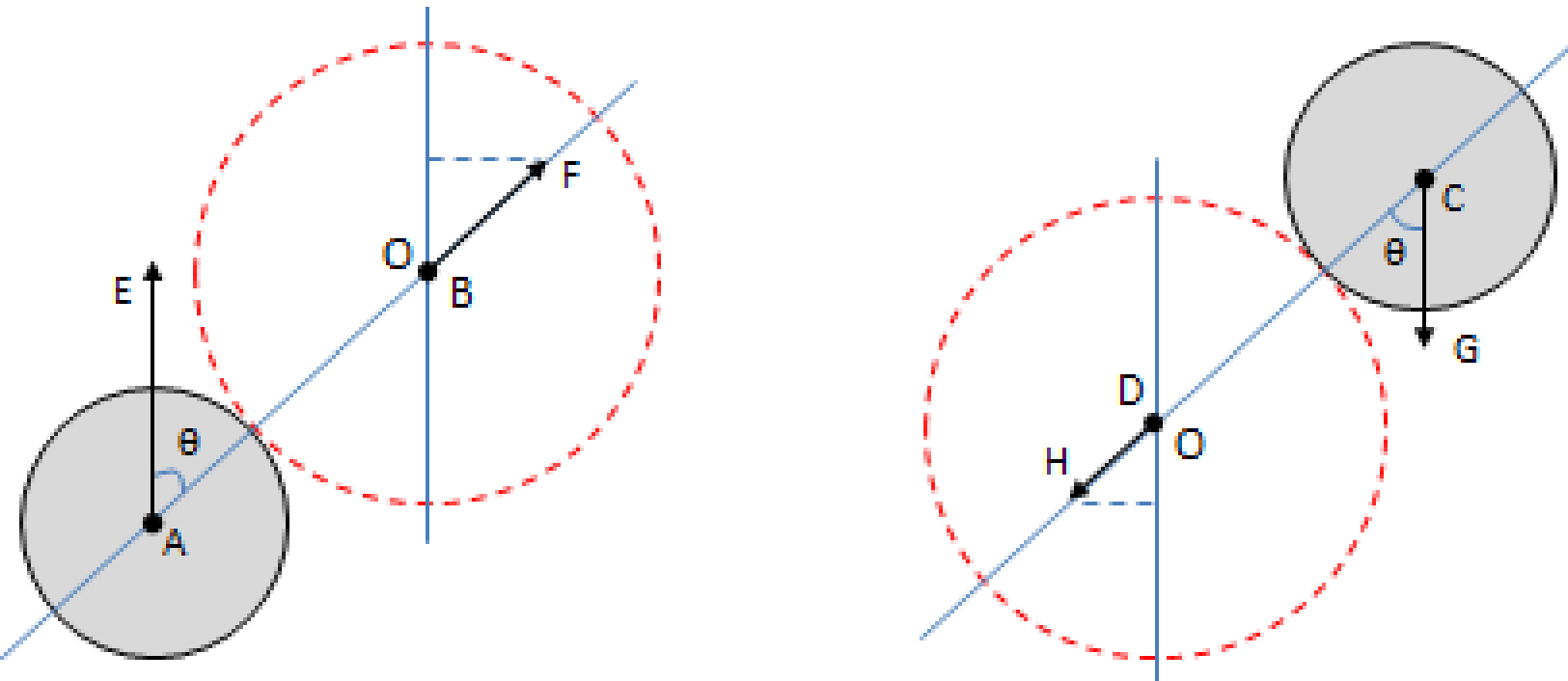

**Fig-B4** Collision between a grey ball and the white ball enlarged

Suppose that the grey ball with centre of A collides with the enlarged white ball B with centre of O, and point B overlaps with point O, as shown in Fig-B4 left, before collision, velocity of grey ball A is $AE$, and white ball B is stationary, and angle between AE and AB is $\theta$, then, after collision, velocity of white ball B is $BF = AE * cos\,\theta$, and the upward component of BF is $BF * cos\,\theta = AE * cos^2\theta$. Since the probability of grey balls moving in all directions is equal, it is always possible to find a grey ball C colliding with a white ball D on the extension line of AB. Grey ball with centre of C collides with the enlarged white ball whose centre is D, as shown in Fig-B4 right, before collision, velocity of grey ball C is $CG$, and white ball D is stationary, and angle between CG and CD is $\theta$, after collision, velocity of white ball D is $DH = CG * cos\,\theta$, and the downward component of DH is $DH * cos\,\theta = CG * cos^2\theta$. After collision between grey ball A and white ball B, grey ball C and white ball D, the sum of upward momentum of white ball B and white ball D is

$$AE * cos^2\theta - CG * cos^2\theta = (AE - CG) * cos^2\theta = v * cos^2\theta$$

Suppose that the lower hemisphere surface of sphere O is Σ, the white ball enters the spherical O after each collision between the grey ball and the white ball on surface of the spherical O. Calculated in spherical coordinates, the macro momentum $\emptyset$ entering the sphere O in unit time is

$$\emptyset = \iint_{\Sigma} m * v\,cos^2\theta * NdS = \int_0^{2\pi} d\varphi \int_0^{\pi/2} mv\,cos^2\theta * NR^2 \sin\theta\, d\theta$$

$$= 2\pi R^2 Nmv \int_0^{\pi/2} cos^2\theta \sin\theta\, d\theta = \frac{2\pi R^2 Nmv}{3}$$

If there is no white ball on surface of the virtual sphere O, after the grey ball randomly collides with the virtual sphere O, it enters directly the virtual sphere O, then, the macro momentum $\emptyset_V$ entering the virtual sphere O in unit time is

$$\emptyset_V = \oiint mv * NdS = 4\pi R^2 Nmv$$

So

$$\frac{\emptyset}{\emptyset_V} = \frac{2\pi R^2 Nmv}{3 * 4\pi R^2 Nmv} = \frac{1}{6}$$

That is, the momentum entering the sphere O in unit time after random collision between grey balls and white balls is equal to 1/6 of the momentum of grey balls entering the virtual sphere O randomly in unit time.

**Appendix C: Three dimensional coordinate system and simplified operation of Hamilton operator**

Explanation: the spatial structure of three-dimensional rectangular coordinate system$(x, y, z)$, cylindrical coordinate system $(r, \varphi, z)$, and spherical coordinate system$(r, \theta, \varphi)$ used in this paper is shown in Fig-B.

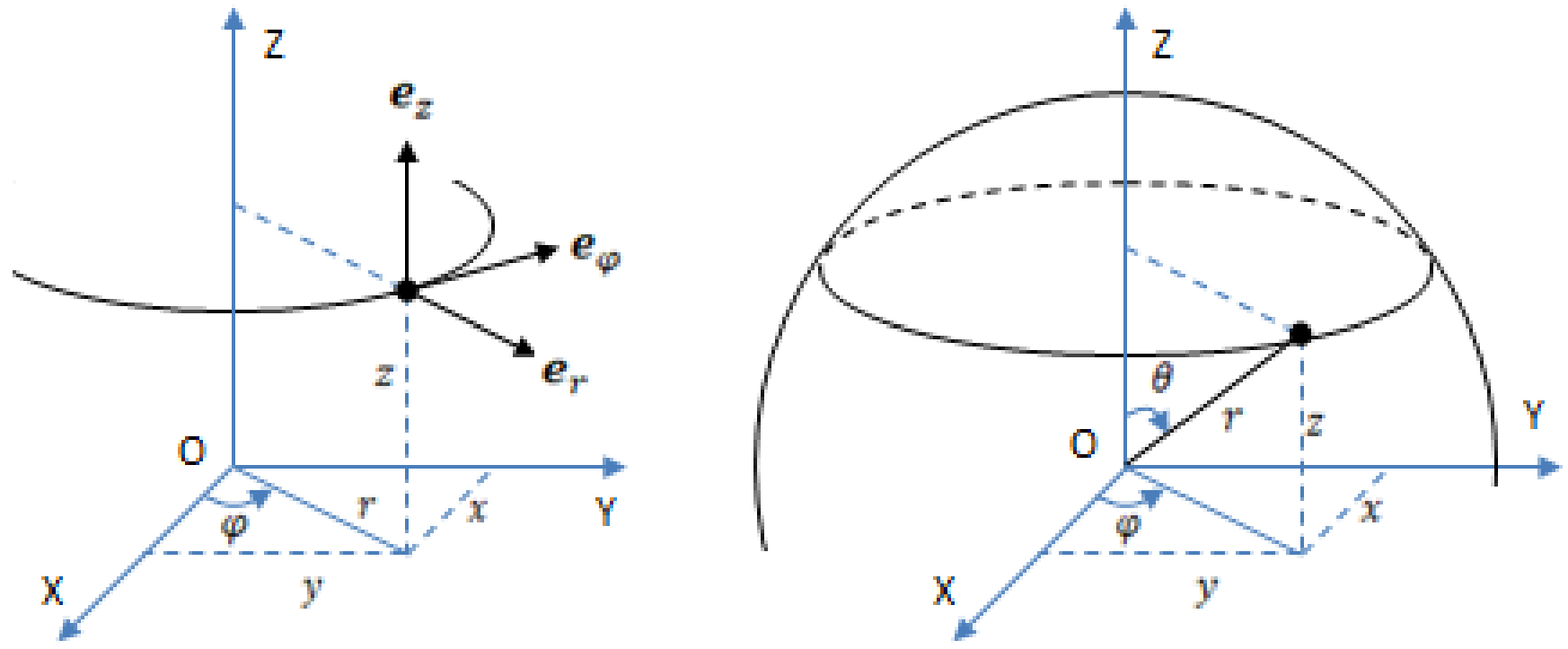

**Fig-C** cylindrical coordinate and spherical coordinate

Vector of cylindrical coordinate system

$$\boldsymbol{F} = F_r(r, \varphi, z) * \boldsymbol{e}_r + F_\varphi(r, \varphi, z) * \boldsymbol{e}_\varphi + F_z(r, \varphi, z) * \boldsymbol{e}_z$$

$$\boldsymbol{\nabla} \cdot \boldsymbol{F} = \frac{1}{r} * \frac{\partial (r * F_r)}{\partial r} + \frac{1}{r} * \frac{\partial F_\varphi}{\partial \varphi} + \frac{\partial F_z}{\partial z}$$

$$\boldsymbol{\nabla} \times \boldsymbol{F} = \frac{1}{r} * \begin{vmatrix} \boldsymbol{e}_r & r * \boldsymbol{e}_\varphi & \boldsymbol{e}_z \\ \frac{\partial}{\partial r} & \frac{\partial}{\partial \varphi} & \frac{\partial}{\partial z} \\ F_r & r * F_\varphi & F_z \end{vmatrix}$$

When $F_r(r, \varphi, z) = 0$ and $F_\varphi(r, \varphi, z) = 0$, the calculation of $\boldsymbol{\nabla} \cdot \boldsymbol{F}$ in cylindrical coordinate system can be simplified as follows

$$\boldsymbol{\nabla} \cdot \boldsymbol{F} = \frac{\partial F_z(r, \varphi, z)}{\partial z}$$

When $F_r(r, \varphi, z) = 0$ and $F_\varphi(r, \varphi, z) = 0$ and $F_z(r, \varphi, z)$ is independent of $\varphi$, the calculation of $\boldsymbol{\nabla} \times \boldsymbol{F}$ in cylindrical coordinate system can be simplified as follows

$$\boldsymbol{\nabla} \times \boldsymbol{F} = \frac{1}{r} * \begin{vmatrix} \boldsymbol{e}_r & r * \boldsymbol{e}_\varphi & \boldsymbol{e}_z \\ \frac{\partial}{\partial r} & \frac{\partial}{\partial \varphi} & \frac{\partial}{\partial z} \\ 0 & 0 & F_z \end{vmatrix} = \frac{1}{r} * \frac{\partial F_z}{\partial \varphi} * \boldsymbol{e}_r - \frac{1}{r} * r * \boldsymbol{e}_\varphi * \frac{\partial F_z}{\partial r} = 0 - \frac{\partial F_z}{\partial r} * \boldsymbol{e}_\varphi$$

So

$$\boldsymbol{\nabla} \times \boldsymbol{F} = -\frac{\partial F_z(r, \varphi, z)}{\partial r} * \boldsymbol{e}_\varphi$$

When $F_r(r, \varphi, z) = 0$ and $F_z(r, \varphi, z) = 0$ and $F_\varphi(r, \varphi, z)$ is independent of $z$, the calculation of $\boldsymbol{\nabla} \times \boldsymbol{F}$ in cylindrical coordinate system can be simplified as follows

$$\boldsymbol{\nabla} \times \boldsymbol{F} = \frac{1}{r} * \begin{vmatrix} \boldsymbol{e}_r & r * \boldsymbol{e}_\varphi & \boldsymbol{e}_z \\ \frac{\partial}{\partial r} & \frac{\partial}{\partial \varphi} & \frac{\partial}{\partial z} \\ 0 & r * F_\varphi & 0 \end{vmatrix} = \frac{1}{r} * \frac{\partial}{\partial r}\left(r * F_\varphi\right) * \boldsymbol{e}_z - \frac{1}{r} * \frac{\partial}{\partial z}\left(r * F_\varphi\right) * \boldsymbol{e}_r$$

$$= \frac{1}{r} * \frac{\partial}{\partial r}\left(r * F_\varphi\right) * \boldsymbol{e}_z$$

So

$$\boldsymbol{\nabla} \times \boldsymbol{F} = \frac{1}{r} * \frac{\partial}{\partial r}\left[r * F_\varphi(r, \varphi, z)\right] * \boldsymbol{e}_z$$

## Appendix D: Integral calculation related to electrostatic force

**D-1:**

$$F_e = \frac{\pi R^2 \rho C_U^2}{3} \int_0^\pi \left(1 + \frac{K_U R}{\sqrt{L^2 + R^2 - 2LRcos\theta}}\right) sin\,\theta\, cos\,\theta\, d\theta$$

$$= \frac{\pi R^3 K_U \rho C_U^2}{3} \int_0^\pi \frac{sin\,\theta\, cos\theta d\theta}{\sqrt{L^2 + R^2 - 2LR\,cos\,\theta}} = \frac{\pi R^3 K_U \rho C_U^2}{3} \int_{-1}^1 \frac{udu}{\sqrt{L^2 + R^2 - 2LRu}}$$

According to the integral table

$$\int \frac{xdx}{\sqrt{ax + b}} = \frac{2(ax - 2b)\sqrt{ax + b}}{3a^2} + C$$

So

$$\int \frac{udu}{\sqrt{L^2 + R^2 - 2LRu}} = \frac{2(-2LRu - 2L^2 - 2R^2)\sqrt{L^2 + R^2 - 2LRu}}{3 * 4L^2R^2} + C$$

$$= -\frac{(L^2 + R^2 + LRu)\sqrt{L^2 + R^2 - 2LRu}}{3L^2R^2} + C$$

$$\int_{-1}^1 \frac{udu}{\sqrt{L^2 + R^2 - 2LRu}} = -\frac{(L^2 + R^2 + LR)(L - R) - (L^2 + R^2 - LR)(L + R)}{3L^2R^2}$$

$$= -\frac{(L^3 - R^3) - (L^3 + R^3)}{3L^2R^2} = \frac{2R^3}{3L^2R^2} = \frac{2R}{3L^2}$$

$$F_e = \frac{\pi R^3 K_U \rho C_U^2}{3} * \frac{2R}{3L^2} = \frac{2\pi R^4 K_U \rho C_U^2}{9} * \frac{1}{L^2}$$

**D-2:**

$$hF_e = \frac{\pi R^2 \rho C_U^2}{3} \int_0^{\pi/2} \left(1 + \frac{K_U R}{\sqrt{L^2 + R^2 - 2LR\,cos\,\theta}}\right) sin\,\theta\, cos\,\theta\, d\theta$$

$$= \frac{\pi R^2 \rho C_U^2}{3} \int_0^{\pi/2} \frac{sin\,2\theta\, d\theta}{2} + \frac{\pi R^2 \rho C_U^2}{3} \int_0^{\pi/2} \frac{K_U R\, sin\,\theta\, cos\theta d\theta}{\sqrt{L^2 + R^2 - 2LR\,cos\,\theta}}$$

$$= \frac{\pi R^2 \rho C_U^2}{6} + \frac{\pi R^3 K_U \rho C_U^2}{3} \int_0^{\pi/2} \frac{sin\,\theta\, cos\theta d\theta}{\sqrt{L^2 + R^2 - 2LR\,cos\,\theta}}$$

$$= \frac{\pi R^2 \rho C_U^2}{6} + \frac{\pi R^3 K_U \rho C_U^2}{3} \int_0^1 \frac{udu}{\sqrt{L^2 + R^2 - 2LRu}}$$

According to D-1

$$\int \frac{udu}{\sqrt{L^2 + R^2 - 2LRu}} = -\frac{(L^2 + R^2 + LRu)\sqrt{L^2 + R^2 - 2LRu}}{3L^2R^2} + C$$

$$\int_0^1 \frac{udu}{\sqrt{L^2+R^2-2LRu}} = -\frac{(L^2+R^2+LR)(L-R)-(L^2+R^2)^{3/2}}{3L^2R^2}$$
$$= -\frac{(L^3-R^3)-L^3(1+R^2/L^2)^{3/2}}{3L^2R^2}$$

When L $\gg R$, according to Maclaurin's formula

$$\left(1+\frac{R^2}{L^2}\right)^{3/2} \approx 1+\frac{3R^2}{2L^2}$$

$$\int_0^1 \frac{udu}{\sqrt{L^2+R^2-2LRu}} \approx -\frac{(L^3-R^3)-(L^3+3LR^2/2)}{3L^2R^2} = \frac{R^3+3LR^2/2}{3L^2R^2} \approx \frac{3LR^2/2}{3L^2R^2} = \frac{1}{2L}$$

$$hF_e \approx \frac{\pi R^2\rho C_U^2}{6} + \frac{\pi R^3 K_U \rho C_U^2}{3} * \frac{1}{2L} = \frac{\pi R^2\rho C_U^2}{6} * \left(1+\frac{K_U R}{L}\right) \approx \frac{\pi R^2\rho C_U^2}{6}$$

**D-3:**

$$F_e = \frac{\pi R^2\rho C_U^2}{3}\int_\alpha^\pi \left(1+\frac{K_U R}{\sqrt{L^2+R^2-2LR\cos\theta}}\right)\sin\theta\cos\theta\, d\theta$$
$$= \frac{\pi R^2\rho C_U^2}{3}\left(\int_\alpha^\pi -\cos\theta * d\cos\theta + \int_\alpha^\pi \frac{-K_U R\cos\theta * d\cos\theta}{\sqrt{L^2+R^2-2LR\cos\theta}}\right)$$
$$= \frac{\pi R^2\rho C_U^2}{3}\left(\int_{\cos\alpha}^{-1} -u * du + \int_{\cos\alpha}^{-1} \frac{-K_U Ru * du}{\sqrt{L^2+R^2-2LRu}}\right)$$

Suppose $x = \cos\alpha = L/2R$, then

$$\int_{\cos\alpha}^{-1} -u * du = \frac{x^2}{2} - \frac{1}{2}$$

According to D-1

$$\int \frac{udu}{\sqrt{L^2+R^2-2LRu}} = -\frac{(L^2+R^2+LRu)\sqrt{L^2+R^2-2LRu}}{3L^2R^2} + C$$

Substitute $x = \cos\alpha = L/2R$

$$\int_{\cos\alpha}^{-1} \frac{udu}{\sqrt{L^2+R^2-2LRu}} = -\frac{L^3+R^3-(L^2+R^2+LR\cos\alpha)\sqrt{L^2+R^2-2LR\cos\alpha}}{3L^2R^2}$$
$$= -\frac{L^3+R^3-(L^2+R^2+LR*L/2R)\sqrt{L^2+R^2-2LR*L/2R}}{3L^2R^2}$$
$$= -\frac{L^3+R^3-(L^2+R^2+L^2/2)R}{3L^2R^2} = -\frac{L^3-3L^2R/2}{3L^2R^2} = \frac{1}{2R} - \frac{L}{3R^2} = \frac{1}{2R} - \frac{2x}{3R}$$

$$\int_{\cos\alpha}^{-1} \frac{-K_U Ru * du}{\sqrt{L^2+R^2-2LRu}} = -\frac{K_U}{2} + \frac{2K_U x}{3}$$

$$F_e = \frac{\pi R^2\rho C_U^2}{3}\left(\frac{x^2}{2} - \frac{1}{2} - \frac{K_U}{2} + \frac{2K_U x}{3}\right) = \frac{\pi R^2\rho C_U^2}{18}(3x^2 + 4K_U x - 3 - 3K_U)$$

## Appendix E: Calculation of the point where the axial magnetic field drops to zero

As shown in Fig-E1, an electron with a radial velocity $V_y$ travels at high speed $V_x$ from left to right through the coil-pair. Each coil of the coil-pair has radius R equal to its length L, and the coil-pair is fixed. The axial magnetic fields of the left and right coils form an odd-symmetric axial field $B_x$, and their radial fields form an even-symmetric radial field $B_r$, with $B_r = 0$ on the X-axis. The intersection of the contact plane between the two coils with the X-axis is point C, and its coordinate on the X-axis is denoted $a$. In the plane $x = a$, the axial field $B_x$ is zero, while the radial field $B_r$ varies smoothly on either side of this plane. The experiment mainly measures the deflection of cathode rays in the region near the X-axis, where $B_r$ approaches zero and the deflection of electrons is primarily dominated by $B_x$.

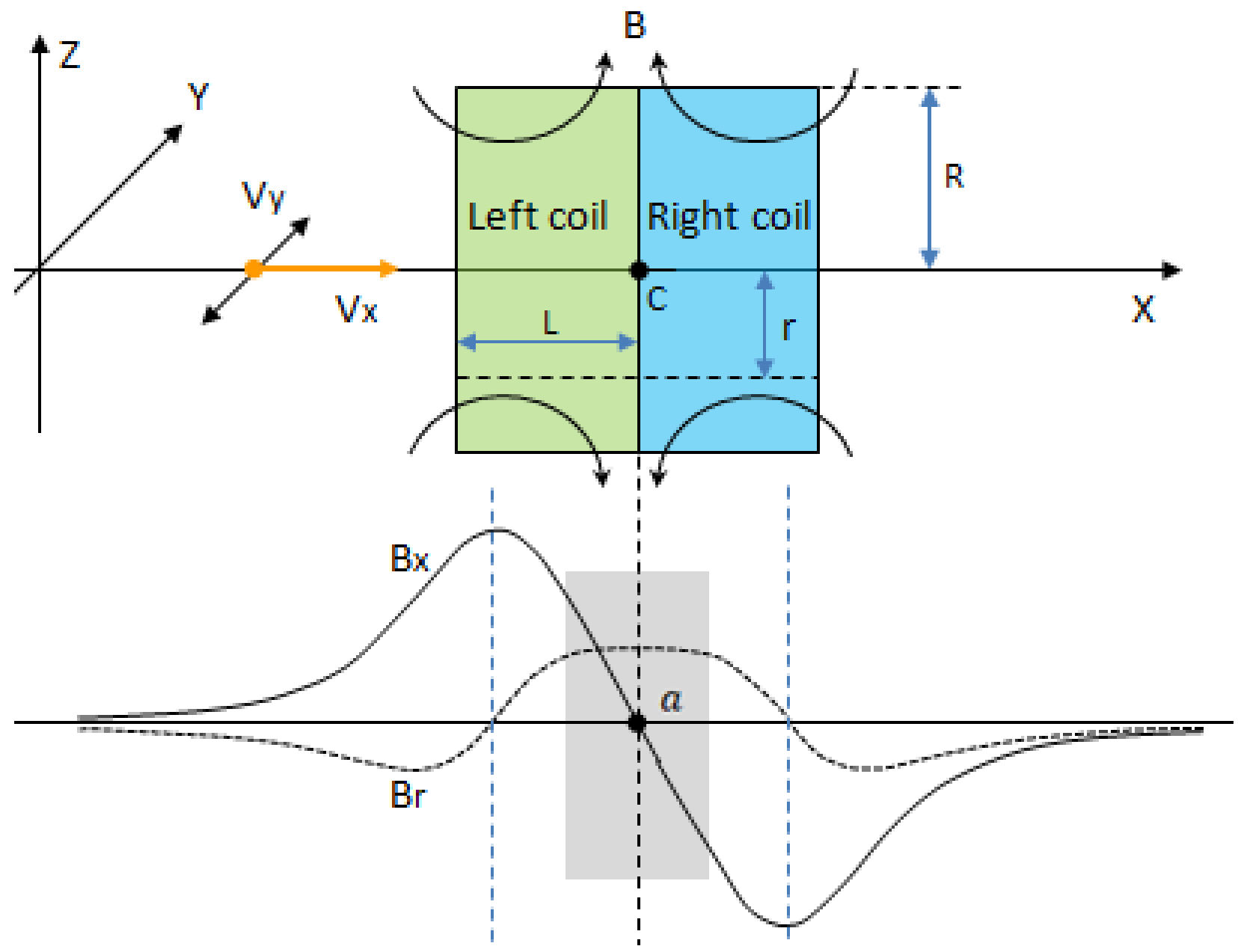

**Fig-E1** Axial and radial magnetic fields

If electrons have an alternating velocity component $V_y$ in the Y-direction, the afterglow from electrons striking the fluorescent screen forms a line pattern. Because the axial magnetic field $B_x(r)$ inside the coil-pair increases with radial distance $r$, the pattern is not a straight line but an odd-symmetric S-shaped curve. We define the fluorescent screen position as the variable x. By moving the screen to different positions $x_i$ and taking photographs, we obtain a series of S-shaped curves $z(x_i, y)$. Their slope curves are $k(x_i, y)$, as shown in Fig-E2.

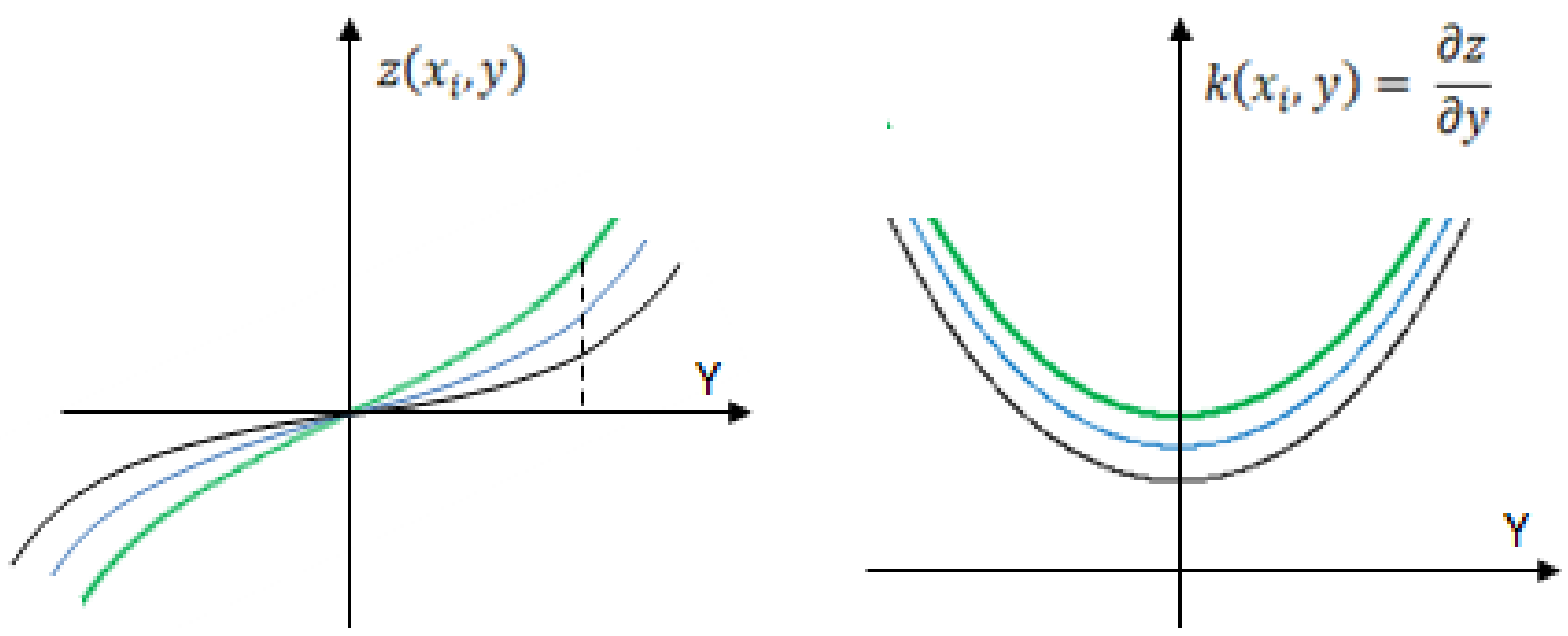

**Fig-E2** line patterns and their slope curves

Since $z(x_i, y)$ is an odd-symmetric curve, the corresponding slope curve $k(x_i, y)$ is an even-symmetric curve. These slope curves form a three-dimensional smooth surface $k(x, y) = \partial z/ \partial y$, as shown in Fig-E2. The minimum of each slope curve occurs at $(x_i, 0, k_i)$. These points form a smooth planar curve (orange) , as shown in Fig-E3. Let the maximum point of this orange curve be at $x = x_1$.

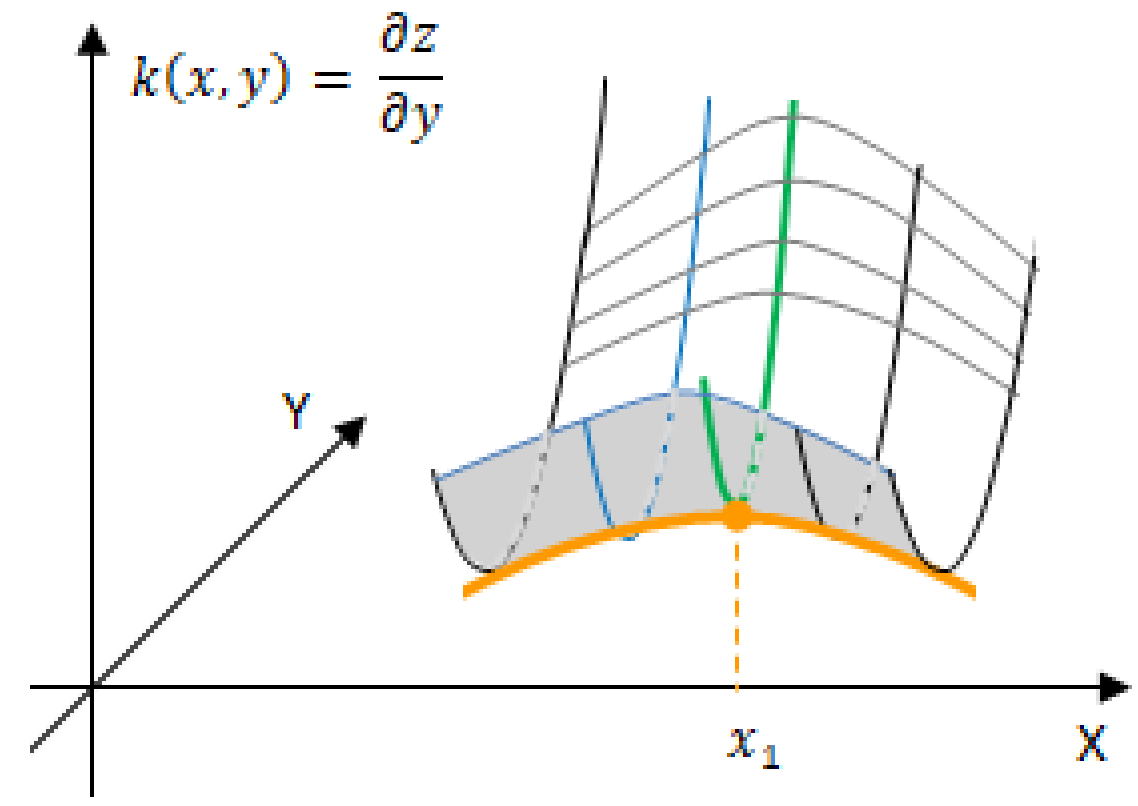

**Fig-E3** Change of slope in X-axis and Y-axis directions

The curves obtained from photographs taken at different screen positions form a three-dimensional smooth surface $z(x,y)$ that describes all electron trajectories. For any x, $z(x,y)$ is odd in y, therefore,

$$\frac{\partial^2 z}{\partial x^2}(x,0) = 0 \qquad \frac{\partial^2 z}{\partial y^2}(x,0) = 0$$

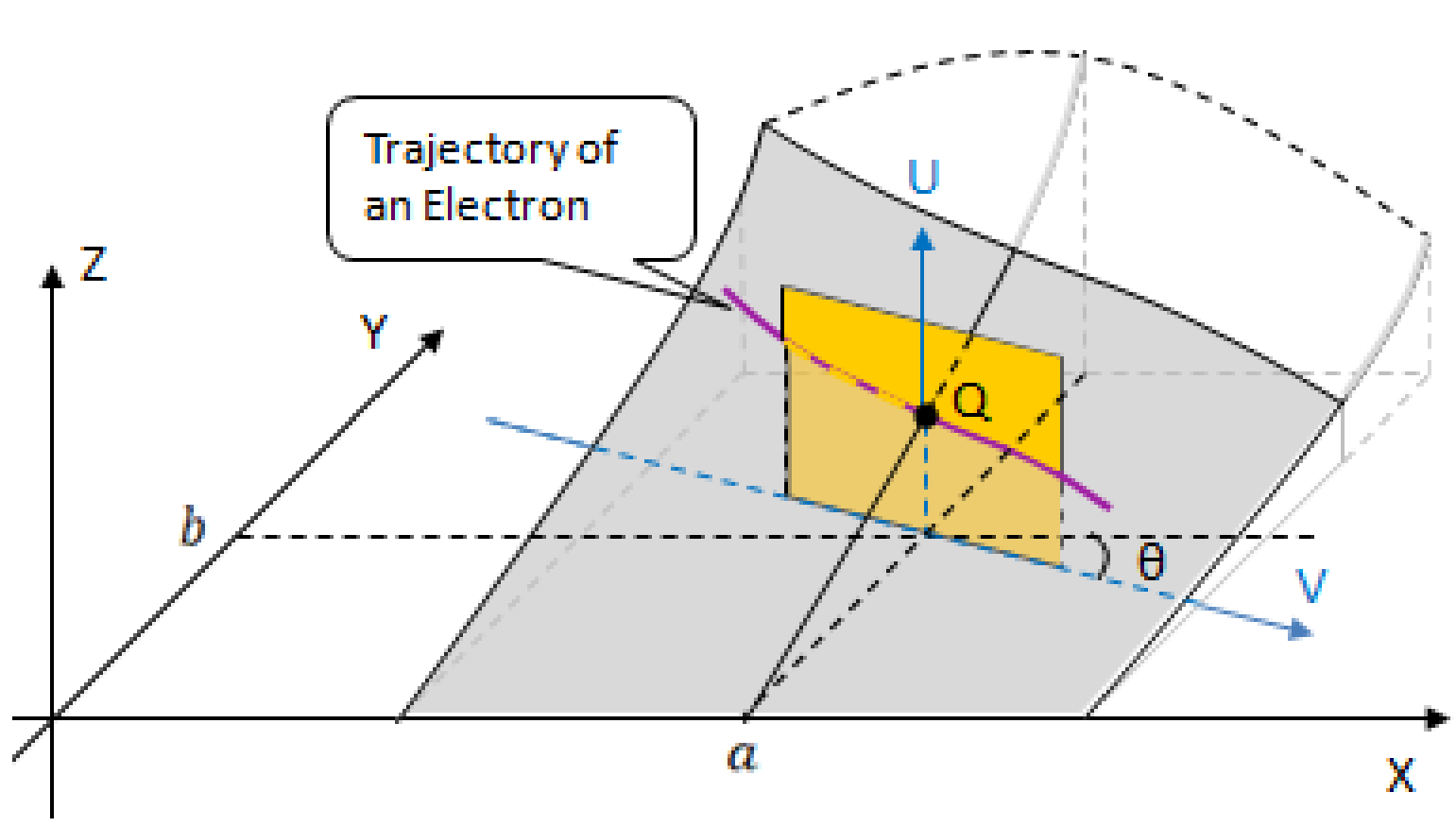

**Fig-E4** Trajectory of a single electron

Fig-E4 shows a schematic diagram of the three-dimensional smooth surface $z(x,y)$ in the region $y \geq 0$, where the purple curve represents the trajectory of a single electron. Point Q on the trajectory has coordinates $(a,b)$ in the XY plane, that is, point Q lies on the contact plane between the two coils in Fig-E1. An appropriate excitation current is selected so that, in Fig-E4, the y-coordinate of a single electron's trajectory decreases as x increases (in which case, the S-shaped curve shortens with increasing x).

Pass point Q to construct a U-axis parallel to the Z-axis. The U-axis and the velocity vector of the electron at point Q define a plane Π. The intersection of plane Π with the XY-plane is the V-axis. The projection of the electron trajectory onto the UV coordinate plane is a function $u(v)$. Let the angle between the V-axis and the X-axis be $\theta > 0$. Then in the UV coordinate plane, we have: $x = a + v\,cos\theta$, $y = b - v\,sin\theta$, $z = u$. At point Q, the magnetic field $B_x = 0$ and the Lorentz force is zero, so the electron moves in uniform linear motion, therefore, the second derivative of function $u(v)$ is zero, that is:

$$\frac{d^2u}{dv^2} = \frac{d^2z}{dv^2} = cos^2\theta \cdot \frac{\partial^2 z}{\partial x^2}(a,b) - 2\sin\theta\cos\theta \cdot \frac{\partial^2 z}{\partial x \partial y}(a,b) + sin^2\theta \cdot \frac{\partial^2 z}{\partial y^2}(a,b) = 0$$

Hence,

$$\frac{\partial^2 z}{\partial x \partial y}(a,b) = \frac{cot\,\theta}{2} \cdot \frac{\partial^2 z}{\partial x^2}(a,b) + \frac{tan\,\theta}{2} \cdot \frac{\partial^2 z}{\partial y^2}(a,b\,)$$

Taking the limit $b \to 0$ on both sides of the above equation gives:

$$\frac{\partial^2 z}{\partial x \partial y}(a,0) = \lim_{b\to 0}\left(\frac{cot\,\theta}{2} \cdot \frac{\partial^2 z}{\partial x^2}(a,b)\right) + \lim_{b\to 0}\left(\frac{tan\,\theta}{2} \cdot \frac{\partial^2 z}{\partial y^2}(a,b)\right)$$

Since

$$\lim_{b\to 0}\theta = 0 \quad \lim_{b\to 0}\frac{\partial^2 z}{\partial y^2}(a,b) = \frac{\partial^2 z}{\partial y^2}(a,0) = 0$$

We have

$$\lim_{b\to 0}\left(\frac{tan\,\theta}{2} \cdot \frac{\partial^2 z}{\partial y^2}(a,b)\right) = 0$$

In the shaded region of Fig-E1, the rotation angle of the S-shaped curve first increases and then decreases, so $z(x,b)$ is curved downward. Thus,

$$\frac{\partial^2 z}{\partial x^2}(a,b) < 0$$

Consequently,

$$\frac{\partial^2 z}{\partial x \partial y}(a,0) = \lim_{b\to 0}\left(\frac{cot\,\theta}{2} \cdot \frac{\partial^2 z}{\partial x^2}(a,b)\right) \le 0$$

From Fig-E3, $\partial z(x,0)/\partial y$ is curved downward, so $\partial^2 z(x,0)/\partial y \partial x$ transitions from positive to negative as x increases. Since the maximum occurs at $x = x_1$, we have $\partial^2 z(x_1,0)/\partial x \partial y = 0$, and therefore $x_1 \le a$. That is, the maximum point of the rotation angle of the line pattern lies to the left of the centre of the coil-pair.

**Appendix F: Calculation of partial derivatives related to induced electric field and displacement current**

According to the integral table

$$\int \frac{dx}{\sqrt{x^2 + a^2}} = ln\left(x + \sqrt{x^2 + a^2}\right) + Constant$$

So

$$v_U(r,t) = \int_{-\sqrt{C_U^2 t^2 - r^2}}^{\sqrt{C_U^2 t^2 - r^2}} \left(\frac{K_U R \lambda a t}{\sqrt{z^2 + r^2}} - \frac{K_U R \lambda a}{C_U}\right) dz$$

$$= K_U R \lambda a t * ln\frac{C_U t + \sqrt{C_U^2 t^2 - r^2}}{C_U t - \sqrt{C_U^2 t^2 - r^2}} - \frac{2K_U R \lambda a}{C_U}\sqrt{C_U^2 t^2 - r^2}$$

$$= -\frac{2K_U R \lambda a}{C_U}\left(\sqrt{C_U^2 t^2 - r^2} + C_U t\, ln\frac{C_U t - \sqrt{C_U^2 t^2 - r^2}}{r}\right)$$

Let

$$PartialE = ln\frac{C_U t - \sqrt{C_U^2 t^2 - r^2}}{r}$$

Then

$$\frac{\partial(PartialE)}{\partial r} = \frac{r}{C_U t - \sqrt{C_U^2 t^2 - r^2}} * \frac{\frac{r * r}{\sqrt{C_U^2 t^2 - r^2}} - C_U t + \sqrt{C_U^2 t^2 - r^2}}{r^2}$$

$$= \frac{1}{r\left(C_U t - \sqrt{C_U^2 t^2 - r^2}\right)} * \frac{r^2 - C_U t\sqrt{C_U^2 t^2 - r^2} + C_U^2 t^2 - r^2}{\sqrt{C_U^2 t^2 - r^2}} =$$

$$= \frac{1}{r\left(C_U t - \sqrt{C_U^2 t^2 - r^2}\right)} * \frac{C_U t\left(C_U t - \sqrt{C_U^2 t^2 - r^2}\right)}{\sqrt{C_U^2 t^2 - r^2}} = \frac{C_U t}{r\sqrt{C_U^2 t^2 - r^2}}$$

So

$$\frac{\partial v_U(r,t)}{\partial r} = -\frac{2K_U R\lambda a}{C_U}\left(\frac{-r}{\sqrt{C_U^2 t^2 - r^2}} + \frac{C_U^2 t^2}{r\sqrt{C_U^2 t^2 - r^2}}\right) = -\frac{2K_U R\lambda a}{C_U} * \frac{\sqrt{C_U^2 t^2 - r^2}}{r}$$

$$\frac{\partial v_U(r,t)}{\partial t} = \frac{-2K_U R\lambda a}{C_U}\left[\frac{C_U^2 t}{\sqrt{C_U^2 t^2 - r^2}} + C_U\, ln\frac{C_U t - \sqrt{C_U^2 t^2 - r^2}}{r} + \frac{C_U t\left(C_U - \frac{C_U^2 t}{\sqrt{C_U^2 t^2 - r^2}}\right)}{C_U t - \sqrt{C_U^2 t^2 - r^2}}\right]$$

The last item in brackets is simplified as

$$\frac{C_U^2 t\left(1 - \frac{C_U t}{\sqrt{C_U^2 t^2 - r^2}}\right)}{C_U t - \sqrt{C_U^2 t^2 - r^2}} = \frac{C_U^2 t\left(\frac{\sqrt{C_U^2 t^2 - r^2} - C_U t}{\sqrt{C_U^2 t^2 - r^2}}\right)}{C_U t - \sqrt{C_U^2 t^2 - r^2}} = \frac{-C_U^2 t}{\sqrt{C_U^2 t^2 - r^2}}$$

So

$$\frac{\partial v_U(r,t)}{\partial t} = \frac{-2K_U R\lambda a}{C_U} * C_U\, ln\frac{C_U t - \sqrt{C_U^2 t^2 - r^2}}{r} = -2K_U R\lambda a\, ln\frac{C_U t - \sqrt{C_U^2 t^2 - r^2}}{r}$$

Since

$$\frac{\partial(PartialE)}{\partial t} = \frac{C_U - \frac{C_U^2 t}{\sqrt{C_U^2 t^2 - r^2}}}{C_U t - \sqrt{C_U^2 t^2 - r^2}} = C_U\frac{1 - \frac{C_U t}{\sqrt{C_U^2 t^2 - r^2}}}{C_U t - \sqrt{C_U^2 t^2 - r^2}} = C_U\frac{\frac{\sqrt{C_U^2 t^2 - r^2} - C_U t}{\sqrt{C_U^2 t^2 - r^2}}}{C_U t - \sqrt{C_U^2 t^2 - r^2}}$$

$$= \frac{-C_U}{\sqrt{C_U^2 t^2 - r^2}}$$

So

$$\frac{\partial^2 v_U(r,t)}{\partial t^2} = -2K_U R\lambda a * \frac{\partial(PartialE)}{\partial t} = \frac{2K_U R\lambda a C_U}{\sqrt{C_U^2 t^2 - r^2}}$$